\PassOptionsToPackage{pdfpagelabels=false}{hyperref} 
\documentclass[useAMS,usenatbib]{mnras}
\pdfoutput=1
\usepackage[pdftex]{graphicx}
\usepackage{amsmath}
\usepackage{natbib}
\usepackage{txfonts}
\usepackage{times}
\usepackage{xspace}
\usepackage{bm}
\renewcommand{\vec}[1]{\mbox{\boldmath$#1$}}
\newcommand{\msun}{\,M$_{\odot}$\xspace}
\newcommand{\msunyr}{M$_{\odot}$\,yr$^{-1}$\xspace}

\volume{{\rm submitted}}

\title[TNG50: Small-scale, cold CGM gas]{Resolving small-scale cold circumgalactic gas in TNG50}
\author[D. Nelson et al.]{Dylan Nelson$^{1}$\thanks{E-mail: dnelson@mpa-garching.mpg.de},
Prateek Sharma$^{2}$, Annalisa Pillepich$^{3}$, Volker Springel$^{1}$, \newauthor
R{\"u}diger Pakmor$^{1}$, Rainer Weinberger$^{4}$, Mark Vogelsberger$^{5}$, Federico Marinacci$^{6}$, \newauthor
Lars Hernquist$^{4}$
\\\\
$^{1}$Max-Planck-Institut f\"{u}r Astrophysik, Karl-Schwarzschild-Str. 1, 85741 Garching, Germany\\
$^{2}$Department of Physics, Indian Institute of Science, Bangalore, 560012, India \\
$^{3}$Max-Planck-Institut f\"{u}r Astronomie, K\"{o}nigstuhl 17, 69117 Heidelberg, Germany\\
$^{4}$Harvard-Smithsonian Center for Astrophysics, 60 Garden Street, Cambridge, MA, 02138, USA\\
$^{5}$Kavli Institute for Astrophysics and Space Research, Department of Physics, MIT, Cambridge, MA, 02139, USA\\
$^{6}$Department of Physics and Astronomy, University of Bologna, Via Gobetti 93/2, I-40129 Bologna, Italy
}

\begin{document}

\maketitle

\begin{abstract}
We use the high-resolution TNG50 cosmological magnetohydrodynamical simulation to explore the properties and origin of cold circumgalactic medium (CGM) gas around massive galaxies ($M_\star > 10^{11}$\msun) at intermediate redshift ($z\,\sim\,0.5$). We discover a significant abundance of small-scale, cold gas structure in the CGM of `red and dead' elliptical systems, as traced by neutral HI and MgII. Halos can host tens of thousands of discrete absorbing cloudlets, with sizes of order a kpc or smaller. With a Lagrangian tracer analysis, we show that cold clouds form due to strong $\delta \rho / \bar{\rho} \gg 1$ gas density perturbations which stimulate thermal instability. These local overdensities trigger rapid cooling from the hot virialized background medium at $\sim 10^7$ K to radiatively inefficient $\sim 10^4$ K clouds, which act as cosmologically long-lived, `stimulated cooling' seeds in a regime where the global halo does not satisfy the classic $t_{\rm cool} / t_{\rm ff} < 10$ criterion. Furthermore, these small clouds are dominated by magnetic rather than thermal pressure, with plasma $\beta \ll 1$, suggesting that magnetic fields may play an important role. The number and total mass of cold clouds both increase with resolution, and the $m_{\rm gas} \simeq 8 \times 10^4$\msun cell mass of TNG50 enables the $\sim$\,few hundred pc, small-scale CGM structure we observe to form. Finally, we make a preliminary comparison against observations from the COS-LRG, LRG-RDR, COS-Halos, and SDSS LRG surveys. We broadly find that our recent, high-resolution cosmological simulations produce sufficiently high covering fractions of extended, cold gas as observed to surround massive galaxies.
\end{abstract}



\section{Introduction}

Observations as well as numerical simulations suggest that the circumgalactic medium (CGM) of galaxies is a complex, multi-phase gaseous reservoir exhibiting a rich variety of gas dynamics.

One of the most interesting puzzles of the observed CGM is the apparent abundance of cold ($\sim 10^4$\,K) gas even in the halos of massive galaxies with virial temperature exceeding $10^6$\,K \citep{prochaska13}. Observations have established the existence, and high column densities, of cold tracers including MgII and HI in samples of intermediate redshift $0.3 < z < 0.8$ luminous red galaxies (LRGs), including the COS-LRG \citep{chen18,zahedy19} and LRG-RDR \citep{berg19} surveys, as well as in the main SDSS LRG sample \citep{zhu14}. 

As these galaxies are largely passive, an origin of this cold gas in recent supernovae-driven winds is disfavored. Non-negligible covering fractions out to large $>$\,100 kpc distances \citep{lan18,lan19} imply that cold absorbing gas must either form in-situ, or else be transported to such distances. In both cases: what is the physics of the formation mechanism(s), and (how) do clouds avoid hydrodynamical disruption and survive for long timescales? 

Although hot circumgalactic gas is in a global quasi-equilibrium, kinematically and thermally balanced, it is susceptible to local perturbations. Most notably, in the form of the classic thermal instability \citep[TI;][]{field65}. This mechanism has been connected to the observed existence of cold gas in galactic halos, for quasar absorption systems \citep{mo96} as well as Milky Way high velocity clouds \citep{maller04}. Runaway local production of cold gas has been seen in simulations \citep[in][albeit with a numerical origin]{kaufmann06}, and TI has been shown to lead to the generation of multi-phase structure in idealized hot halos \citep{sharma12,mccourt12}.

In the case of a realistic circumgalactic/intracluster medium (CGM/ICM), where both gravity and hydrostatic density stratification exist, buoyancy effects couple thermal and convective instabilities \citep{balbus89,malagoli90}. The presence of a non-vanishing magnetic field, as in a realistic CGM, also changes the conditions for TI \citep{balbus91}, affects the role of thermal conduction \citep{koyama04,sharma10}, and shapes the non-linear evolution and morphology of resultant cold gas structures \citep{ji18,cottle20}.

A thermally unstable CGM leads to the concept of condensation, with cold gas arising from a background hot halo atmosphere \citep{voit15}. Such a phenomena may be intermittent in time, coupled to the self-regulation of both star formation and supermassive black hole growth \citep{gaspari17,beckmann19}. A precipitation-unstable state occurs for sufficiently short cooling timescales, relative to dynamical timescales of the halo, although runaway local cooling can occur even above the canonical thresholds of $t_{\rm cool} / t_{\rm ff} < 1 - 10$ \citep{sharma12} if sufficiently strong density perturbations arise \citep{choudhury19}.

Given that the cold-phase CGM which results from such processes is a small scale phenomenon, it has been studied largely in idealized, numerical experiments \citep{fielding20}. The cloud crushing problem -- survival of a cold cloud in relative motion with respect to a hot background media -- is often posed in the opposite direction, i.e. cold gas acceleration in an outflow \citep{scannapieco15,armillotta17,schneider18b}. However, the same physics governs the survivability of filamentary streams \citep{mandelker19} as well as the evolution of accreting, TI-induced cold clouds.

Similarly, common physics also governs cold gas shattering (or breakup) into small-scale structure below the length scale of the original cloud \citep{mccourt18,sparre19,waters19,gronke20b}. The resulting circumgalactic mist is naturally comprised of a large number of small cold cloudlets, with low volume filling factor but high covering fraction \citep{liang20}. The implied size-scale of cold gas is small, from multiply lensed quasar constraints of $\sim$ 10s pc \citep{rauch99} to the theoretically motivated, characteristic $\ell_{\rm cool} = c_{\rm s} t_{\rm cool}$ ($\sim$ pc) length scale for cold gas to shatter \citep{mccourt18}. At the same time, \cite{rubin18b} showed that the coherence scale over which cold MgII absorption in the CGM does not vary is \mbox{$\gtrsim$\,1 kpc}. Individual clouds must either exceed this size scale, or density and kinematic substructure must be coherent on larger scales.

However, fully cosmological galaxy formation simulations typically resolve the CGM only at this $\sim$ kpc scale, or worse \citep{nelson16}, making the study of cold phase gas in the CGM difficult. As a result, this regime is one of the key targets of CGM (super)refinement simulations \citep{suresh19,hummels19,peeples19,vdv19}. However, none of these efforts have yet simulated halos as massive as LRGs, due to the computational cost. Nonetheless, as we show below, the new TNG50 cosmological simulation actually has comparable, or better, gas-dynamical resolution in the CGM regime, offering a unique view on cold circumgalactic gas.

Previously, \cite{oppenheimer18c} demonstrated a relatively significant abundance of cold, low ionization gas in the CGM of EAGLE galaxies, albeit for lower mass halos and with uncertainties on the numerical robustness of the hydrodynamical technique in this regime \citep[see also][]{ford13}. Semi-analytic models have been developed to explain the unique, possibly drag-induced sub-virial kinematics of cold gas flowing through LRG host halos  \citep{afruni19}. Indeed, the strongest absorbers around passive galaxies may be directly connected to their satellite galaxies and the accretion of substructure \citep{lan20}. This underscores the importance of the cosmological context and the inseparability of circumgalactic gas from the baryon cycle of cosmic inflows and outflows, as accessible in cosmological simulations, which also capture the impact of the fluctuating dark matter potential on the thermodynamical evolution of the gaseous component.

In the context of IllustrisTNG, \cite{nelson18b} showed that TNG is consistent with empirical OVI constraints in the CGM, and the cold gas content of galaxies themselves (both neutral HI and molecular H2) has been favorably validated against data in several regimes \citep{stevens19,popping19,diemer19}, but no investigation of cold gas in the circumgalactic medium has yet been undertaken.

In this work, we use the high-resolution TNG50 volume \citep{pillepich19,nelson19b} to study the abundance, properties, and physical origin of cold CGM gas in massive, LRG-host mass halos at $z \simeq 0.5$. In Section \ref{sec_methods} we describe the simulations and our analysis methodology, while Section \ref{sec_results1} explores the abundance of cold CGM gas. We investigate the small-scale structure and properties of this cold-phase in Section \ref{sec_results2}, assess its origin and formation in Section \ref{sec_origin}, and compare to observational data in Section \ref{sec_obs}. In Section \ref{sec_conclusions} we summarize our main conclusions.


\section{Methods} \label{sec_methods}

\subsection{The TNG Simulations} \label{sec_sims}

The IllustrisTNG project\footnote{\url{http://www.tng-project.org}} \citep{pillepich18b, nelson18a, naiman18, marinacci18, springel18} is a series of three large cosmological volumes, simulated with gravo-magnetohydrodynamics (MHD) and incorporating a comprehensive model for galaxy formation physics \citep[detailed in the TNG methods papers:][]{weinberger17,pillepich18a}, which is based on improvements to the original Illustris model \citep{vog13,torrey14}. The third and final simulation of the project is the high-resolution TNG50 volume \citep{nelson19b,pillepich19}, which we use herein. All aspects of the model, including parameter values and the simulation code, are previously described and remain \textit{unchanged} for the production simulations, and we give here only a brief overview.

The TNG project includes three distinct simulation volumes: TNG50, TNG100, and TNG300. The intermediate volume \textbf{TNG100} includes 2$\times$1820$^3$ resolution elements in a $\sim$\,100 Mpc (comoving) box, while \textbf{TNG300} includes 2$\times$2500$^3$ resolution elements in a $\sim$\,300 Mpc box. Both are now publicly released in their entirety \citep{nelson19a}. The smallest volume simulation is \textbf{TNG50}, which includes 2$\times$2160$^3$ resolution elements in a $\sim$\,50 Mpc (comoving) box, giving a baryon mass resolution of $8.5 \times 10^4$\msun, a collisionless softening of 0.3 kpc at $z$\,=\,0, and a minimum gas softening of 74 comoving parsecs.

All adopt a cosmology consistent with the \cite{planck2015_xiii} results, namely $\Omega_{\Lambda,0}=0.6911$, $\Omega_{m,0}=0.3089$, $\Omega_{b,0}=0.0486$, $\sigma_8=0.8159$, $n_s=0.9667$ and $h=0.6774$. 

TNG uses the \textsc{Arepo} code \citep{spr10} which solves for the coupled evolution of dark matter, gas, stars, and black holes under the influence of self-gravity and ideal, continuum MHD \citep{pakmor11,pakmor13}. The former employs the Tree-PM approach, whereas the fluid dynamics uses a Godunov type finite-volume scheme where an unstructured, moving, Voronoi tessellation provides the spatial discretization. The simulations include a physical model for the most important processes relevant for the formation and evolution of galaxies. Specifically: (i) gas radiative processes, including primordial/metal-line cooling and heating from the background radiation field, (ii) star formation in the dense ISM, (iii) stellar population evolution and chemical enrichment following supernovae Ia, II, as well as AGB stars, with individual accounting for the nine elements H, He, C, N, O, Ne, Mg, Si, and Fe, (iv) supernova driven galactic-scale outflows or winds \citep[see][for details]{pillepich18a}, (v) the formation, coalescence, and growth of supermassive blackholes, (vi) and dual-mode blackhole feedback operating in a thermal `quasar' state at high accretion rates and a kinetic `wind' state at low accretion rates \citep[see][for details]{weinberger17}.

Dark matter halos and their substructures -- subhalos and galaxies -- are identified and characterised with the \textsc{Subfind} algorithm \citep{spr01}. Unless noted otherwise, our analysis typically includes all gas in halos, including satellite gas.

\begin{figure}
\centering
\includegraphics[angle=0,width=3.3in]{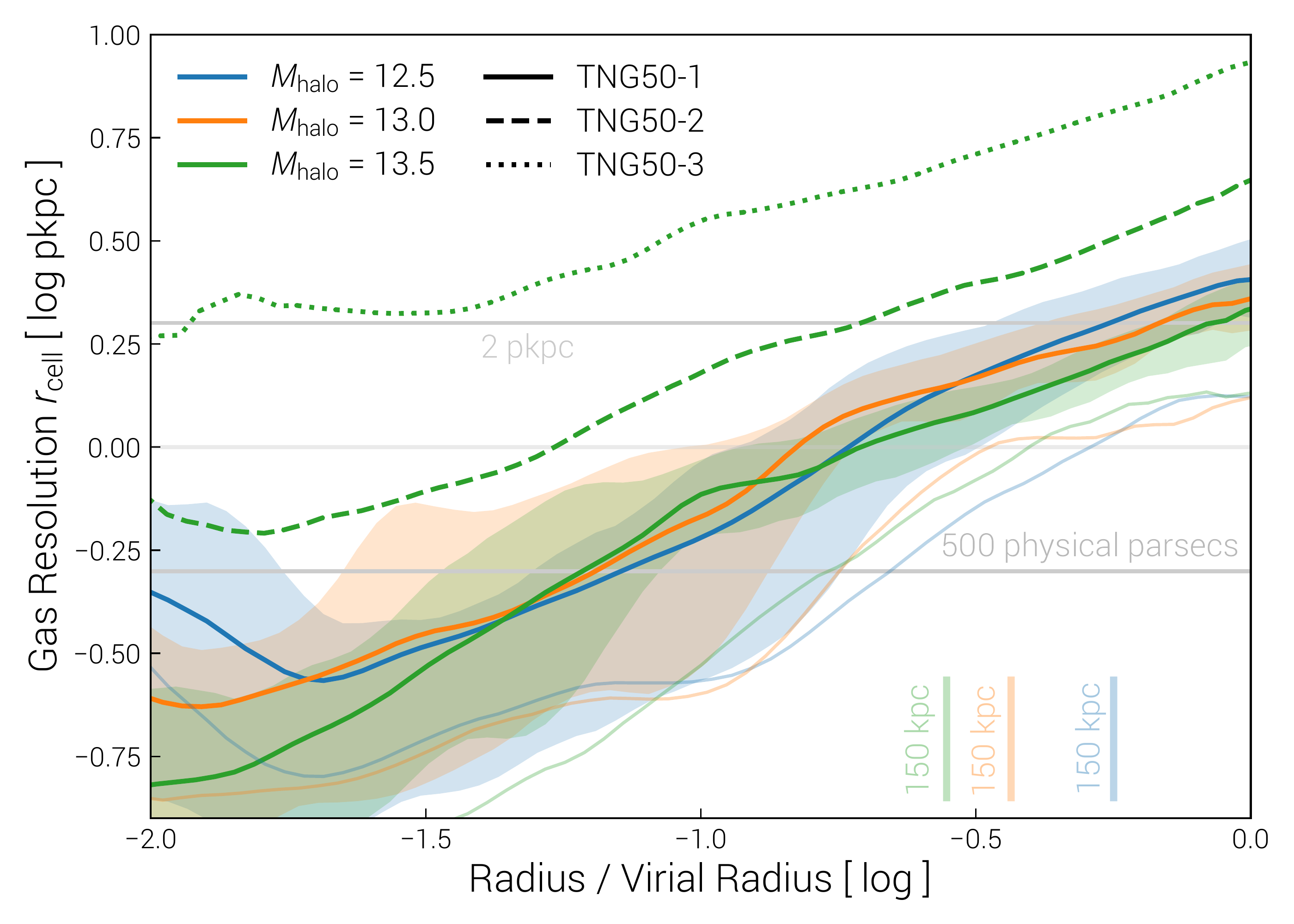}
\caption{ Spatial resolution available with the TNG50 simulation in the CGM, shown at $z\,=\,0.5$ for reference. The three colored solid lines indicate the median size of gas cells, as a function of distance, for three different halo mass bins (halo-to-halo variation is shown by 10-90th percentiles in shaded bands). All gas cells in the halo are included. For the most massive bin, we also show two progressively lower resolution simulations, TNG50-2 and TNG50-3. The resolutions of TNG100 and original Illustris are both intermediate between these two. In contrast, TNG50-1 reaches, on average, sub-kiloparsec resolution within a third of the virial radius. At any distance, dense substructures are much better resolved than this value -- the lower 10th percentile is shown by the faint solid lines.
 \label{fig_res}}
\end{figure}

The resolution we achieve with TNG50 in the circumgalactic medium is demonstrated in Figure \ref{fig_res}. The \textit{average} gas resolution is better than 1 physical kpc in the inner third of the halo, and lower outside. Furthermore, because of the quasi-Lagrangian nature of our simulation technique, denser gas is always resolved better. The lower 10th percentile (thin solid lines) show that the densest 10 percent of gas, at any distance, has a spatial resolution half an order of magnitude better than the average. As we show below, this allows us to resolve cold-phase CGM gas with a hydrodynamical resolution of $\sim$\,100$-$200 parsecs.

\subsection{Modeling Metal Ionization States} \label{sec_ions}

The technique for computing MgII follows the analysis methodology of \cite{nelson18b} previously used for oxygen ions including OVI, and we briefly review it here.

To compute MgII we take the total magnesium mass per cell as tracked during the simulation, and then derive ionization states using \textsc{Cloudy} \citep[][v13.03]{ferland13} including both collisional and photo-ionization in the presence of a UV + X-ray background \citep[the 2011 update of][]{fg09}. This makes the calculation self-consistent with the TNG simulations themselves. We use \textsc{Cloudy} in single-zone mode and iterate to equilibrium, accounting for a frequency dependent shielding from the background radiation field (UVB) at high densities, using the fitting function of \cite{rahmati13}. Any physical mechanism producing metal ions in gas structure with very small physical scales ($\ll\,$100 pc) would not be resolved in our current simulations, and not accounted for herein. We do not consider the impact of local sources of radiation beyond the UVB.

To compute neutral hydrogen we adopt the per-cell modeling of \cite{popping19} for the TNG simulations, taking the \cite{gnedin11} model to estimate and remove the contribution from molecular H$_2$, following the self-shielding approximation of \cite{rahmati13}. This provides us an estimate of the total mass, spatial distribution of that mass, and kinematics, of both neutral HI and MgII throughout the entire simulation volume.

\subsection{Measuring Column Densities} \label{sec_columns}

To obtain column densities around a simulated galaxy, ionic mass is projected onto a regular grid with a given transverse size and projection depth. Grid sizes are chosen to cover the range of observed impact parameters, and projection depths are approximately equivalent to the velocity interval over which the observational search for absorption is undertaken. For comparison to COS-LRG, the velocity search window is $\pm\,$500 km/s, corresponding to an adopted projection depth of $\sim\,$5.5 pMpc (assumed at $z\,=\,0.5$), while for LRG-RDR the velocity window is twice as large at $\pm\,$1000 km/s. All gas in the Hubble flow of that velocity interval is included. The pixel scale is always 2 physical kpc (pkpc), and projection is always along the fixed z-axis of the simulation volume (and therefore random with respect to the orientation of the galaxy).

We assume the ion mass in a given gas cell is spatially distributed following the standard cubic-spline kernel with support $h = [3V/(4\pi)]^{1/3}$ where $V$ is the volume of the Voronoi gas cell.

\subsection{Identifying Small-Scale Gas Structures} \label{sec_segmentation}

To identify, separate, count, and characterize the discrete gas structures we find in the CGM of simulated halos, we have developed a new watershed-type segmentation algorithm on the Voronoi tessellation. First, we compute the natural neighbor connectivity of the Voronoi gas cells by reconstructing the mesh from its generating sites in post-processing (following \textcolor{blue}{Byrohl et al., in prep}). A threshold on a gas property is then defined, for instance on the physical number density of MgII ions, $n_{\rm MgII} > 10^{-8}$ cm$^{-3}$ (see below). We then identify all spatially contiguous sets of naturally connected Voronoi cells which satisfy this threshold. As each Voronoi cell is a convex polyhedron, and Voronoi natural neighbors share one face, unions of naturally connected cells likewise form extended polyhedra. Each has a connected interior and boundary, forming a coherent structure defined by its constituent gas cells.

We refer to these structures are `clouds.' Every cloud has a well-defined volume, mass, position, and so on, as defined by accumulation over its member cells. By construction, all clouds are disjoint; they do not touch. Note that we do not carry out a hierarchical segmentation: each cloud is monolithic and could contain, for instance, two local density maxima, which would occur during the coalescence of two cloudlets.


\begin{figure*}
\centering
\includegraphics[angle=0,width=7.0in]{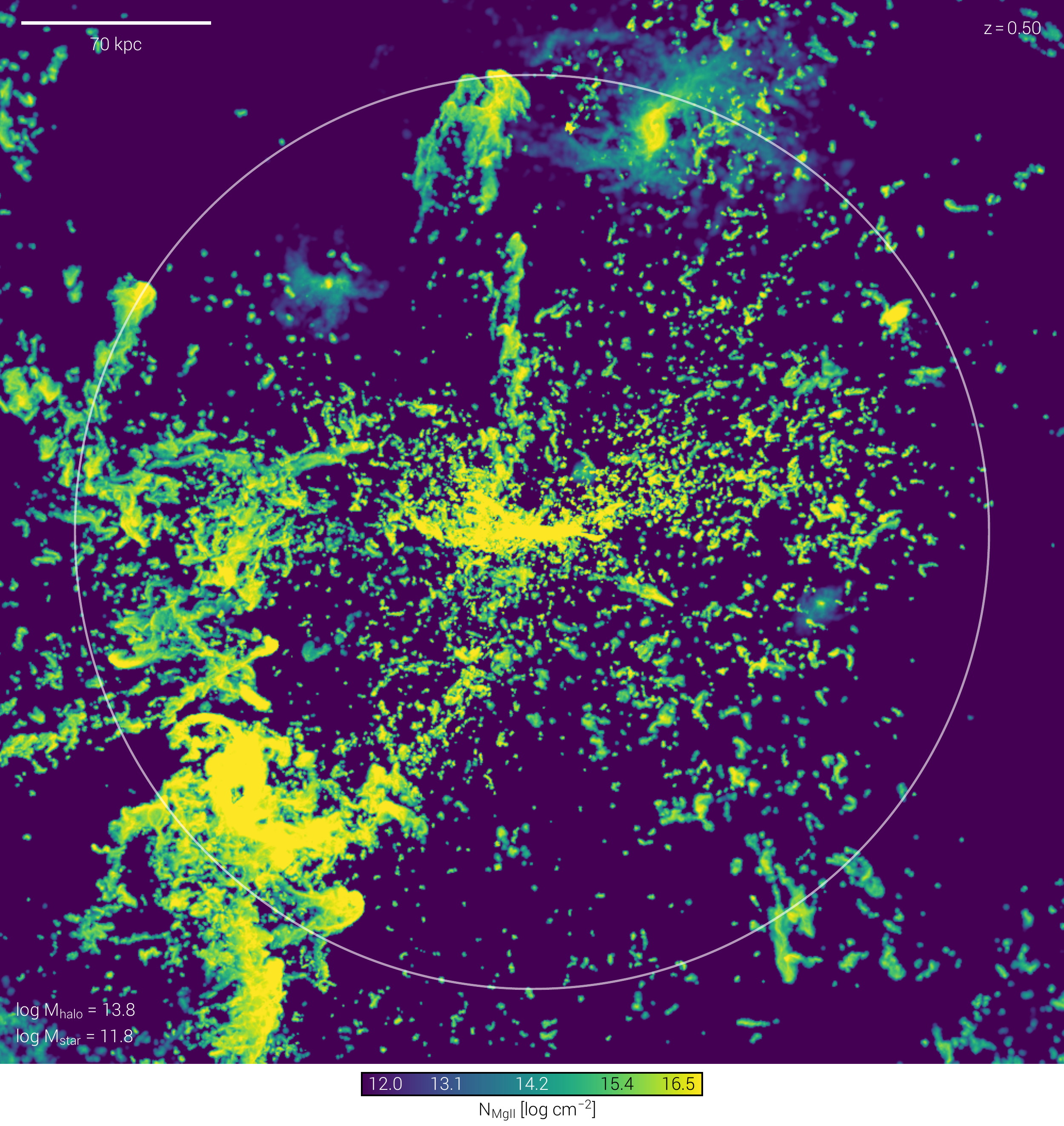}
\caption{ Visualization of the MgII column density distribution around a single massive halo at $z\,=\,0.5$ in TNG50, projected through $\Delta v = \pm\,$500 km/s. The halo has a total mass of $10^{13.8}$\msun and a stellar mass of $10^{11.8}$\msun, where a quarter of the virial radius is indicated by the white circle. In the inner halo, impact parameters $\lesssim$ a few hundred kpc, the covering fraction of dense clouds is high -- all observable as `strong' MgII absorption systems with $N_{\rm MgII} > 10^{13}$ cm$^{-2}$. 
 \label{fig_vis_mg2_single}}
\end{figure*}

\begin{figure*}
\centering
\includegraphics[angle=0,width=7.0in]{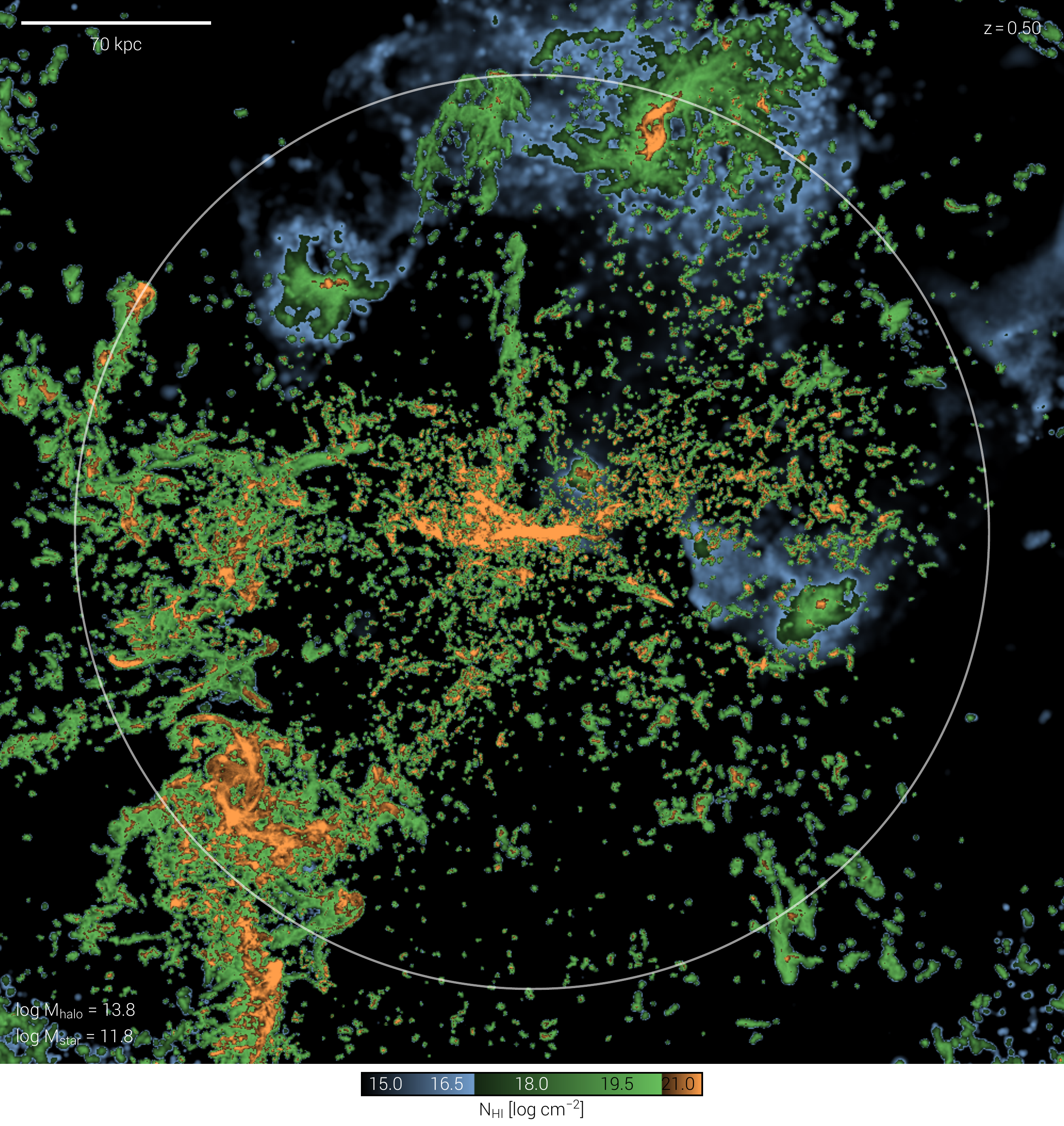}
\caption{ Visualization of the neutral HI column density distribution around the same massive TNG50 halo as in Figure \ref{fig_vis_mg2_single} at $z\,=\,0.5$. Three distinct observationally motivated regimes of $N_{\rm HI}$ are separated by color: sub-LLS (blue), LLS (green), and DLAs (orange).
 \label{fig_vis_hi_single}}
\end{figure*}

\section{Abundance of Cold Gas in Massive Halos} \label{sec_results1}

We begin by studying the amount of cold gas in the CGM of massive TNG50 galaxies. Figure \ref{fig_vis_mg2_single} visualizes the distribution of MgII, tracing $\sim$\,$10^4$\,K gas surrounding a massive, luminous red galaxy (LRG) analog, with total halo mass $10^{13.8}$\msun at $z=0.5$. The central galaxy has $M_\star \sim 10^{11.9}$\msun, log(sSFR) $-10.5$\,yr$^{-1}$, and hosts a black hole with mass $M_{\rm SMBH} \sim 10^{9.2}$\,\msun. The dark matter halo is filled with a hot, virialized plasma with a characteristic temperature $T_{\rm vir} \simeq 10^7$\,K. Nonetheless, the inner halo is populated with an enormous amount of cold gas ($\sim 10^{11.5}$\msun, five percent of the total halo gas mass), and a sightline piercing this halo would have a non-negligible probability of intersecting a strong absorber with $N_{\rm MgII} > 10^{13}$ cm$^{-2}$. Moreover, the MgII-traced cold gas is in the form of thousands of small, discrete clouds. This CGM, both locally and globally, is a multi-phase gaseous reservoir.

Figure \ref{fig_vis_hi_single} shows, for the same halo and the same view, the neutral hydrogen column density. Detectable MgII absorption is commonly if not always associated with high HI columns, and there is a correspondingly high covering fraction of neutral gas in the halo. Sightlines which would be classified as damped Lyman-alpha absorbers (DLAs) with $\log{(N_{\rm HI})} > 20.3$ cm$^{-2}$ (orange) are found not only in the center of the halo and surrounding infalling satellites, but also in regions occupied by smaller, discrete cold gas clouds.

\begin{figure*}
\centering
\includegraphics[angle=0,width=7.0in]{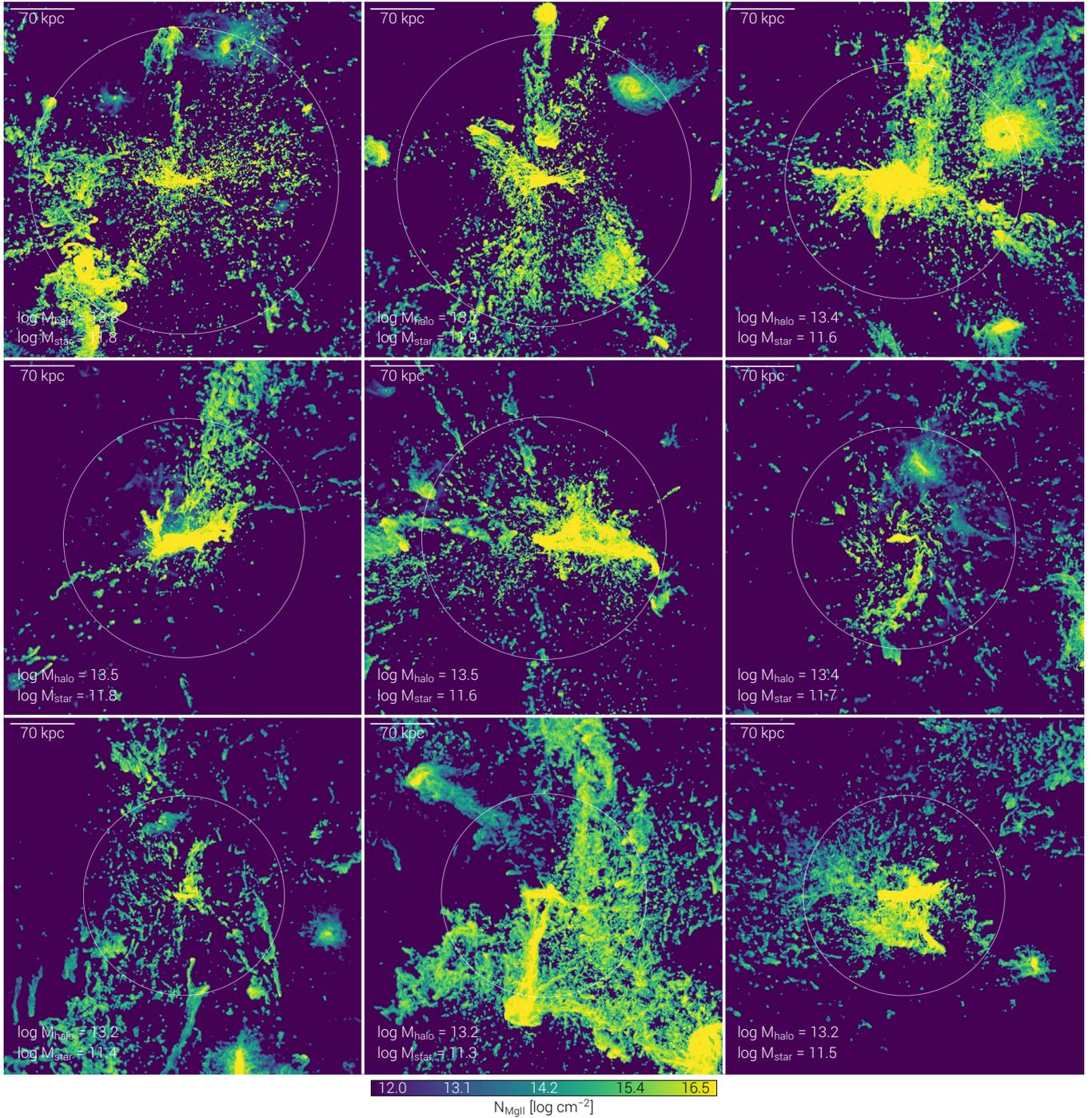}
\caption{ Visualization of the diversity of MgII column density distributions around nine massive halos in TNG50 at $z\,=\,0.5$, all selected within the halo mass range from $10^{13.2}$\msun to $10^{13.8}$\msun. The configuration is the same as in Figure \ref{fig_vis_mg2_single}, with $r_{\rm vir}/4$ shown by the white circles.
 \label{fig_vis_mg2_gallery}}
\end{figure*}

In general, although some of the dense gas has clear associations with disrupting/stripped satellites and their tidal gas tails (lower left), much of the small-scale structure has no obvious origin (lower right) and may arise from a more in-situ type formation mechanism, as we explore below. There is also a non-negligible contribution from projection effects along the line of sight. For example, the spatially extended, diffuse galaxy (upper right; blue envelope) lies outside the virial radius of the host halo itself.

\begin{figure*}
\centering
\includegraphics[angle=0,width=3.45in]{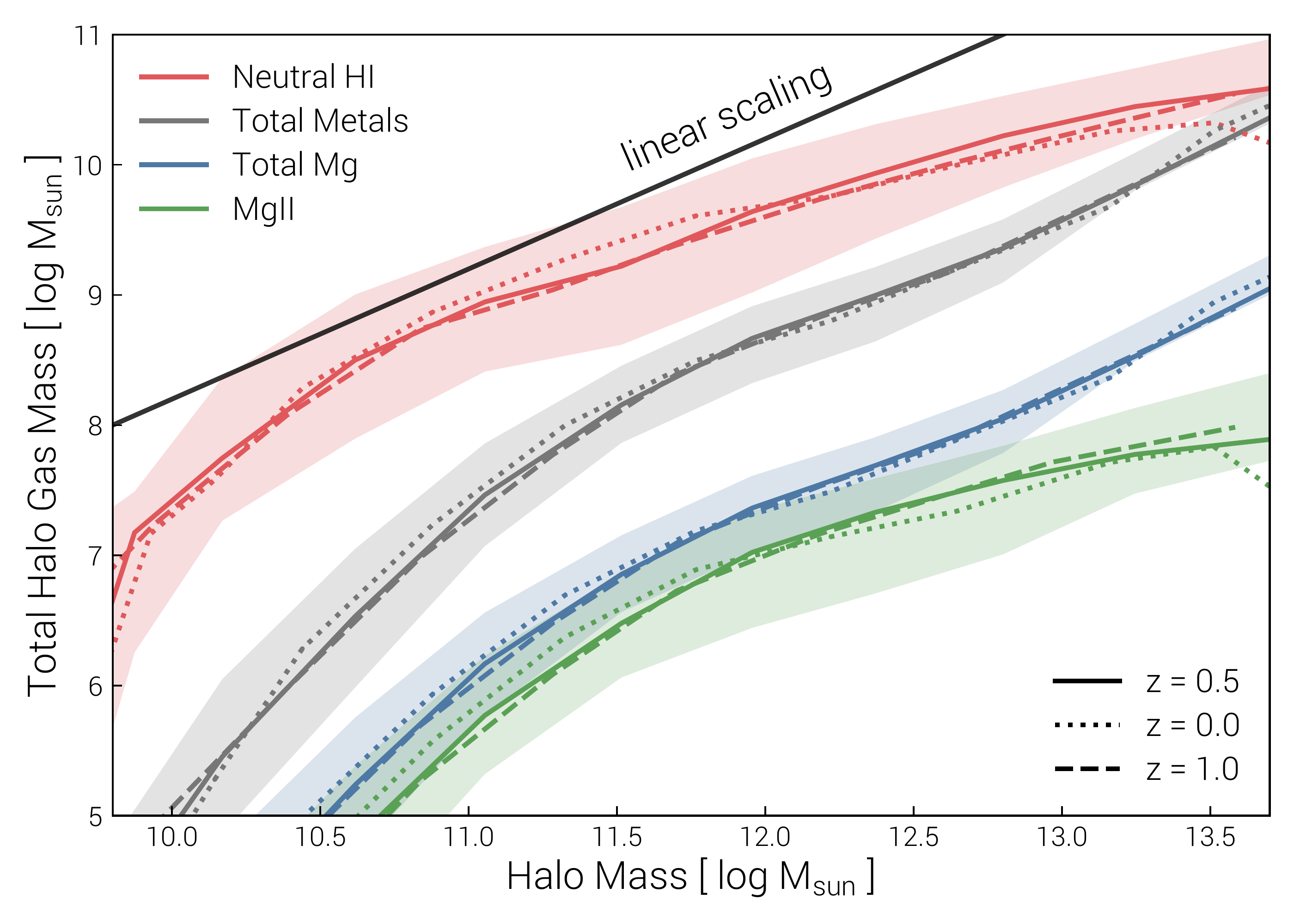}
\includegraphics[angle=0,width=3.45in]{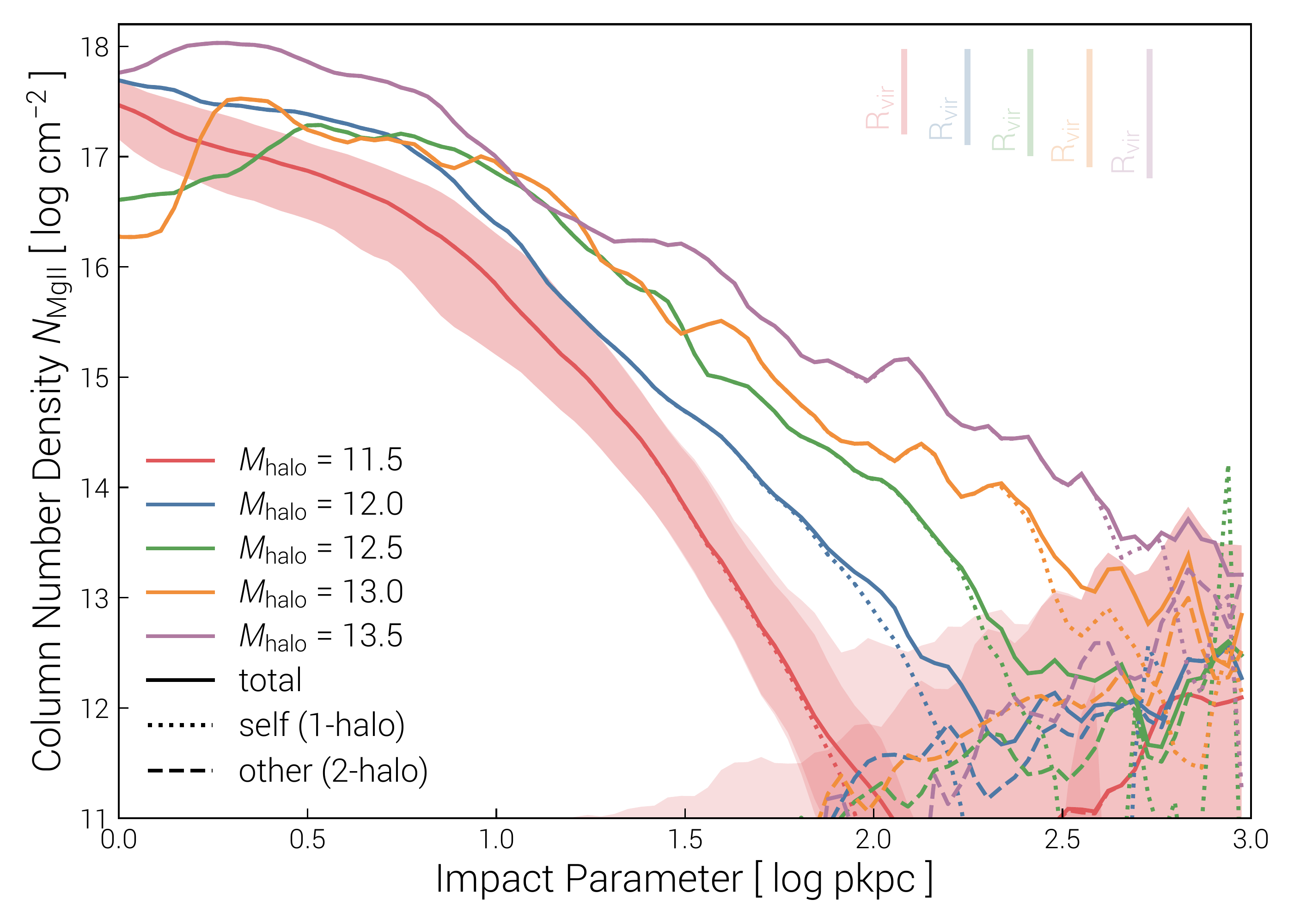}
\caption{ \textbf{Left:} Total halo gas mass in different phases as a function of halo mass at $z=0.5$, showing total metal mass (gray), magnesium (blue) , MgII (green), and neutral HI (red). In TNG50 all increase with halo mass: there is always more cool gas in more massive halos, at least up to the regime of massive groups ($M_{\rm halo} \sim 10^{13.5}$\msun). \textbf{Right:} TNG50 stacked radial profiles of MgII column density for different halo masses (16-84th percentile shaded band). The largest columns are always reached in the halo center where densities are the highest. Towards higher halo mass, $N_{\rm MgII}$ profiles flatten, such that column densities at large impact parameters ($>$\,100 kpc) increase appreciably. This is also true when normalizing distances by $r_{\rm vir}$ (not shown). The majority of this gas is physically associated with the halo, and the contribution from the projected 2-halo term (dashed lines) is small, dominating only beyond $\gtrsim 500$\,kpc.
 \label{fig_ions_vs_mass}}
\end{figure*}

In Figure \ref{fig_vis_mg2_gallery} we highlight the large diversity in the morphology of cold gas in LRG host halos, showing a gallery of nine random, massive systems. A large amount of cold, spatially distributed gas is common in the inner CGM of massive halos. All nine examples show large areas covered by high $N_{\rm MgII}$ column densities, although there is a tendency that the most massive halos have the highest covering fractions. The degree of uniformity versus angular variability differs among halos. Here we have placed each central galaxy edge-on, to the extent possible, by aligning with cold star-forming gas and/or the shape of the inner stellar body. An overabundance of MgII in the vertical direction would therefore indicate preferential alignment with the minor axis and possible association with galactic-scale outflows \citep{nelson19b}. We note that this is a sizeable fraction of the massive halos available in TNG50 at $z=0.5$ -- due to the relatively small volume there are only ten halos above $10^{13.2}$\msun, which we would consider LRG-mass analogs, and twenty-seven halos above $10^{12.8}$\msun, in the group mass regime.

Many although not all halos have significant accumulations of cold gas in their centers, which fall into rotationally supported disk structures. Although stellar mass dominates, large central reservoirs of neutral or molecular gas have been observed in the central ellipticals of large groups and clusters \citep{edge01}, including kinematics consistent with rotating disks \citep{rose19} and more complex inflow/outflow motions consistent with an episodic interplay of accretion and feedback \citep{tremblay18,olivares19}. In TNG, a diversity of central cooling and thermodynamical states is present in the centers of both intermediate \citep{truong20} and high-mass halos \citep{barnes18}. Here we have not made any selection on the cooling state of these LRG-mass halos, nor on the star formation rate activity of their central galaxies; although TNG is broadly consistent with observational constraints on SFRs and quiescent fractions in this regime \citep{donnari19}, massive galaxies can rejuvenate back towards appreciable SFRs at higher resolution \citep{nelson18a}, implying that the cold reservoirs in these LRG hosts may be somewhat too highly star forming.

Figure \ref{fig_ions_vs_mass} shows that the total amount of cool gas increases with increasing galaxy or halo mass (left panel). While the neutral HI gas (red) and cool MgII components (green) start to flatten as we approach the cluster-mass regime of $M_{\rm halo} > 10^{13.5}$\msun, the total amount of metals (gray) and magnesium (blue) continue to rise, roughly linearly with halo mass. This implies that for Milky Way mass halos of $\sim 10^{12}$\msun, the metal fraction is $f_{\rm Z} = M_{\rm Z} / M_{\rm halo} \sim 4 \times 10^{-4}$, while the MgII fraction is $f_{\rm MgII} \sim 10^{-5}$. For LRG-mass hosts with $M_{\rm halo} \sim 10^{13.5}$\msun we see that $f_{\rm Z}$ remains largely unchanged, while $f_{\rm MgII} \sim 10^{-6}$, decreasing by an order of magnitude \citep[see][who show that the mean halo mass hosting strong MgII absorbers is $\sim 10^{12}$\msun]{bouche06}. At fixed mass, the cold gas content of halos does not evolve with redshift from $0 < z < 1$. 

In TNG, a significant fraction of spatially extended metals in the outskirts of larger halos have an ex-situ origin, meaning that they are accreted from previously enriched environments and/or with satellite galaxies, rather than being expelled from the central itself \citep{vog18a}. Similarly, we see that the stacked radial profiles of $N_{\rm MgII}$ (right panel) flatten significantly for increasingly massive halos. This is also true when normalizing by $r_{\rm vir}$ (not shown). As a consequence, observably high columns are found at larger impact parameters, $>$\,100 kpc for the more massive systems considered here. On average, this strong MgII absorption arises from gas physically within the halo (dotted lines), rather than due to line-of-sight contributions from other halos (dashed lines). This two-halo contribution begins to dominate only for $b \gtrsim$\,500 kpc ($\gtrsim$\,650 kpc) for $M_{\rm halo} \gtrsim 10^{13}$\msun ($\gtrsim 10^{13.5}$\msun), slightly beyond their respective virial radii.


\section{Cold Gas Structure in Massive Halos} \label{sec_results2}

\begin{figure*}
\centering
\includegraphics[angle=0,width=3.45in]{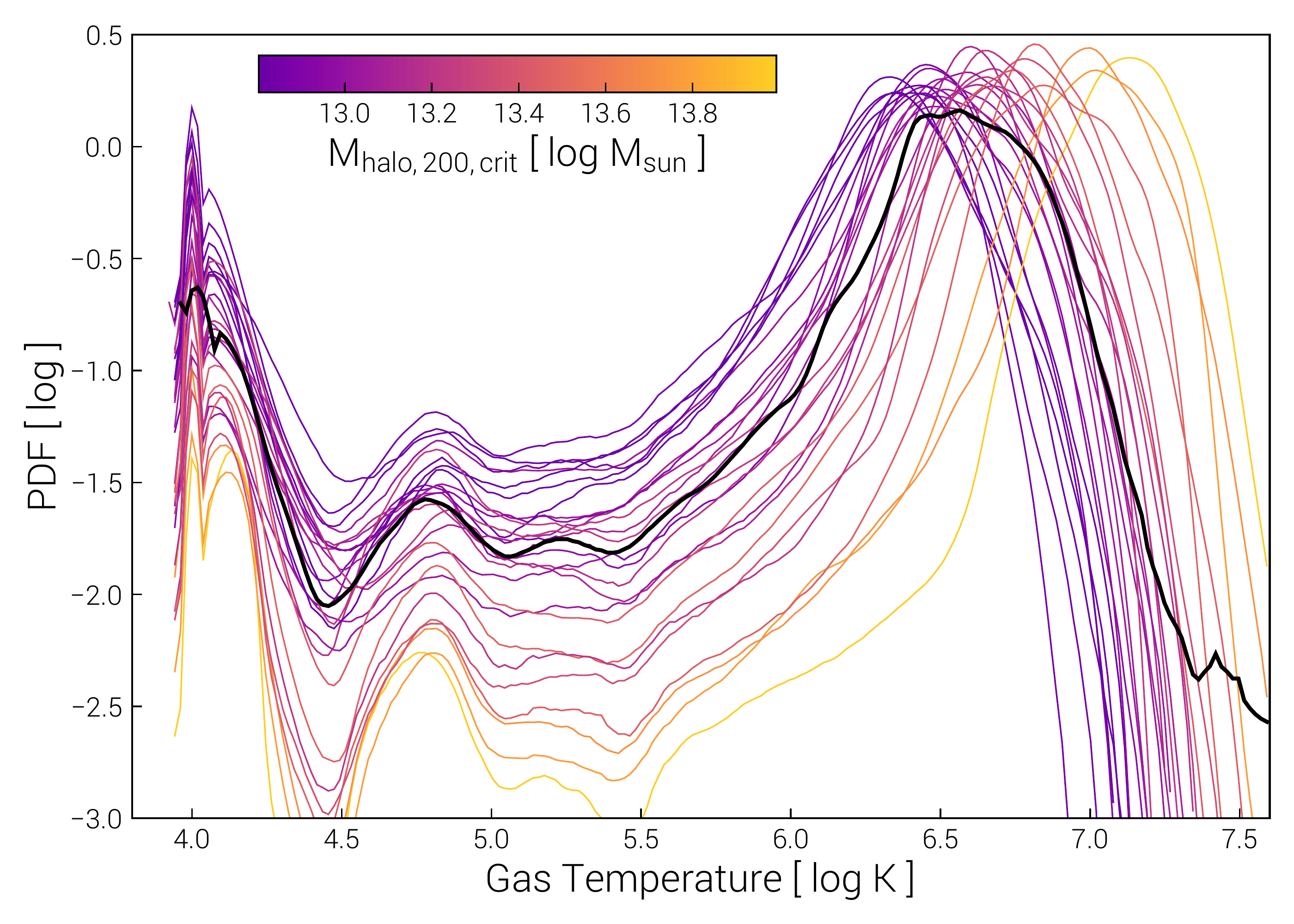}
\includegraphics[angle=0,width=3.45in]{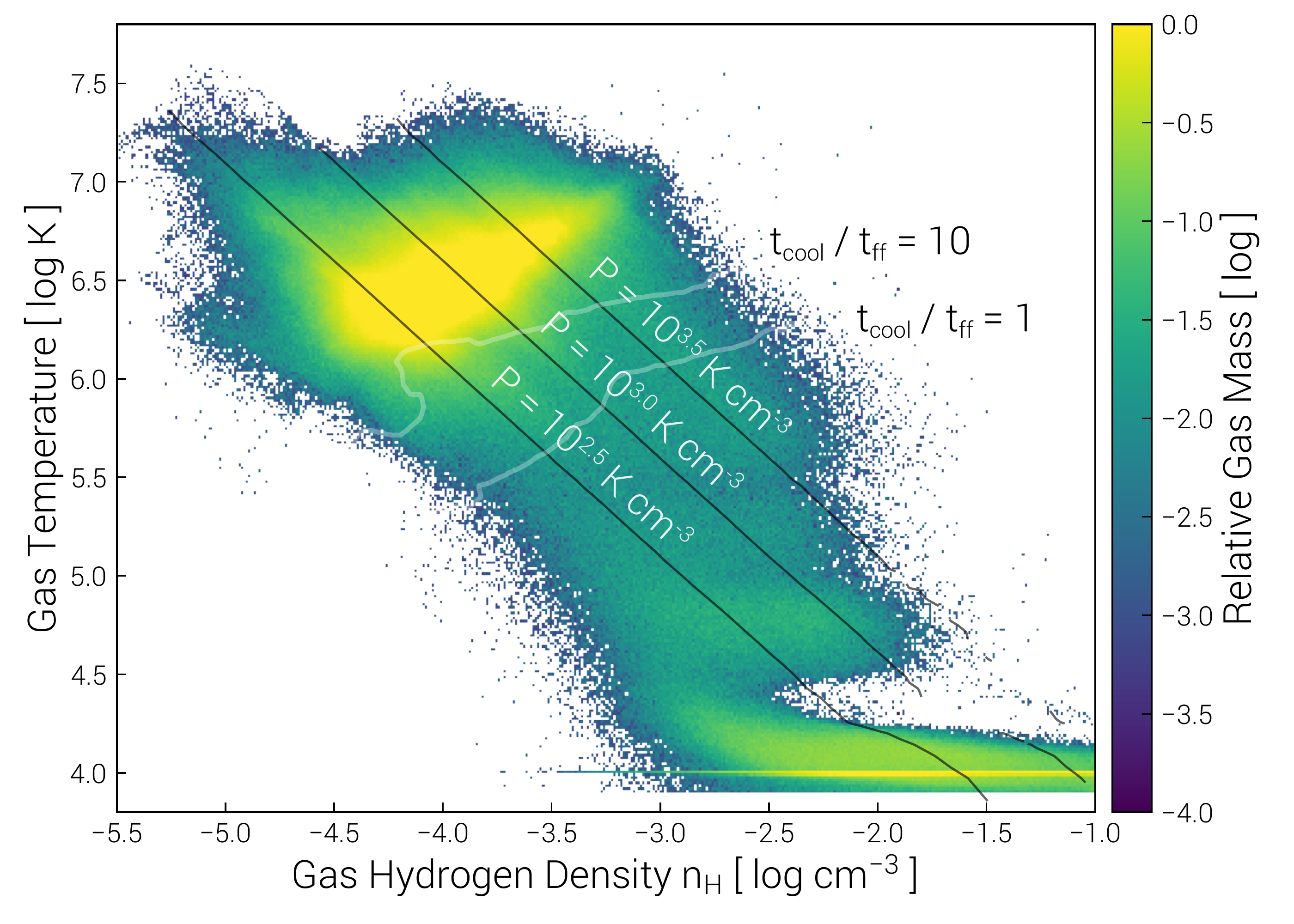}
\caption{ Phase structure of halo gas between $0.1 < r/r_{\rm vir} < 1$. \textbf{Left:} gas temperature distributions in all TNG50 halos above $10^{12.8}$\msun, where the black line shows the median. The primary peak at $\sim 10^{6.5} - 10^{7.5}$ K traces the virialized, hot halo gas. The other features at lower temperatures are independent of halo mass, reflecting the shape of the cooling curve. \textbf{Right:} temperature-density phase diagram for a single halo ($M_{\rm halo} = 10^{13}$\msun). Overplotted are three black contours of constant gas pressure, as indicated, as well as two white contours denoting thresholds of $t_{\rm cool} / t_{\rm ff} < 1$ (below the lower line) and $t_{\rm cool} / t_{\rm ff} > 10$ (above the upper line). Globally in the CGM of halos, multi-phase gas is seen across a broad range of temperature and density.
 \label{fig_phase}}
\end{figure*}

\subsection{Multi-phase gas in the CGM}

There is an abundance of cold gas, but what is its morphology and properties? Figure \ref{fig_phase} shows the temperature distribution (left panel) and temperature-density phase diagram (right panel) of all halo gas between $0.1 < r/r_{\rm vir} < 1$. In the former, we show all halos in TNG50 with $M_{\rm halo} > 10^{12.8}$\msun, color corresponding to mass. The dominant peak in the temperature distribution reveals the hot halo: virialized gas, primarily heated through the accretion shock virialization which accompanies the formation of the halo \citep{wr78}. The location of this peak scales in temperature as $T_{\rm vir} \propto M_{\rm halo}^{2/3}$. Gas in this temperature regime is ionized, and rather inefficient in radiatively cooling. The secondary features towards lower temperatures are independent of halo mass, reflecting the shape of the cooling curve $\Lambda(T)$. Particularly for metal-enriched gas, radiative cooling losses are significant for $T \lesssim 10^6$\,K until the effective cooling floor at $\sim 10^4$\,K is reached. Although most of the CGM mass is in the hot component, a non-negligible fraction is cool (or cold; we use these terms interchangeably) at $\sim 10^4$\,K, a temperature effectively traced by both MgII and HI.

The temperature-density diagram shows that the hot gas is diffuse, with characteristic densities of $n_{\rm H} \sim 10^{-4}$ cm$^{-3}$, whereas the cold phase is dense, with $n_{\rm H} \gtrsim 10^{-2}$ cm$^{-3}$. The three black diagonal lines indicate curves of constant gas (thermal) pressure: $P = 10^{2.5}$, $10^{3.0}$, and $10^{3.5}$ K\,cm$^{-3}$, respectively.

Note that for an overdense cloud undergoing isobaric contraction (i.e. constantly in pressure equilibrium with the background, neglecting magnetic fields), the ratio of temperature between the background and the cloud should essentially equal the density ratio

\begin{equation}
T_{\rm cloud} = \frac{ T_{\rm CGM} }{1 + \delta}
\end{equation}

\noindent where $\delta = (\rho_{\rm cloud} - \rho_{\rm CGM}) / \rho_{\rm CGM}$ is the density contrast. This expectation is largely true, as the temperature contrast is roughly $10^{6.5} / 10^4 \sim 300$, matching a density contrast of $\delta \simeq \rho_{\rm cloud} / \rho_{\rm CGM} \sim 10^{-1.5} / 10^{-4} \sim 300$. This is the characteristic average temperature of the cold-phase for $M_{\rm halo} \simeq 10^{13.5}$\msun, and we show below that this density decreases towards lower halo masses, as expected by pressure confinement within a cooler ambient medium.

In the right panel of Figure \ref{fig_phase} we also show two contours of the ratio of gas cooling to dynamic free-fall time, $t_{\rm cool} / t_{\rm ff}$ (thin white lines). The cooling time is calculated as the cell internal energy divided by the net cooling rate, as derived within the simulation. The free-fall time is $t_{\rm ff} = (3\pi / (32 G \bar{\rho}))^{1/2}$ where $\bar{\rho}$ is the mean total mass density enclosed within the halocentric distance of the gas cell. Note that this is $\sim \sqrt{2}$ times larger than $t_{\rm ff} = (2 r^3 / GM)^{1/2}$ as commonly defined in ICM studies. These two contours are roughly perpendicular to the lines of constant pressure, and reveal that almost the entirety of the hot-phase has $t_{\rm cool} / t_{\rm ff} > 10$. However, almost all intermediate temperature gas -- by which we mean $T \sim 10^{5.0-5.5}$\,K, near the cooling curve peak -- has $t_{\rm cool} / t_{\rm ff} <$ a few, and so can rapidly cool down to $\sim 10^4$\,K in this thermally unstable regime.

\subsection{Physical properties of cold-phase clouds}

\begin{figure*}
\centering
\includegraphics[angle=0,width=6.9in]{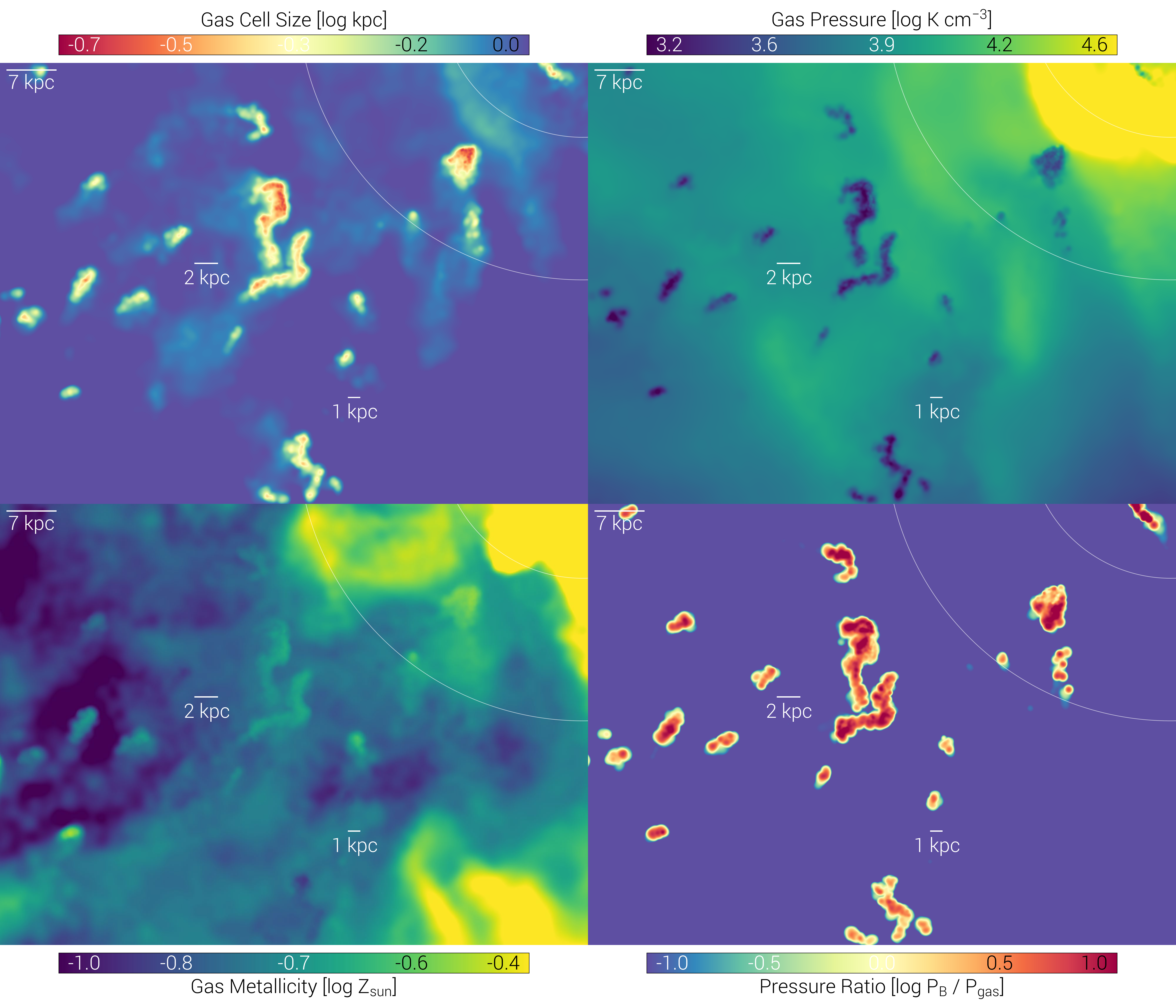}
\caption{ Visualization of the small-scale structure of cold-phase gas structures in the CGM of a TNG50 galaxy. Circles show 5\% and 10\% of the virial radius -- the central LRG is located in the upper right corner of each panel. This is a halo with a total mass of $10^{13.0}$\,\msun (stellar mass of $10^{11.4}$\,\msun) at $z=0.5$, which hosts a large number of small-scale, cold, dense clouds with typical sizes of one to a few kiloparsecs. \textbf{Upper left:} the numerical resolution available in the gas, in terms of gas cell size. Cold overdensities (yellow, orange) are much better resolved than the background hot halo (in purple). \textbf{Upper right:} thermal gas pressure, where the cold clouds stand out as severally underpressurized with respect to the surrounding hot gas. \textbf{Lower left:} gas metallicity, where most clouds are metal enriched with respect to their local background hot halo gas. \textbf{Lower right:} the magnetic to thermal pressure ratio $\beta^{-1} = P_{\rm B} / P_{\rm gas,thermal}$. CGM clouds are magnetically dominated with $\beta \ll 1$, whereas the background halo gas is instead dominated by thermal pressure ($\beta > 10$).
 \label{fig_vis_zoomin_clouds_h19}}
\end{figure*}

This intermediate temperature gas is not distributed uniformly throughout the hot halo, but rather forms the interface (and accretion) regions of the cold gas clouds. Figure \ref{fig_vis_zoomin_clouds_h19} shows four views of a small, zoomed in portion of a LRG-host halo, focusing on the small-scale, physical properties of several individual cold gas clouds. Here, the central galaxy is located towards the upper right of each panel, and the outermost circle shows $0.1\,r_{\rm vir}$. Each panel shows a thin slice projected just $\sim 5$ kpc along the line of sight, to avoid confusion of overlapping clouds.

First, the upper left visualizes the hydrodynamical resolution itself, given as the gas cell size. The background hot halo is resolved with only modest $>$ 1 kpc resolution (purple color) owing to its low density. In contrast, small-scale, cold clouds are highly resolved: the cores of the larger clouds by gas cells as small as 200 physical parsecs in size. The intermediate density interface regions are captured by cells roughly 500 pc in size. We are clearly pushing the limits of resolvable structure in TNG50, and these cold gas structures are only marginally resolved, e.g. by $\sim$ 10s of cells.

The morphology of these cold clouds shows clear signatures of infall: head-tail structure, with gas tails or wakes extending downstream and so roughly radially away from the halo center. The cold/dense blobs are also commonly surrounded by warmer regions -- this gas is cooling and so accreting onto the dense seeds, as we show below. As a result, cooling and dense gas formation will continue in these portions of the halo because they are susceptible to runaway cooling, independent of the presence of the inner cold cores. For intermediate temperatures $T < 10^{6.5}$\,K gas, $t_{\rm TI} \sim 0.5 t_{\rm cool}$ such that $t_{\rm TI} / t_{\rm ff}$ is a factor of 2 or so smaller than $t_{\rm cool} / t_{\rm ff}$. If we consider these interface regions as large density fluctuations, then they would roughly satisfy the condensation criterion of $t_{\rm TI} / t_{\rm ff} \lesssim$ few 10s \citep{choudhury19}.

The upper right panel shows gas pressure. The hot halo is highly pressurized, particularly towards the center, with sonic waves propagating outwards due to the ongoing energy injection from the central supermassive black hole. The small clouds pop out in darker color, indicating that they are not in thermal pressure equilibrium with their surroundings. Rather, they are severely thermally under-pressurized, by roughly an order of magnitude with respect to the background medium.

The lower left panel shows gas-phase metallicity: we see that cold clouds are metal enriched with respect to their locally surrounding hot halo. We return to this important point later, as it argues against an in-situ formation mechanism for the majority of clouds. Instead, enhanced cloud metallicities are more similar to the gas-phase metallicity of infalling satellite galaxies, emphasizing their connection to stripping and mixing with accreting interstellar medium gas. Remaining metal inhomogeneity points to incomplete chemical mixing as cold clouds transit the parent halo.

The lower right panel shows the plasma $\beta^{-1} = P_{\rm B} / P_{\rm gas,thermal}$, the ratio of magnetic to thermal pressure. CGM clouds are magnetically dominated, with typical values of $\beta \lesssim 0.1$, implying that a significant amount of their pressure support arises from the compression and amplification of magnetic fields during their formation. The background halo gas is, in contrast, dominated by thermal pressure ($\beta > 10$), and magnetic fields have a negligible dynamical impact. Cold clouds have higher $P_{\rm B} = B^2 / (8\pi)$ than the background: while the hot halo at these distances and densities has a typical magnetic field strength of $B \sim 0.1-0.5$\,$\mu$G, co-spatial cold clouds have $B \sim 3-10$\,$\mu$G, up to a hundred times larger.

Because of flux freezing we would expect that the gas becomes magnetically dominated in over-dense clouds, even if the original hot media has a large $\beta \gtrsim 100$. For instance, if gas cools from $10^7$ K to $10^4$ K during the non-linear phase of the thermal instability, we expect the density to increase by a factor of one thousand, and the B field strength to increase by a factor of $\sim$\,100. After this point, the high magnetic pressure helps prevent further compression of the cold gas, as a lower gas density can be offset by the non-negligible magnetic support \citep{sharma10}.

\begin{figure}
\centering
\includegraphics[angle=0,width=3.25in]{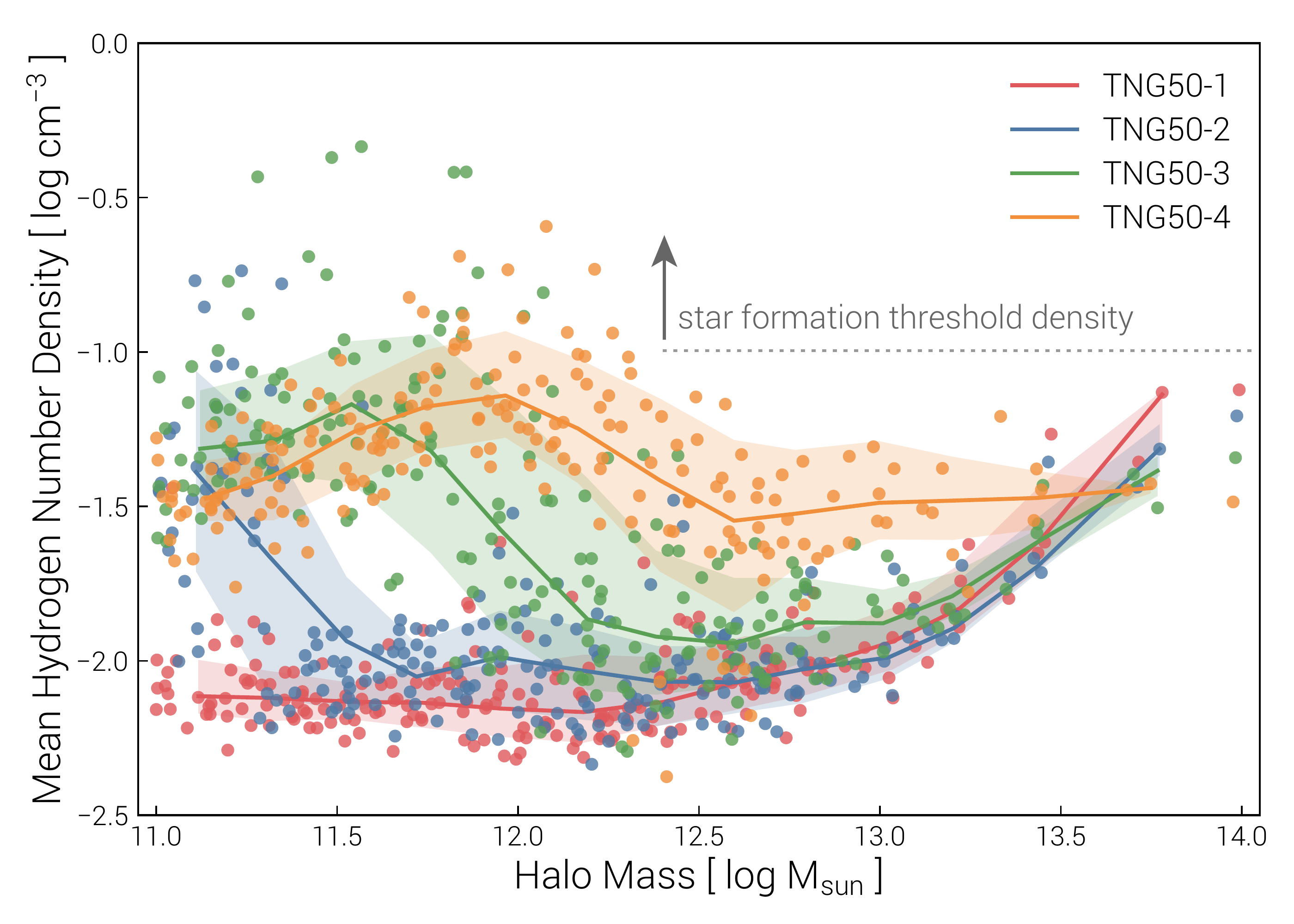}
\includegraphics[angle=0,width=3.25in]{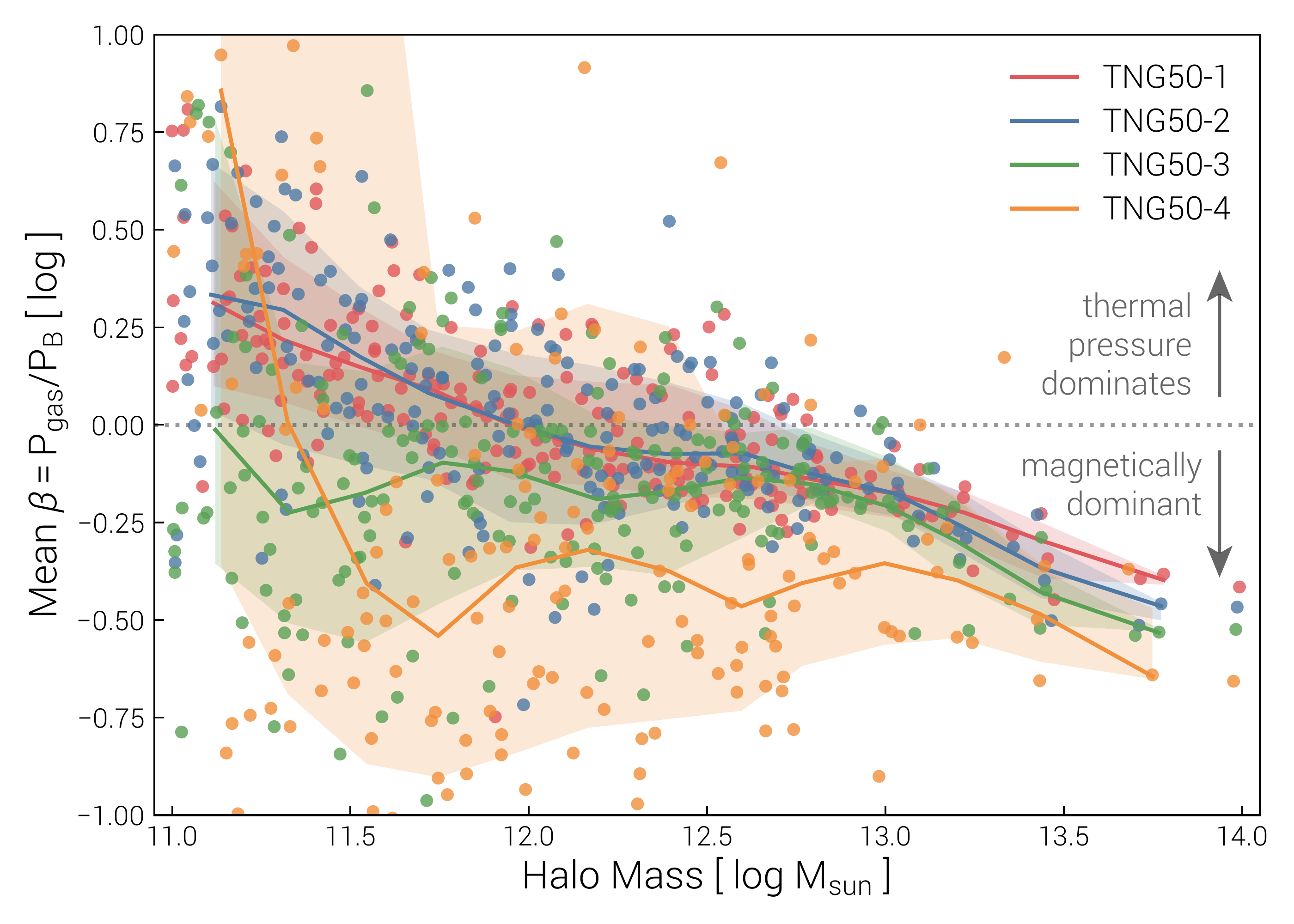}
\includegraphics[angle=0,width=3.25in]{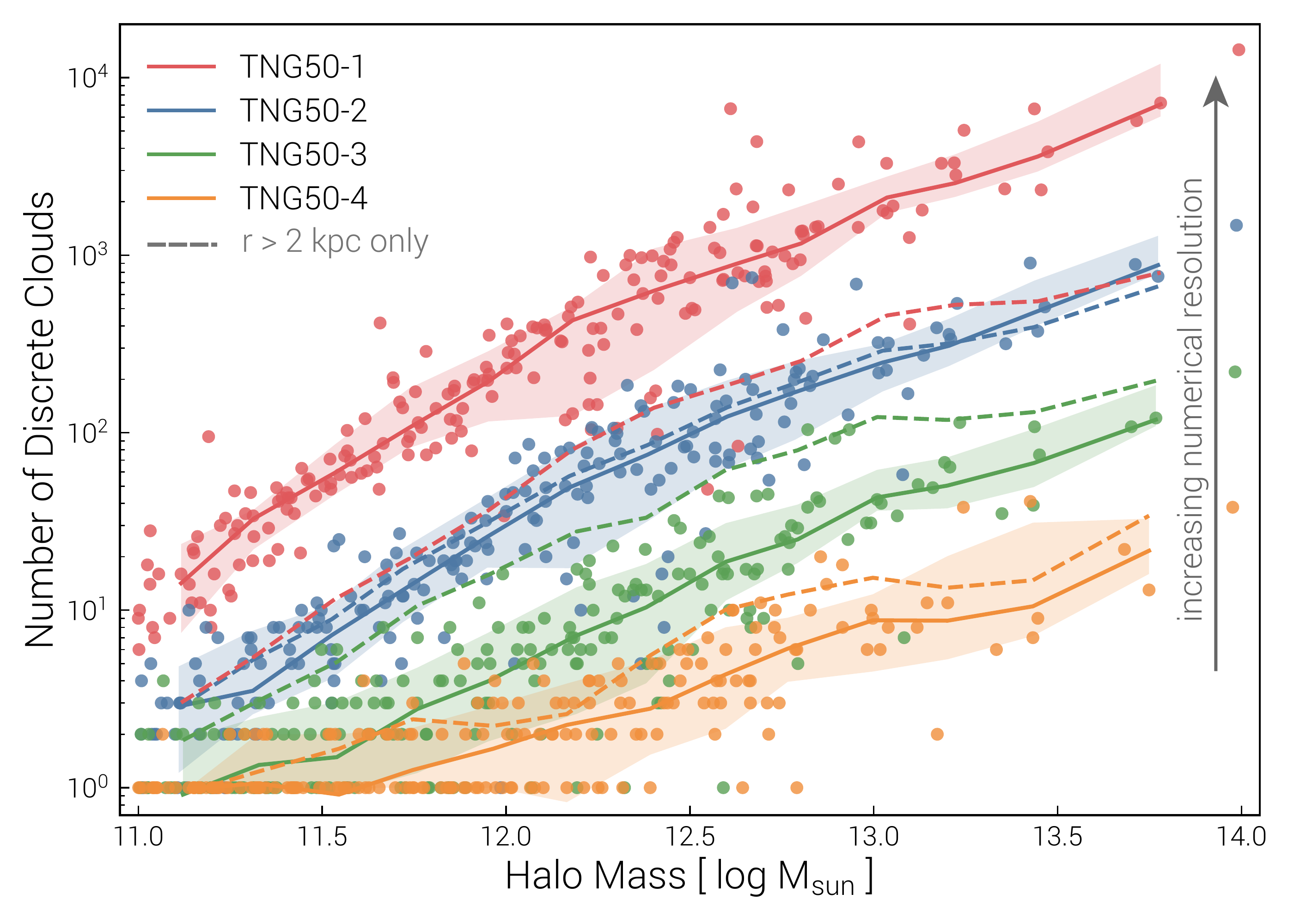}
\caption{ Cold cloud properties as a function of halo mass at $z=0.5$, contrasting the four resolution levels of TNG50. \textbf{Top:} Mean gas density of clouds, which grows with halo mass due to increasing compression from the background hot halo. \textbf{Middle:} Average internal $\beta = P_{\rm gas} / P_{\rm B}$. \textbf{Bottom:} The number of cold clouds in the CGM, defined as a minimum of 10 gas cells satisfying the threshold $n_{\rm MgII} > 10^{-8}$ cm$^{-3}$ (dashed lines: `large' clouds only). While internal properties are well converged, the total number of \textit{small} discrete clouds is not: more are always resolved at better resolution.
 \label{fig_cloudprops_vs_halomass}}
\end{figure}

What is the impact of these magnetic fields on the dynamics of the cold-phase clouds? In general, we expect that magnetic fields do not affect the linear thermal instability for a plasma $\beta \gtrsim 10$, although its nonlinear evolution can be. For instance, \cite{ji18} show that the cold fraction as a function of background $t_{\rm cool}/t_{\rm ff}$ is affected only for initial $\beta \lesssim 30$, while in general the value of $\beta$ can alter the resulting structure of cold gas \citep{berlok19}. To assess the role of the MHD we run two test re-simulations of a single halo. One case is evolved to $z=0.5$ at TNG50-1 resolution and with the fiducial TNG model, whereas in the second case we remove the magnetic fields for the $\sim$\,500\,Myr prior to $z=0.5$. This test is described in detail in Appendix \ref{sec_appendix}. Overall, we observe little qualitative difference in the population wide properties of the small-scale cold clouds, and likewise for their overall abundance. We conclude that, at TNG50-1 resolution, with the included physics, and over a halo dynamical timescale, magnetic fields do not appear to have a strong impact on the formation, evolution, or disruption of the cold gas cloud population.

To better understand the properties of the CGM cold-phase, we consider how they depend on the total mass of the system. Figure \ref{fig_cloudprops_vs_halomass} shows the mean cold cloud density (top), average $\beta$ (middle), both mass-weighted, and the total number of cold clouds (bottom) as a function of halo mass. The cloud density increases with halo mass, particularly above $10^{12.5}$\msun, and this internal property is nicely converged with numerical resolution -- the four resolution levels fall on top of each other at sufficiently high halo masses. Although the total gas mass, MgII mass, and HI mass also increase, the overall size of clouds actually decreases for more massive halos (not shown). We speculate this is due to stronger compression by a hotter, higher pressure background medium. Because of the higher densities, the mean magnetic field strength in clouds increases from $\sim$ 1 $\mu$G for all halos $\lesssim 10^{13}$\msun to $\sim$ 10 $\mu$G for the most massive.

As a consequence, the average $\beta = P_{\rm gas} / P_{\rm B}$ decreases with halo mass, as shown in the middle panel of Figure \ref{fig_cloudprops_vs_halomass}, and this results in a critical halo mass where the importance of magnetic fields changes. Specifically, a crossover of $\beta \simeq 1$ occurs at $M_{\rm halo} \sim 10^{12}$\msun. For halos below this limit, clouds are thermally supported ($\beta > 1$), while only for halos more massive than this limit are clouds magnetically dominated ($\beta < 1$).

The bottom panel of Figure \ref{fig_cloudprops_vs_halomass} shows that the abundance of cold clouds increases rapidly with halo mass: while $10^{12}$\msun halos have of order $\sim 100-200$ clouds in \mbox{TNG50-1}, $10^{13}$\msun halos have $\sim$ 1000, and typical LRG host halos typically have several thousand (up to $\sim$ ten thousand) MgII rich clouds. We note that the size, distances, kinematics, and HI masses of the clouds in Milky Way mass halos resemble the observed population of high velocity clouds \citep[HVCs;][]{muller63}. The same phenomena we explore here for LRGs may also provide a natural explanation for neutral clouds around our own Milky Way galaxy, with abundances similar to the compact HVC population \citep{putman02}.

A caveat is, however, that the number of discrete clouds is a strong function of numerical resolution, and the number of small, cold structures at LRG host masses increases from $\sim 10$ in TNG50-4 to $\sim 10,000$ in TNG50-1. Clearly, the small-scale morphology of the cold gas is not converged between our last two resolution levels. Without an even more highly resolved simulation (i.e. TNG50-0), we cannot show that the results of the current TNG50-1 simulation represent a converged state, and there is little to no indication that the number of clouds is converging. This may be expected in ideal MHD, since these clouds are susceptible to disruption by Kelvin-Helmholtz instabilities. The dashed lines show, in contrast, the total number of `large' cold clouds with sizes $\geq 2$\,kpc only. In this case the resolution dependence is much less, and TNG50-1 represents a converged state: higher resolution produces most small clouds, while the large-size tail is stable for TNG50-2 and better.

\begin{figure*}
\centering
\includegraphics[angle=0,width=3.45in]{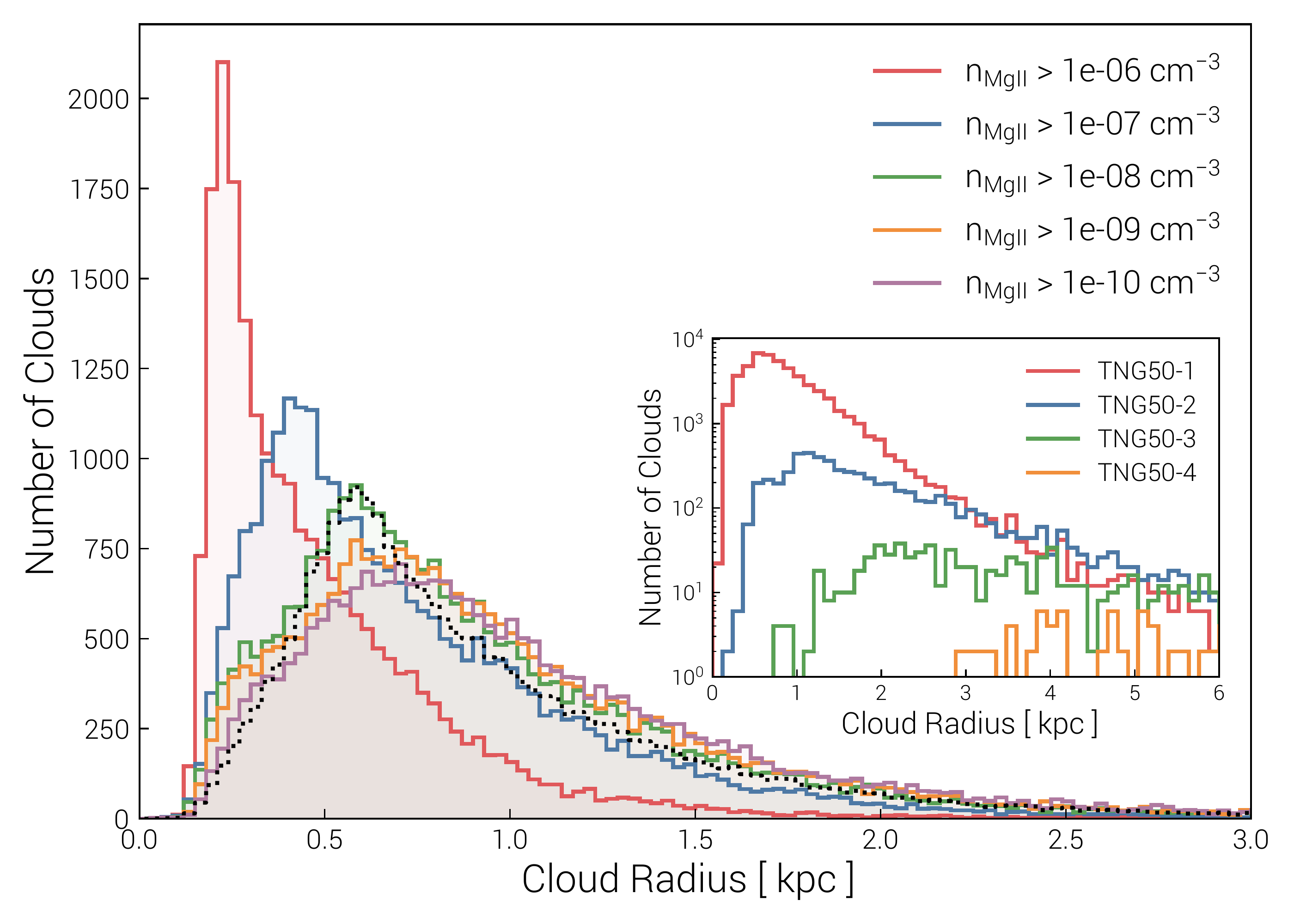}
\includegraphics[angle=0,width=3.45in]{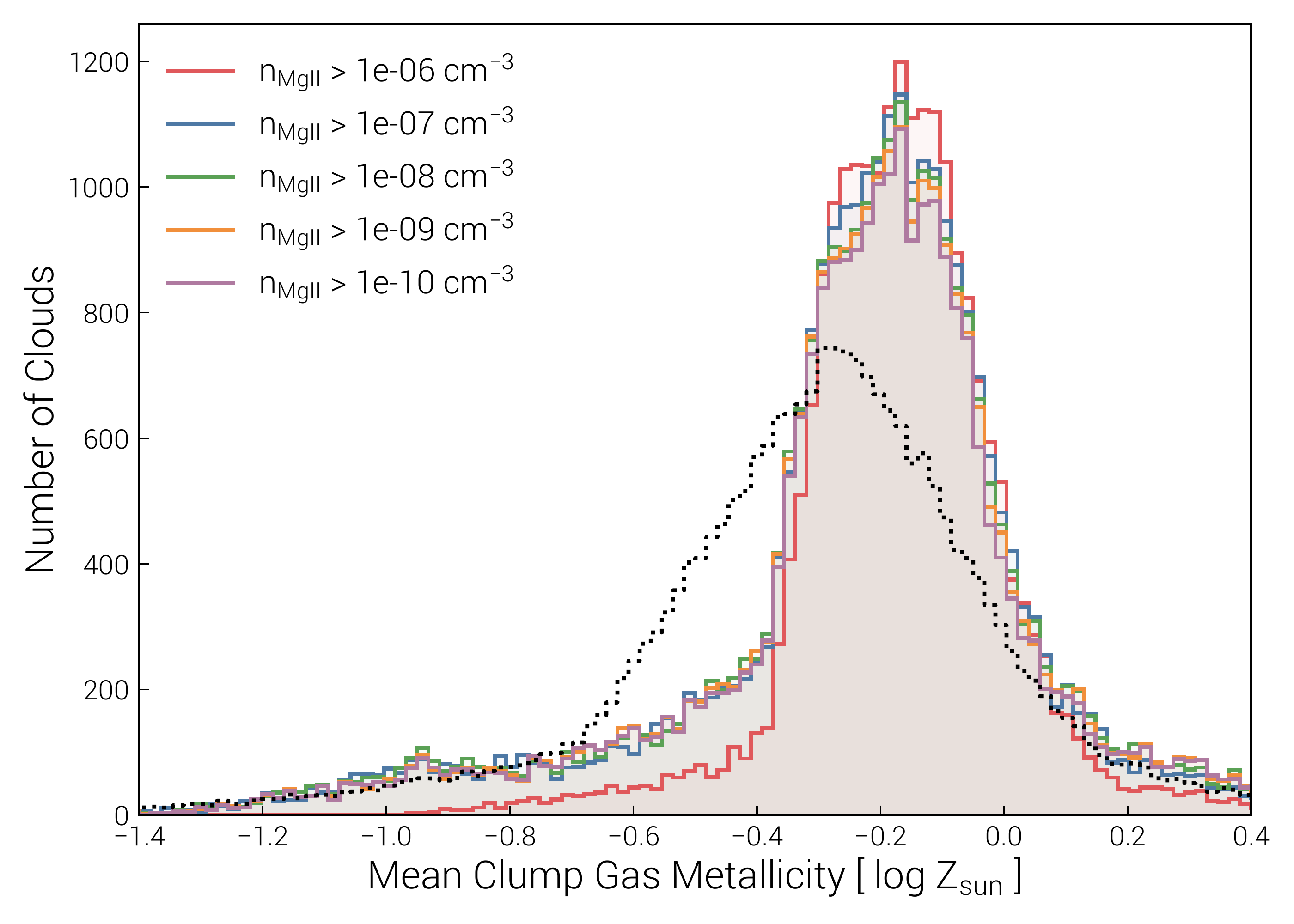}
\includegraphics[angle=0,width=3.45in]{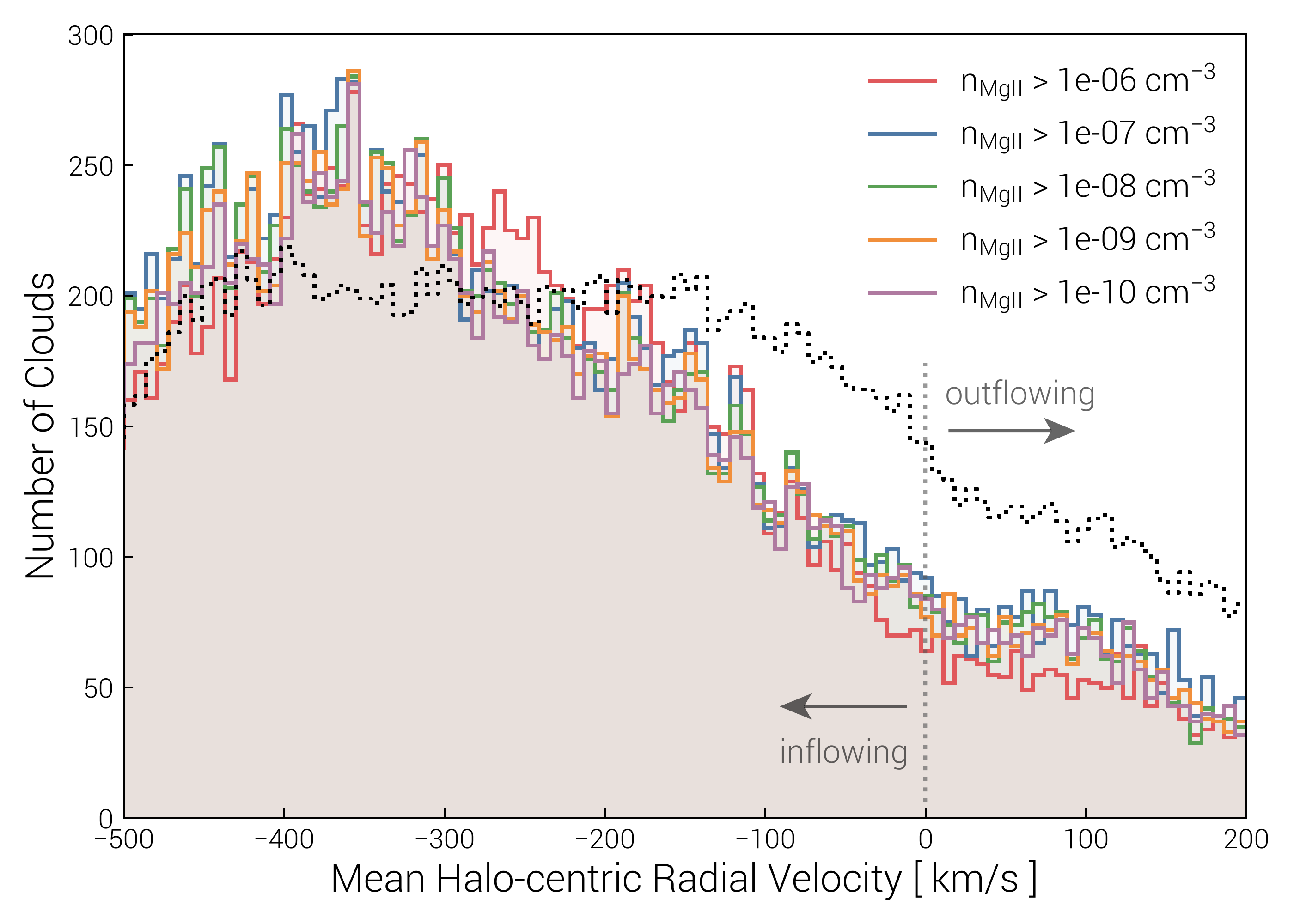}
\includegraphics[angle=0,width=3.45in]{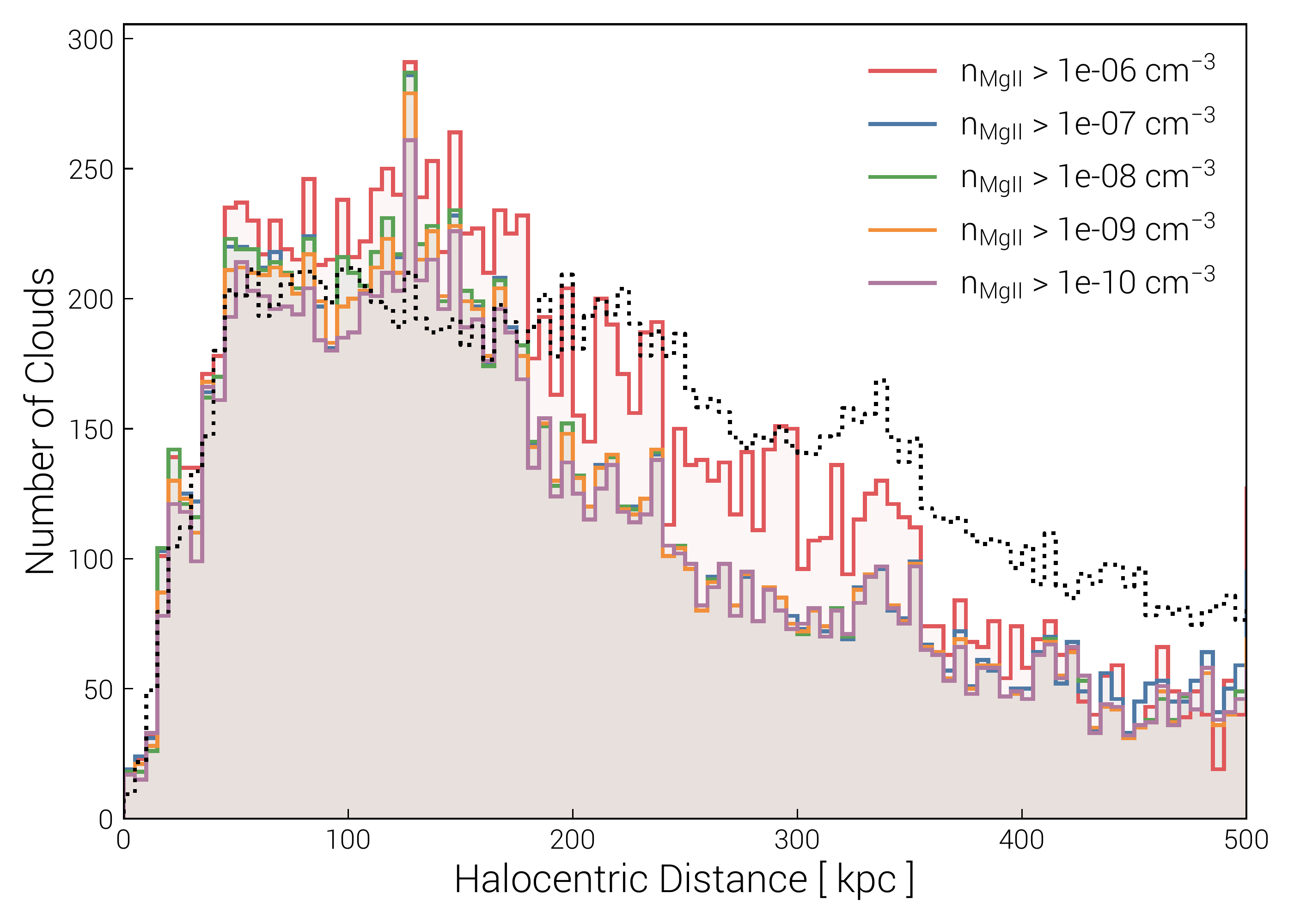}
\caption{ Demographic properties of CGM clouds in TNG50. We show results of the segmentation algorithm run on a single halo, the most massive at $\sim 10^{13.9}$\msun, for cloud size (top left), metallicity (top right), radial infall velocity (lower left), and distance from halo center (lower right). Every panel uses 100 bins, and shows five different threshold values for the identification: beyond the highest (red), the results are largely insensitive to this choice. Black dotted lines show the average result across all halos $> 10^{13.2}$\msun (for $n_{\rm MgII} >$ 1e-08 cm$^{-3}$), showing that these distributions are broadly representative. Clouds have a size distribution peaking at $\sim$ kpc diameters -- the inset shows the resolution convergence in the large-size tail. Clouds are metal enriched, with typical metallicities of $\sim$50 percent solar. The majority of clouds are infalling ($v_{\rm rad} < 0$), and are found throughout the inner halo, from $30 \sim 200$ kpc and beyond.
 \label{fig_cloud_demographics}}
\end{figure*}

In Figure \ref{fig_cloud_demographics} we explore the properties of the cold-phase cloud population. Here we show the outcome of five different segmentation runs, taking a series of threshold values in MgII number density from $10^{-6}$ cm$^{-3}$ to $10^{-10}$ cm$^{-3}$ (different line colors). Excluding the first, which selects only the highest density, inner cores of clouds, the others give reasonably similar results (i.e. cell membership) for cloud properties. This demonstrates the rather sharp boundary (in $n_{\rm MgII}$, temperature, and several other properties) between over-dense clouds and the ambient background medium, and motivates our fiducial choice of $n_{\rm MgII} = 10^{-8}$ cm$^{-3}$. 

In total there are $\sim\,$20,000 discrete cold structures detected in this halo, with a fairly narrow size distribution which peaks at $\sim$ kpc diameters (upper left panel). The integrated number of clouds is fairly constant with the density threshold, ranging from 22,000 to 26,000, implying that different thresholds largely select different portions of the same structures. Cloud sizes extend to the smallest allowed by the simulation resolution, and also have a tail towards larger sizes up to several kpc. The inset panel shows the resolution convergence of the cloud size distribution, demonstrating that the large-size tail converges, while the number of small clouds continues to increase. Note that we define cloud radius by the volume equivalent sphere, $r_{\rm cloud} = [3 V_{\rm cloud} / (4\pi)]^{1/3}$, summing over all member cells. In actuality, more massive clouds become less spherical, forming elongated and asymmetric shapes (see Figure \ref{fig_vis_zoomin_clouds_h19}). Clouds also tend to be larger further from the halo center. For the middle threshold, half (90 per cent) of the total mass in cold gas is contained in clouds with sizes $\gtrsim 0.82$ (0.38) kpc.

\begin{figure*}
\centering
\includegraphics[angle=0,width=2.3in]{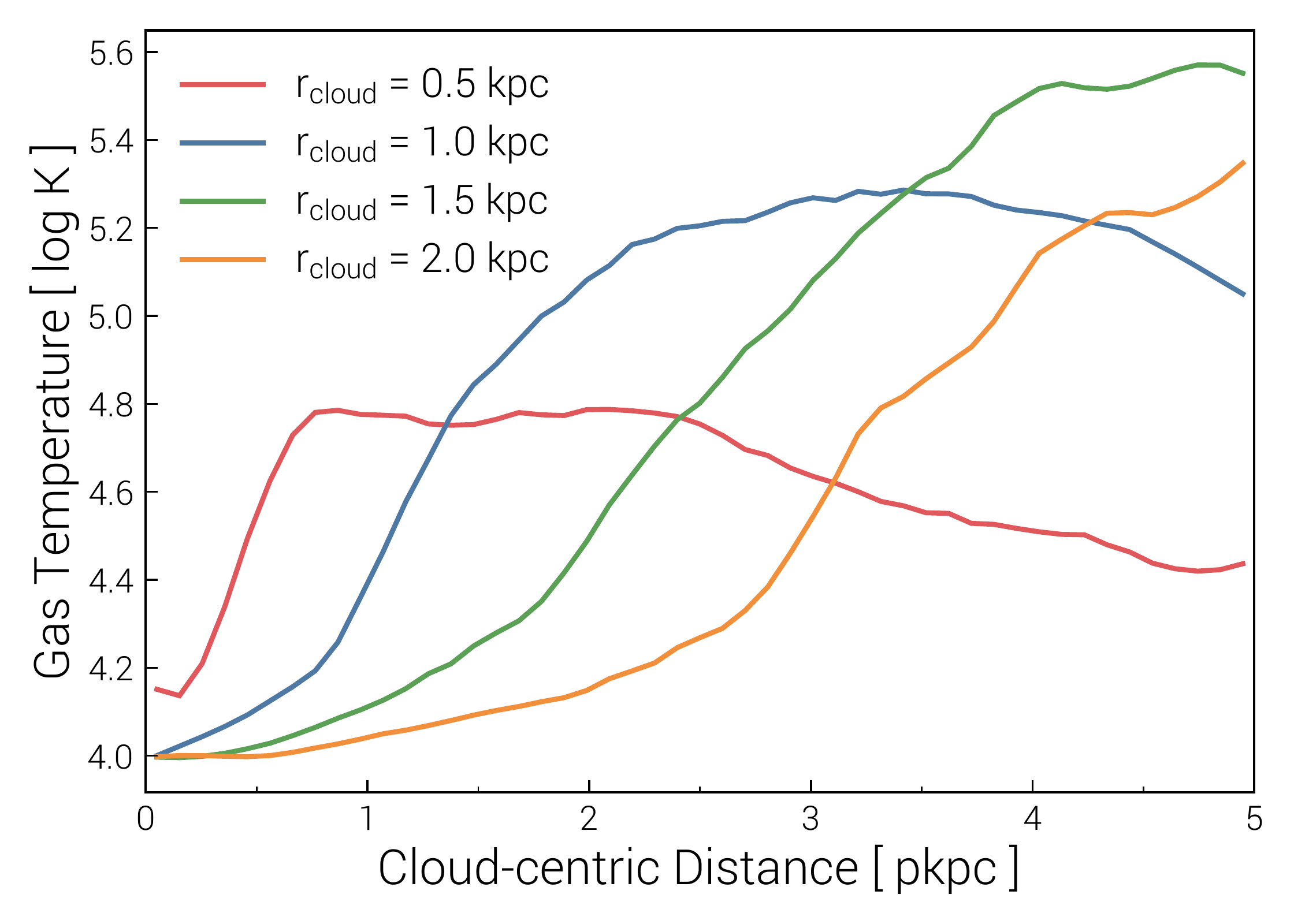}
\includegraphics[angle=0,width=2.3in]{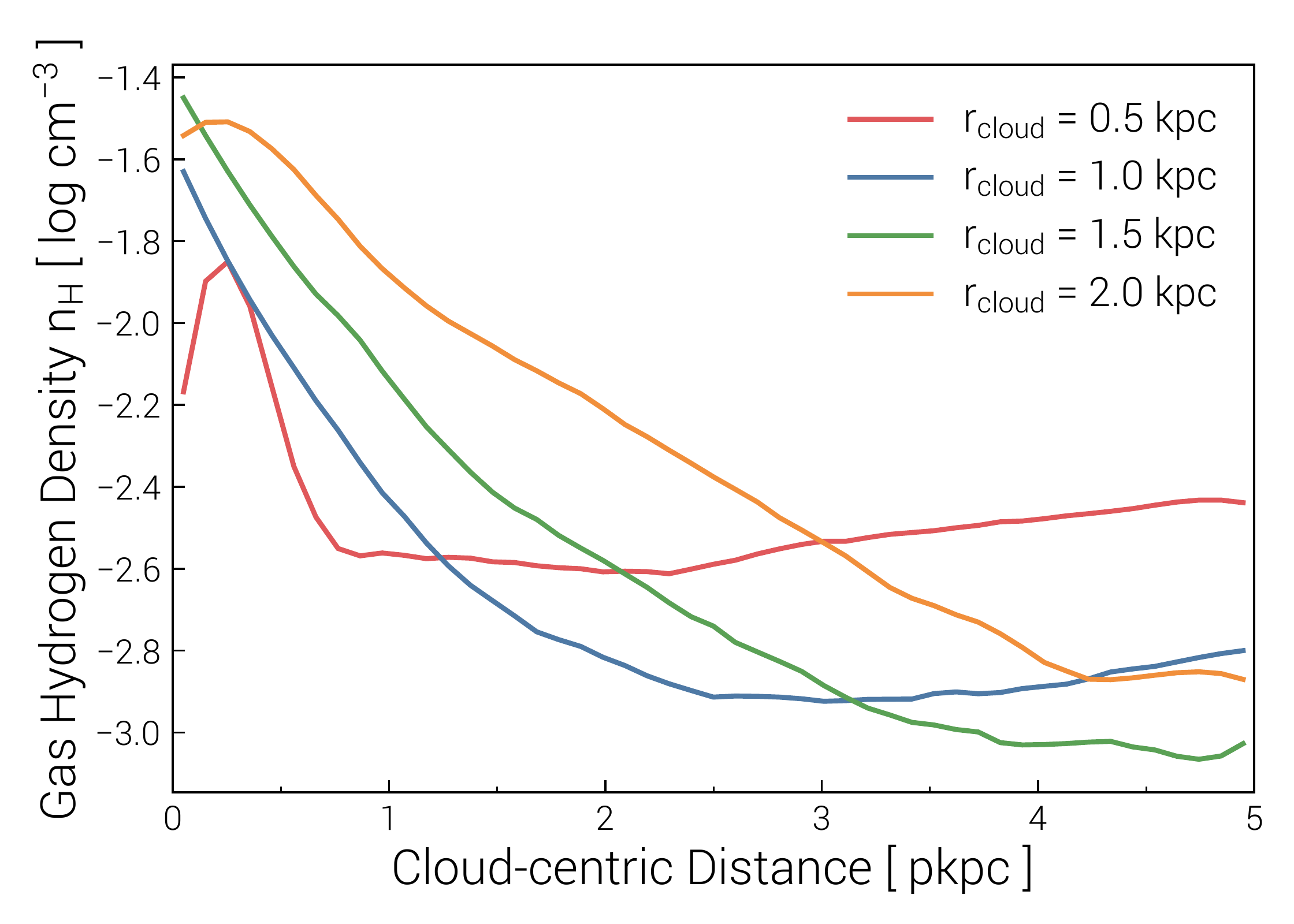}
\includegraphics[angle=0,width=2.3in]{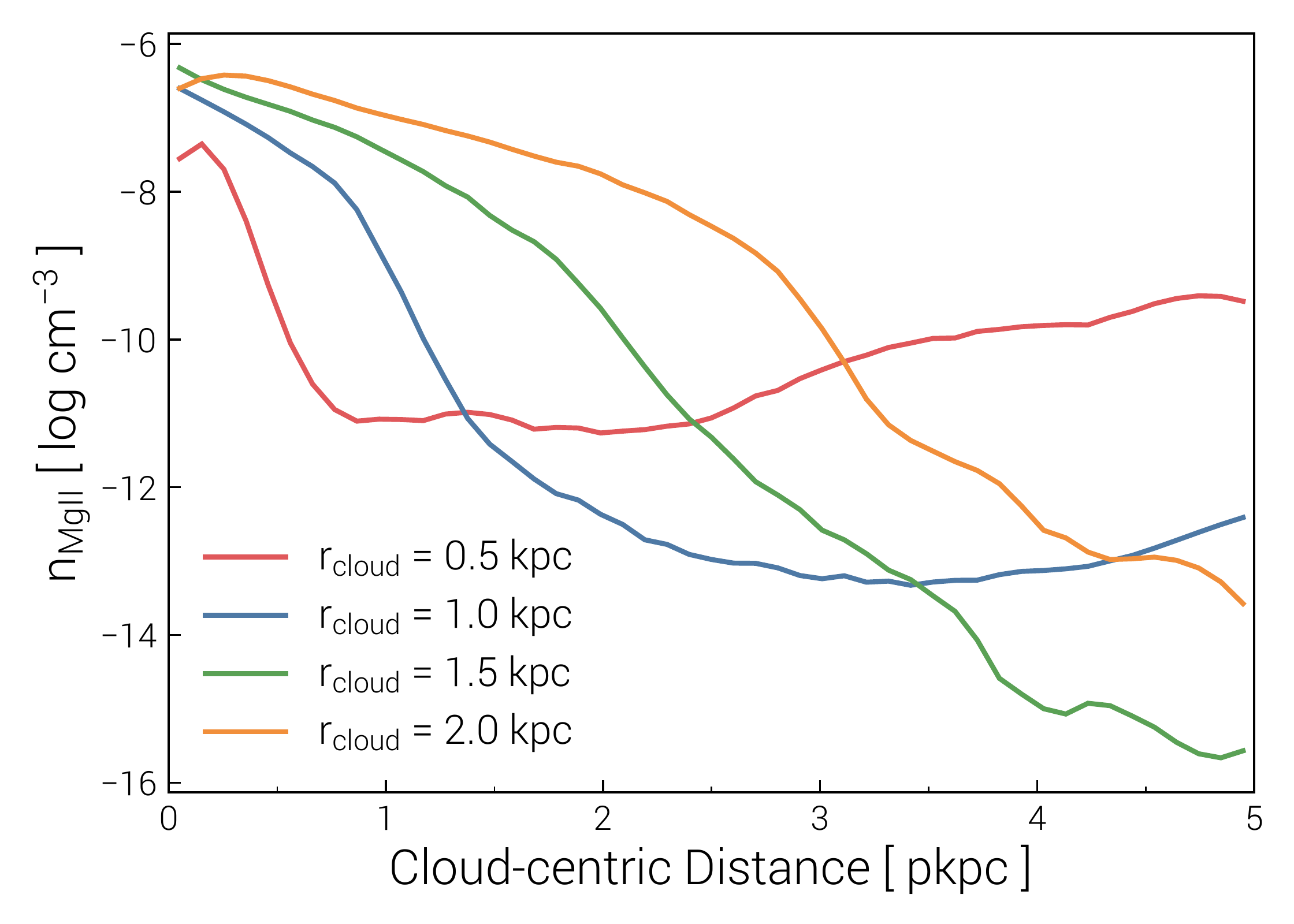}
\includegraphics[angle=0,width=2.3in]{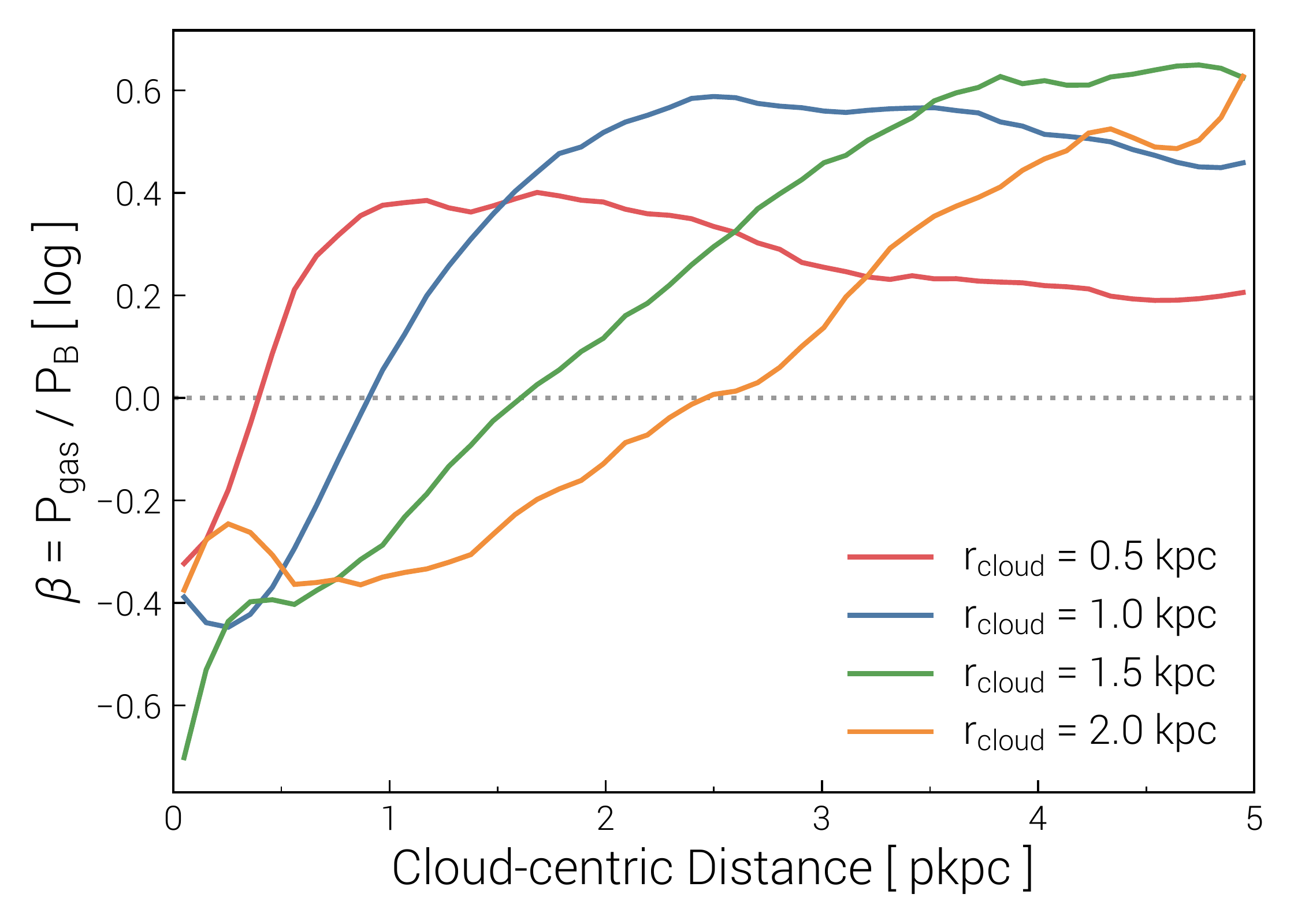}
\includegraphics[angle=0,width=2.3in]{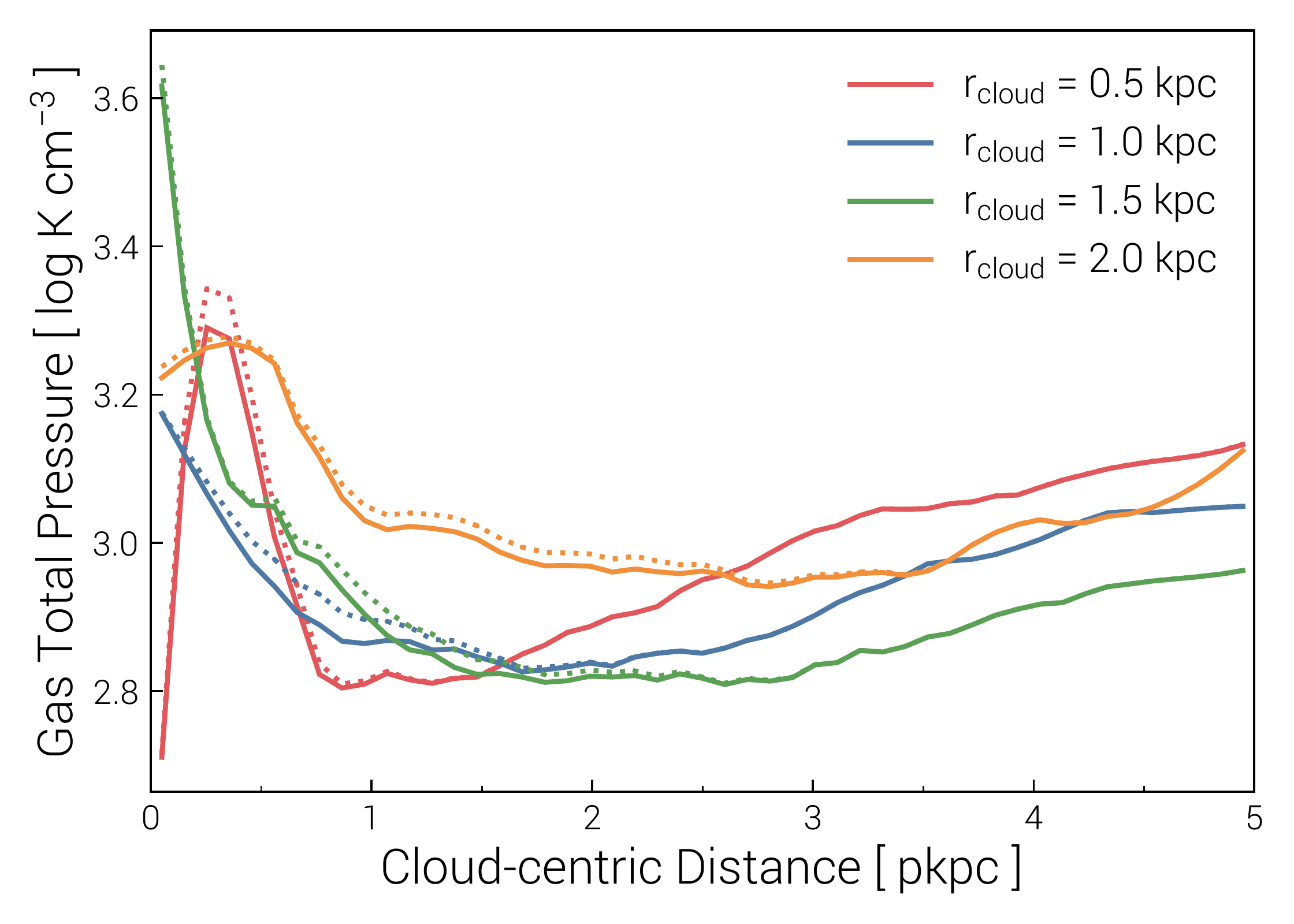}
\includegraphics[angle=0,width=2.3in]{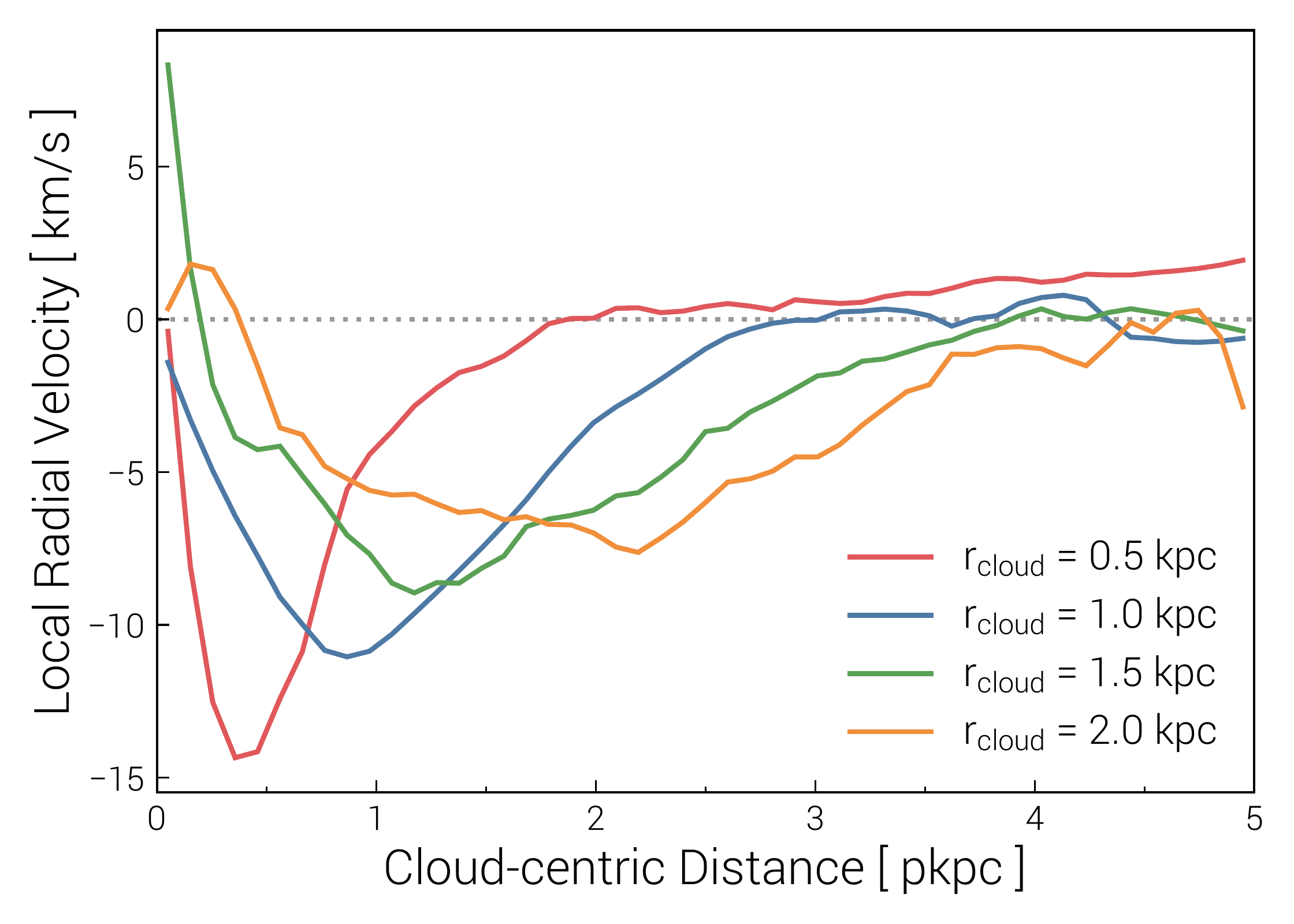}
\caption{ Internal radial structure of TNG50 cold CGM clouds and the transition from cloud to background properties. We stack clouds in four narrow bins of spatial size, from 0.5 kpc to 2 kpc, as shown (different line colors; in total, $\sim$ 2000 stacked objects). \textbf{Upper left:} temperature. \textbf{Upper middle:} total gas density. \textbf{Upper right:} number density of MgII.  \textbf{Lower left:} plasma $\beta$. \textbf{Lower middle:} total pressure $P_{\rm tot} = P_{\rm B} + P_{\rm gas}$, where dotted lines include the $P = \rho v_{\rm rad}^2$ momentum flux of the cooling flow (see text). \textbf{Lower right:} radial velocity, with respect to the cloud center of mass motion itself. We see that cold clouds are surrounded by an intermediate temperature, mixed gas phase which condenses onto these pre-existing cold seeds through a rapidly cooling, local inflow.
 \label{fig_cloud_radprofiles}}
\end{figure*}

Clouds are highly metal enriched, and have typical metallicities of $\sim 0.7$\,$Z_{\odot}$ (upper right panel). Variation is small, roughly $\pm$\,0.2 dex, and few clouds have either super-solar or $<$10 percent solar values. Cloud metallicities decline mildly with increasing distance away from the halo center, decreasing from a median of $\simeq 0.8 Z_\odot$ at $r \sim 100$\,kpc ($\sim$\,$r_{\rm vir}/10$) to a median of $\simeq 0.5 Z_\odot$ at $r \sim 400$\,kpc ($\sim$\,$r_{\rm vir}/2$). The gas-phase metallicity profile of the background hot halo also decreases, somewhat more rapidly over the same distance ($\sim$\,0.1 dex/100 kpc). Cold clouds have, overall, higher metallicities than the hot phase which surrounds them (see Figure \ref{fig_vis_zoomin_clouds_h19}). This contrast is larger at larger distances, suggesting that (i) most distant cold clouds do not form in-situ from the hot gas, and (ii) mixing throughout the infalling orbits of cold clouds is not complete, i.e. chemical inhomogeneity remains. We address the first point and explore cold-phase formation mechanisms in the following section.

Cold clouds are typically infalling (lower left panel): the halo-centric radial velocity distribution peaks at $-200$ km\,s$^{-1}$ (inwards), and only a small fraction have moderate outgoing velocities ($0 < v_{\rm rad} \rm{[km/s]} < 200$), in comparison to the virial velocity $v_{\rm 200} \sim 630$\,km\,s$^{-1}$ of the halo. Radial velocity is correlated with distance from the halo center: at $r \gtrsim 400$ kpc ($\sim$\,$r_{\rm vir}/2$), infall velocities are small and often consistent with zero. Moving inwards these velocities increase, reaching a maximum speed of roughly 250 km\,s$^{-1}$ at $r \sim 100$ kpc. Inside of this radius, the median velocity begins to decrease, presumably because of the mixture of inflowing and (dynamically) outflowing clouds, as a result of non-radial accretion orbits. Clouds are found over a large volume of the halo (lower right panel), out to $\gtrsim 500$ kpc, exceeding half the virial radii of these halos -- they are more abundant at smaller distances, between $50 - 200$ kpc. An exclusion zone exists in the inner 10s of kpc: the total volume drops, clouds conglomerate into a larger disk-like structure, and the background temperature is maximal.

We mention several additional cloud properties which are not shown explicitly. First, the average physical gas densities of clouds span from $n_{\rm H} = 10^{-3} - 10^{-0.5}$ cm$^{-3}$ \citep[see also][]{lan17}. Due to the $n_{\rm MgII}$ threshold, clouds are cold: mean temperatures are always just above the cooling curve floor at $10^4$ K. With total gas masses of $10^5 - 10^7$\msun, clouds contain $\sim 10^4 - 10^5$\msun of neutral HI on average, and only $\sim 10 - 100$\msun of total MgII. The cloud population spans a wide range of mean plasma $\beta$ values, peaking at $\beta \sim 1/3$. The vast majority have no ongoing star formation; the remainder have a residual $\dot{M}_\star < 10^{-3}$\,\msunyr or less.

\subsection{Internal structural properties of cold CGM clouds}

Beyond integral properties, internal cloud structure is evident. Figure \ref{fig_cloud_radprofiles} shows cloud-centric radial profiles of gas properties, where we stack together clouds in four different size bins from $r_{\rm cloud} = 0.5$\,kpc to $2.0$\,kpc. This allows us to construct the average structure of (i) the interior of clouds, (ii) the cloud-halo interface region, and (iii) the surrounding background halo gas. Note that stacking will always tend to artificially broaden otherwise sharp boundaries.

Clouds have a common central temperature of $\sim 10^4$\,K (upper left), and all but the smallest form an isothermal core with an extent of $\sim r_{\rm cloud}$ near this temperature. Beyond this distance, temperature increases rapidly to a plateau value which is much lower than $T_{\rm vir}$ of the halo, indicating that we are probing the intermediate temperature regime out to $< 5$\,kpc, and not reaching unperturbed background media. We note that $\sim 10^4$\,K is our temperature floor, so we expect some degree of `pile-up' at this value; low-temperature cooling would allow some fraction of this gas to cool further and possibly form stars if the physical conditions so allowed.

The total gas density (upper middle) and MgII number density (upper right) both decline rapidly with distance away from clouds, the latter also being modulated by the magnesium mass and temperature (ionization) state. For the two smallest cloud sizes, the gas densities plateau above the background value, indicating that cold gas is surrounded by other cold and cool gas: clouds are not isolated. We note that the average cooling times in the centers of clouds are short, $\lesssim 10$\,Myr, rising to $t_{\rm cool} \sim 100$\,Myr in their outskirts. As a result, these regions have small $t_{\rm cool} / t_{\rm ff} \lesssim 1/100$ (not shown), far below the threshold for atmospheric thermal instability and cold-phase growth \citep{mccourt12}.

\begin{figure*}
\centering
\includegraphics[angle=0,width=3.45in]{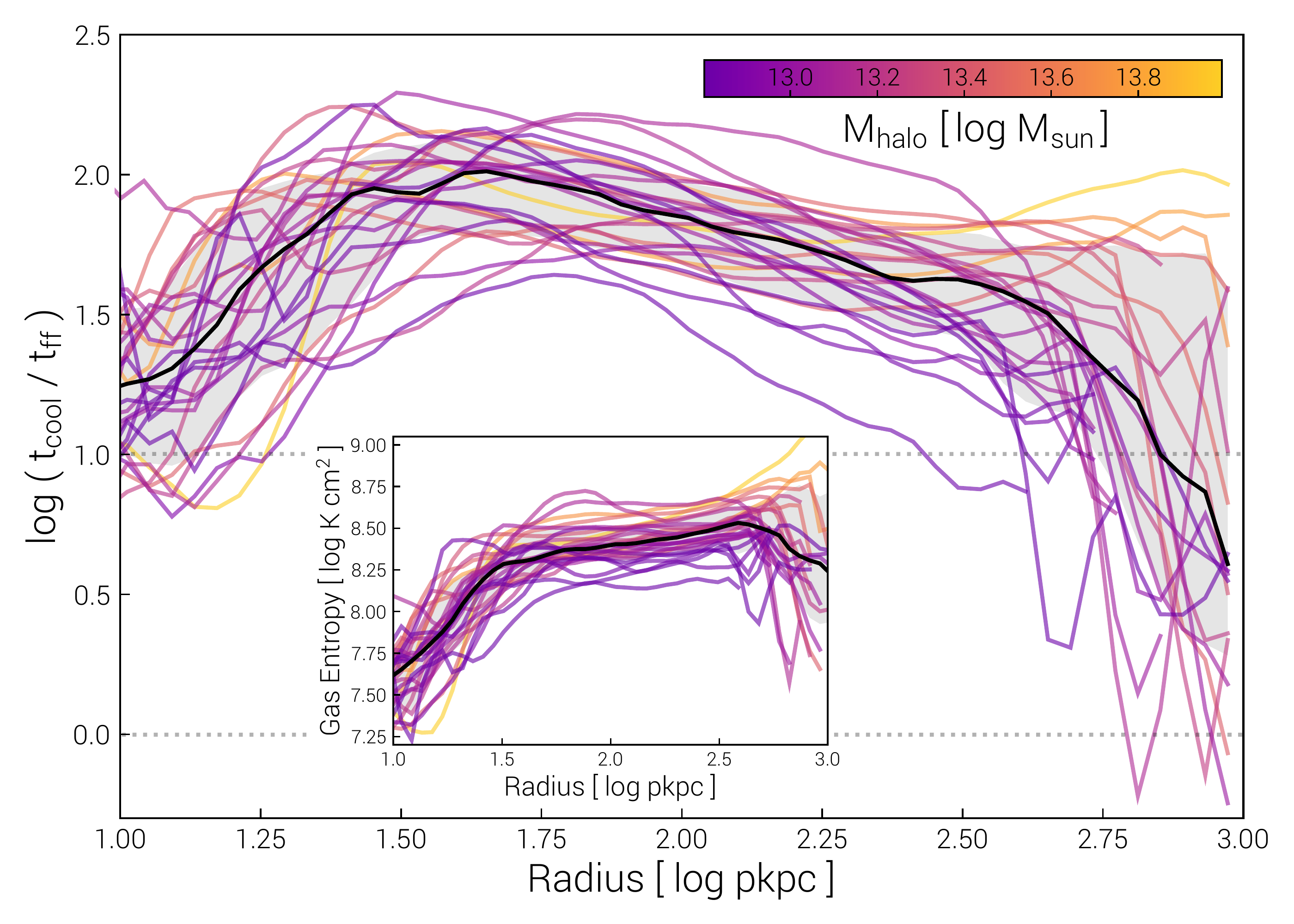}
\includegraphics[angle=0,width=3.45in]{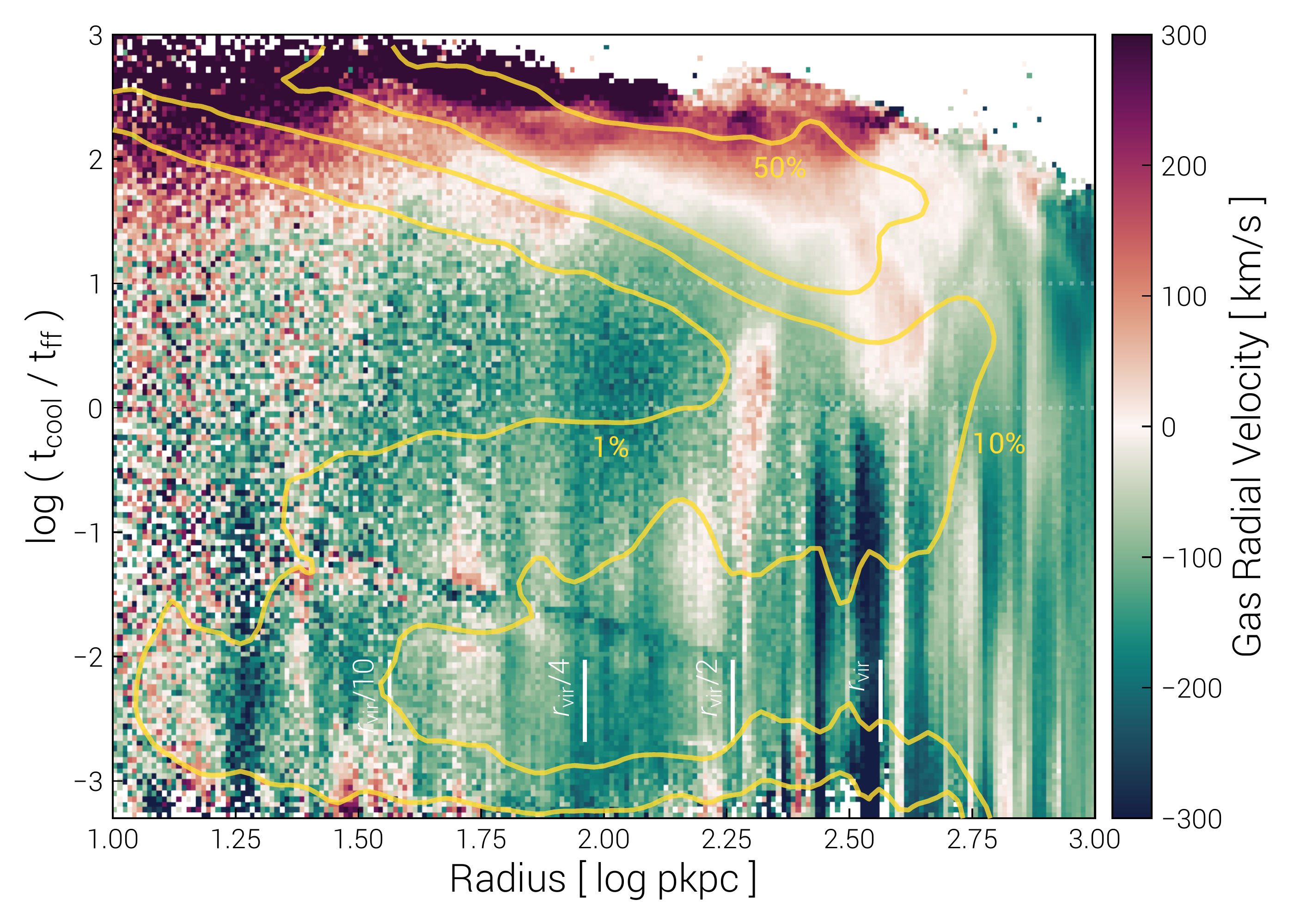}
\caption{ \textbf{Left:} Radial profiles of $t_{\rm cool} / t_{\rm ff}$ (inset: entropy) for individual halos at $z=0.5$ in TNG50. Horizontal lines are shown at 1 and 10, for reference. The median profile is given in black, while the gray band shows the halo-to-halo scatter. Only the hot gas phase is shown in order to evaluate its susceptibility to TI. In the median profile the ratio $t_{\rm cool} / t_{\rm ff}$ has a maxima at $20-50$ kpc, decreasing rapidly towards small radii and dropping below $t_{\rm cool} / t_{\rm ff} < 10$ in the inner $10-20$ kpc. The outer halo volume, however, where the majority of small-scale cold gas is found has median cooling times exceeding this canonical threshold. \textbf{Right:} two-dimensional phase diagram of $t_{\rm cool} / t_{\rm ff}$ as a function of distance, for a single LRG halo, where each pixel is colored by the mean radial velocity; $v_{\rm rad} < 0$ denotes inflow (green), while $v_{\rm rad} > 0$ denotes outflow (red), and $v_{\rm rad} \sim 0$ is indicated by white. At all regions within the virial radius, gas with $t_{\rm cool} / t_{\rm ff} \lesssim 10-50$ is generically inflowing, while gas with larger cooling times is generically outflowing. The three yellow curves show conditional contours enclosing $\{1\%, 10\%, 50\%\}$ of the gas mass at each radius: there is a small, though non-zero amount of gas with low $t_{\rm cool} / t_{\rm ff} < 1$ at intermediate radii. 
 \label{fig_radprof_tcool_tff}}
\end{figure*}

The total pressure (lower middle) and radial velocity (lower right) profiles are particularly informative. Within a distance of $\lesssim 3 r_{\rm cloud}$ gas flows inward ($v_{\rm rad} < 0$). Because the cooling time increases towards lower density, any density differential produces a temperature difference between a cloud and its surrounding medium. It is expected that, across such an interface in a two-phase medium, differential cooling will produce a pressure gradient, causing a gas flow from the low-density background medium into the high-density, perturbed region \citep{field65,burkert00}. The inflow has an associated `ram pressure' momentum flux of $\rho \vec{v} \otimes \vec{v} = \rho v_{\rm rad}^2 / \rm{k_B}$ which we include in $P_{\rm tot}$ (dotted lines), although the contribution is small and the resulting total pressure profiles are not appreciably flatter. This local cooling flow, driven by the pressure gradient rather than gravity, accretes mass on to the overdense seeds \citep[as opposed to shear-driven cooling;][]{gronke18}, resulting in cold clouds which can grow rather than shrink with time.

Alternatively, \cite{fielding20} has recently proposed that cold-phase growth can occur due to a radiative (rapidly cooling) mixing layer. In this case the inflow of cool gas arises not from pressure gradients, but instead from turbulence which replenishes the interface by mixing in hot gas from the background. We would not explicitly resolve this process in TNG50 due to the minute physical scales involved -- there are simply not enough gas cells to capture complex velocity structure in the cloud interfaces, and without sub-grid scale turbulence, mixing at the grid scale will be largely due to numerical diffusion \citep{schmidt10,teyssier15}.


\section{The origin of cold circumgalactic gas} \label{sec_origin}

We move on to consider the physical origin and formation mechanisms of the large population of small-scale, cold gas clouds observed to populate massive TNG50 halos.

\subsection{Conditions for cold-phase condensation}

First, we explore the connection to the idea that the cold-phase can condense or precipitate out of the hot background medium if its cooling time is sufficiently short \citep{sharma12,sharma12b,voit17}. The left panel of Figure \ref{fig_radprof_tcool_tff} shows individual, median radial profiles of $t_{\rm cool} / t_{\rm ff}$ for the LRG host halos considered herein. Individual lines are colored according to their virial mass at $z=0.5$, and horizontal gray lines demarcate $t_{\rm cool} / t_{\rm ff} = \{1,10\}$. We see that the $t_{\rm cool} / t_{\rm ff} \lesssim 10$ regime becomes viable only in the very center of most halos. There is, however, diversity, and we expect that systems with shorter central cooling times will be at different points of the feedback-star formation regulation cycle \citep{prasad15}, with varying amounts of cold interstellar gas present in their centers \citep{werner14}. Importantly, at intermediate radii where the abundance of cold gas clouds peaks ($\sim 50 - 200$ kpc), $t_{\rm cool} / t_{\rm ff}$ is large, and in-situ cold-phase growth would be (globally) inhibited.

In this case in-situ multi-phase condensation is only expected in the presence of rather strong density perturbations with $\delta \rho / \bar{\rho} >$ a few. We note that the CGM of these LRG host halos is thermally unstable according to the classical isobaric linear thermal instability criterion: the only strict requirement is $\partial \ln{\Lambda} / \partial \ln{T} < 2$ \citep{sharma10}, such that a decrease in temperature leads to an increased net cooling rate, and thus runaway loss of internal energy via radiative cooling \citep{field65}. However, it has been shown that this condition is not sufficient for multi-phase condensation in the CGM/ICM. In presence of gravity (and entropy stratification), the ratio of $t_{\rm cool}/t_{\rm ff}$ should be smaller than a threshold of $\sim 10$ \citep{mccourt12}, although this value can be higher for large relative  density fluctuations $\delta \rho / \bar{\rho} \gtrsim 1$ \citep{choudhury19}.

Finally, the value of this ratio tends to again drop in the halo outskirts, enabling a scenario whereby cold clouds form due to perturbations at large distances ($>$ several 100 kpc), and then fall inwards. Note that cold-phase condensation depends on the background entropy profile \citep{binney09}. If entropy is weakly stratified, buoyancy oscillations are absent and condensation is somewhat easier and extends to larger radii where $t_{\rm cool} / t_{\rm ff}$ reaches a minimum \citep{voit17}. The hot gas entropy profiles of LRG host halos in TNG50 are relatively flat (shown in inset). They rise sharply from the center outwards only to $\sim$ tens of kpc, and are then nearly constant, \mbox{$\delta S / \delta r < 0.01$\,dex\,(100 kpc)$^{-1}$}, all the way to the virial radii. These entropy profiles are much flatter than in higher mass halos, where galaxy clusters have a steep, power-law entropy profile at large radii. Only the single most massive halo in TNG50 at $z=0.5$ starts to satisfy this condition, with a mass slightly below $10^{14}$\msun. We speculate that this behavior predominantly reflects the different halo mass scales, together with the strong heating mechanism of the TNG black hole feedback model, which efficiently couples its kinetic energy input to the CGM through shock-mediated thermalization \citep{weinberger17,nelson18b,kauffmann19,terrazas20,zinger20}.

\begin{figure*}
\centering
\includegraphics[angle=0,width=3.45in]{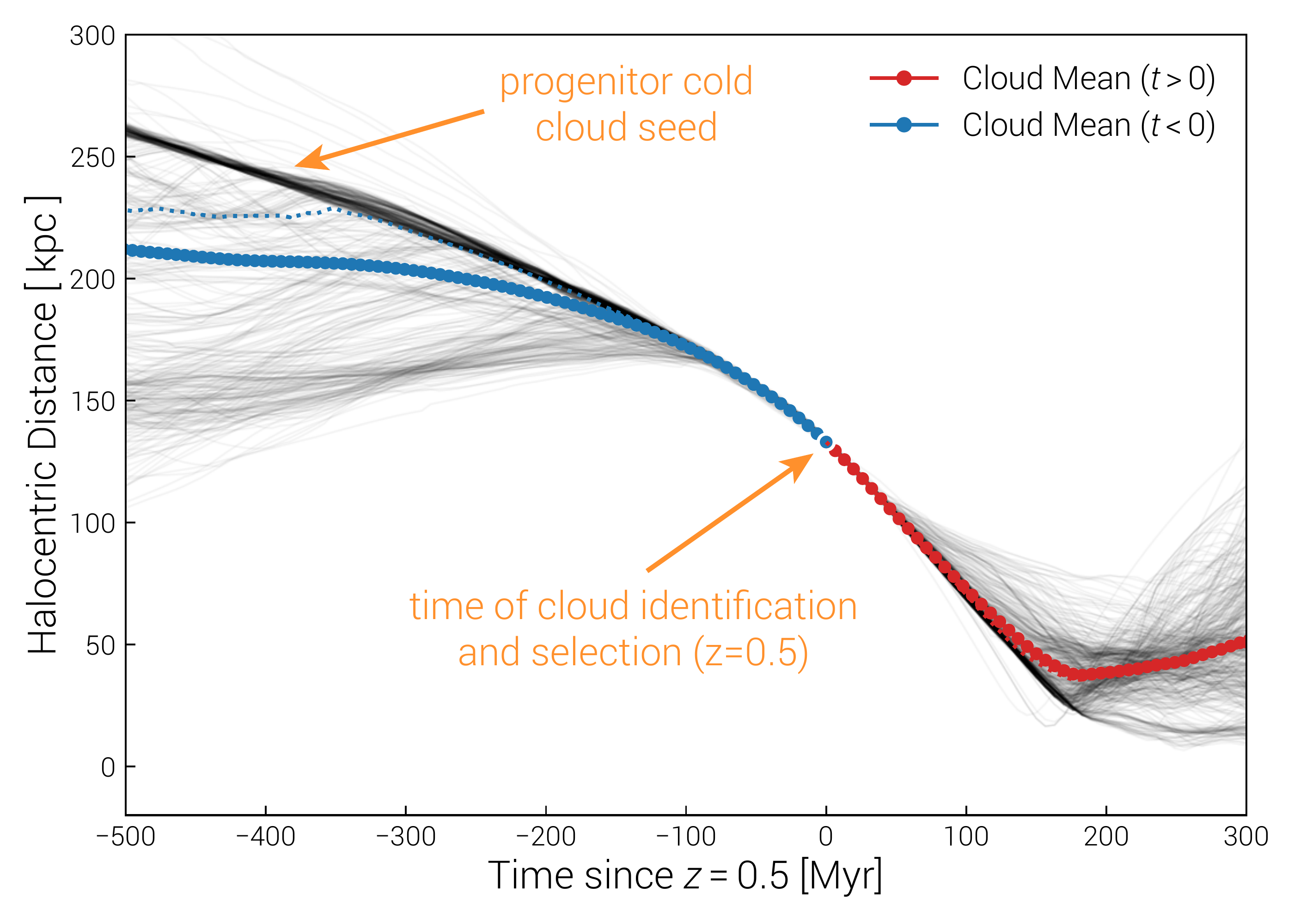}
\includegraphics[angle=0,width=3.45in]{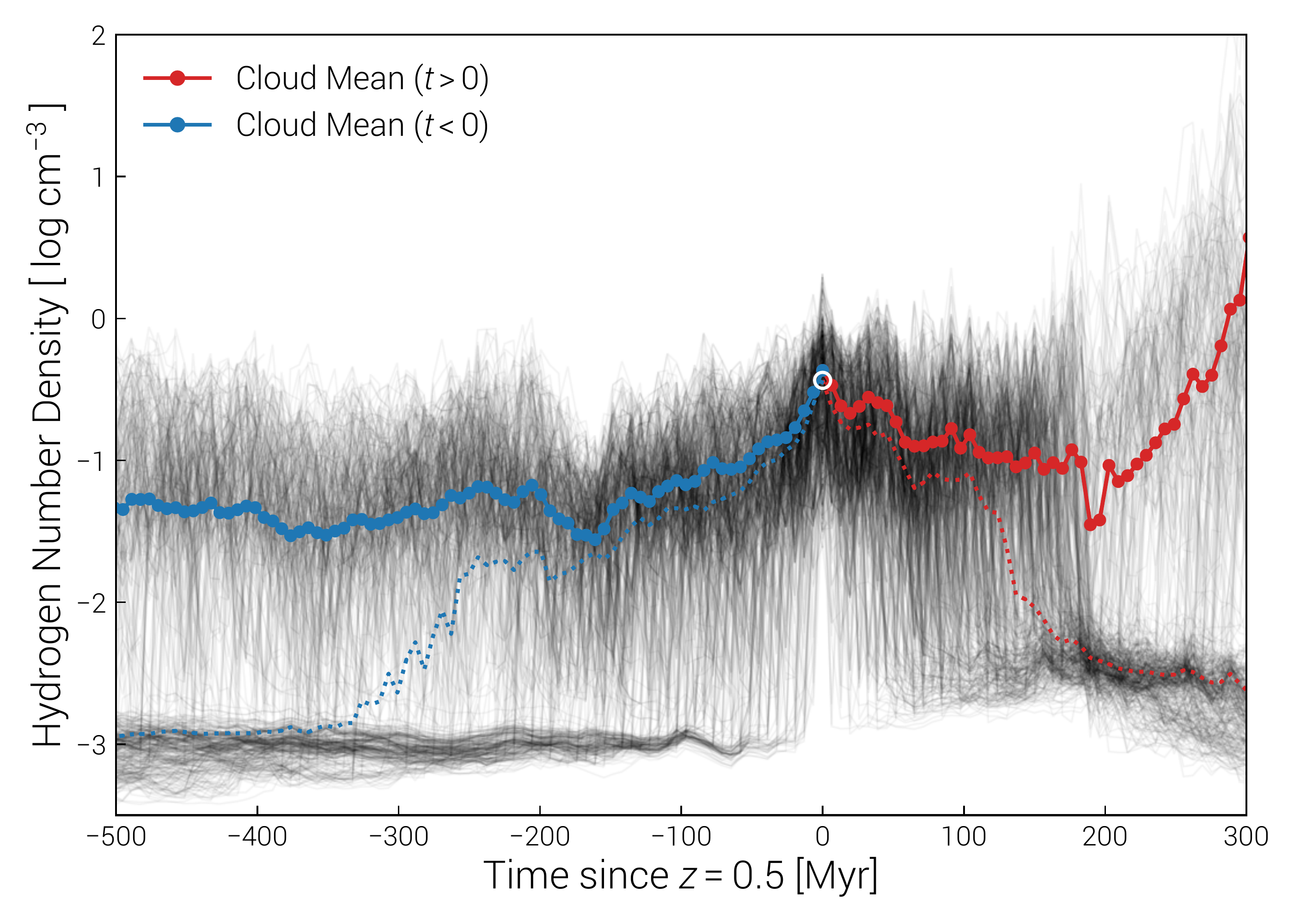}
\includegraphics[angle=0,width=3.45in]{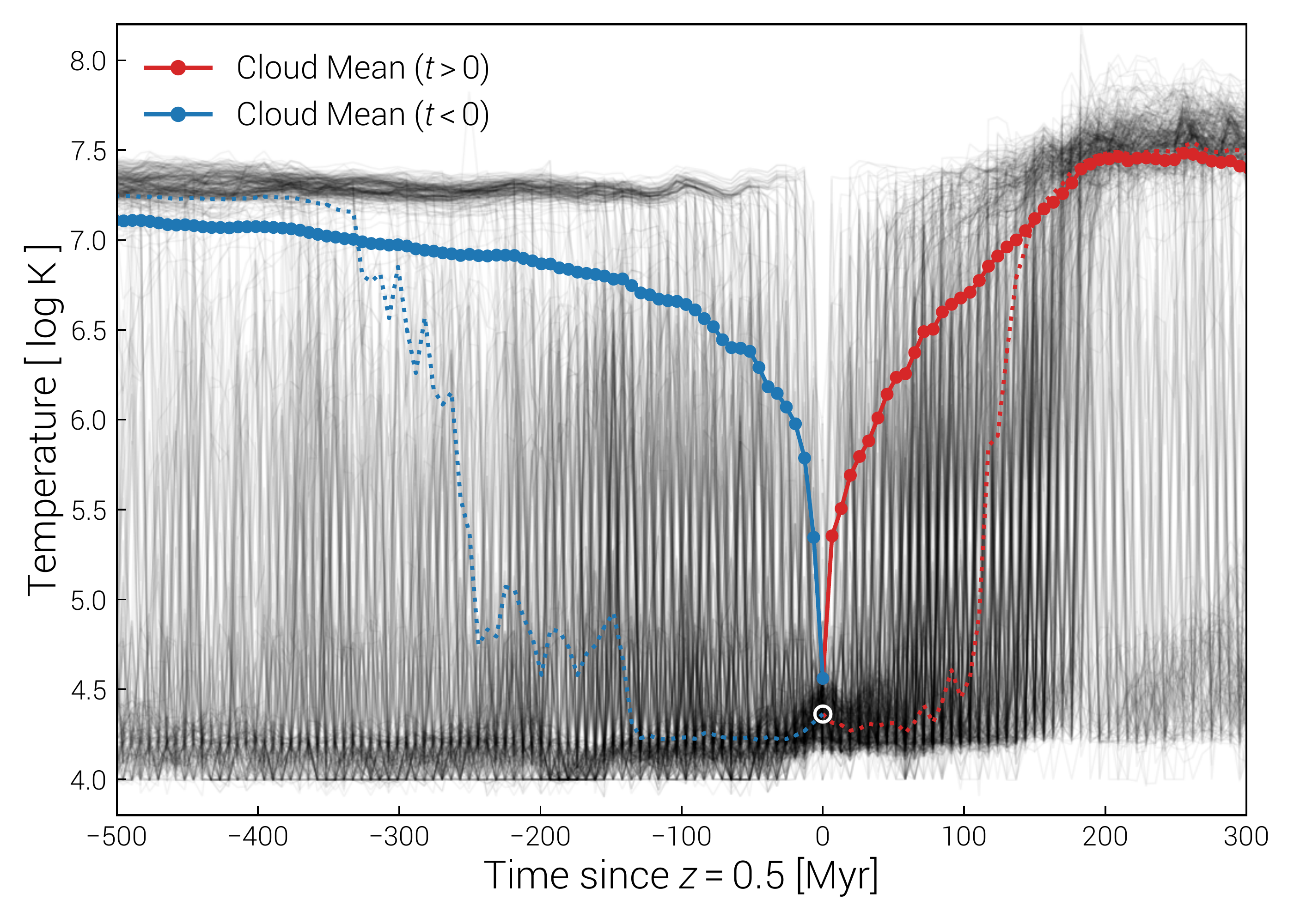}
\includegraphics[angle=0,width=3.45in]{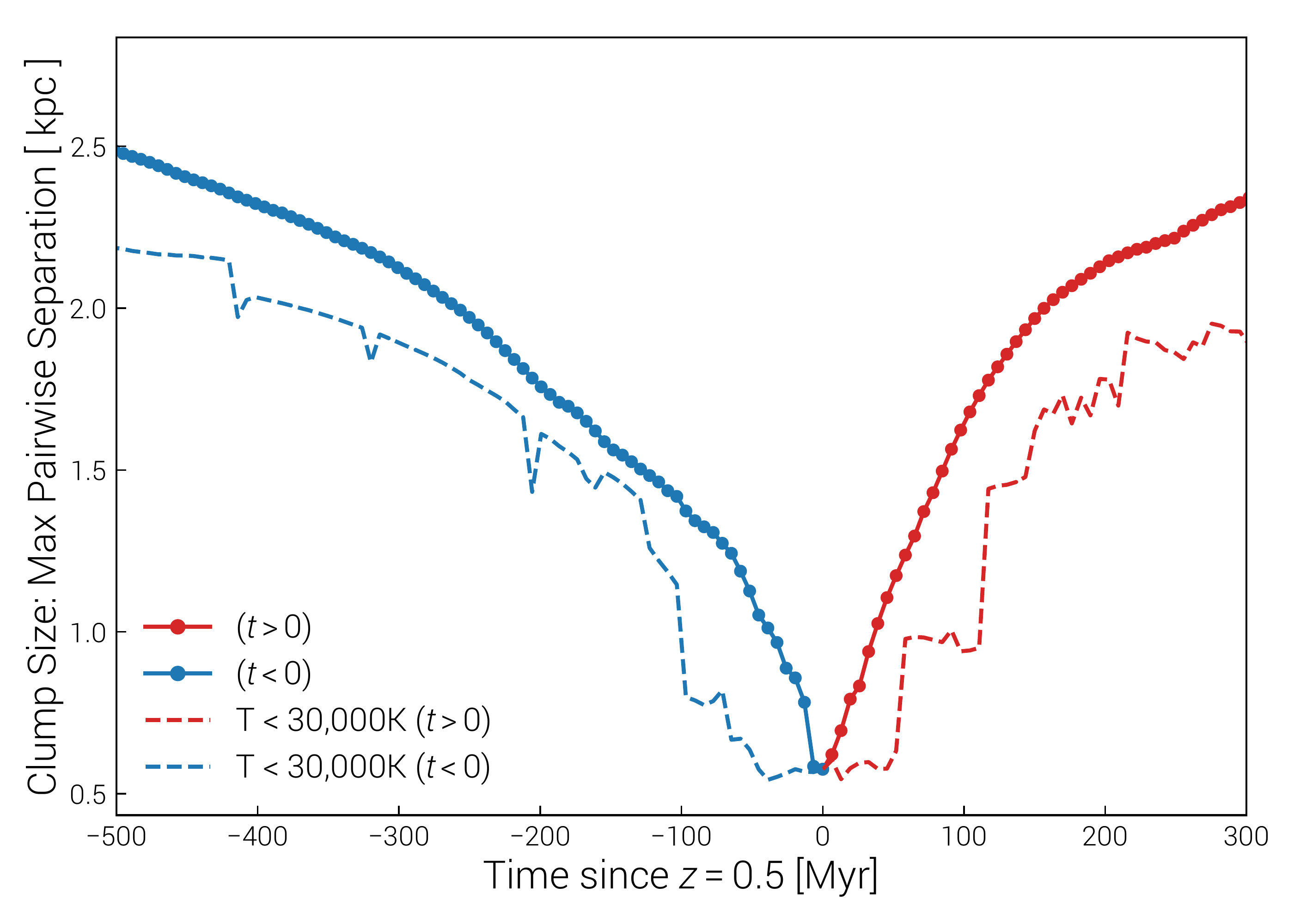}
\includegraphics[angle=0,width=3.45in]{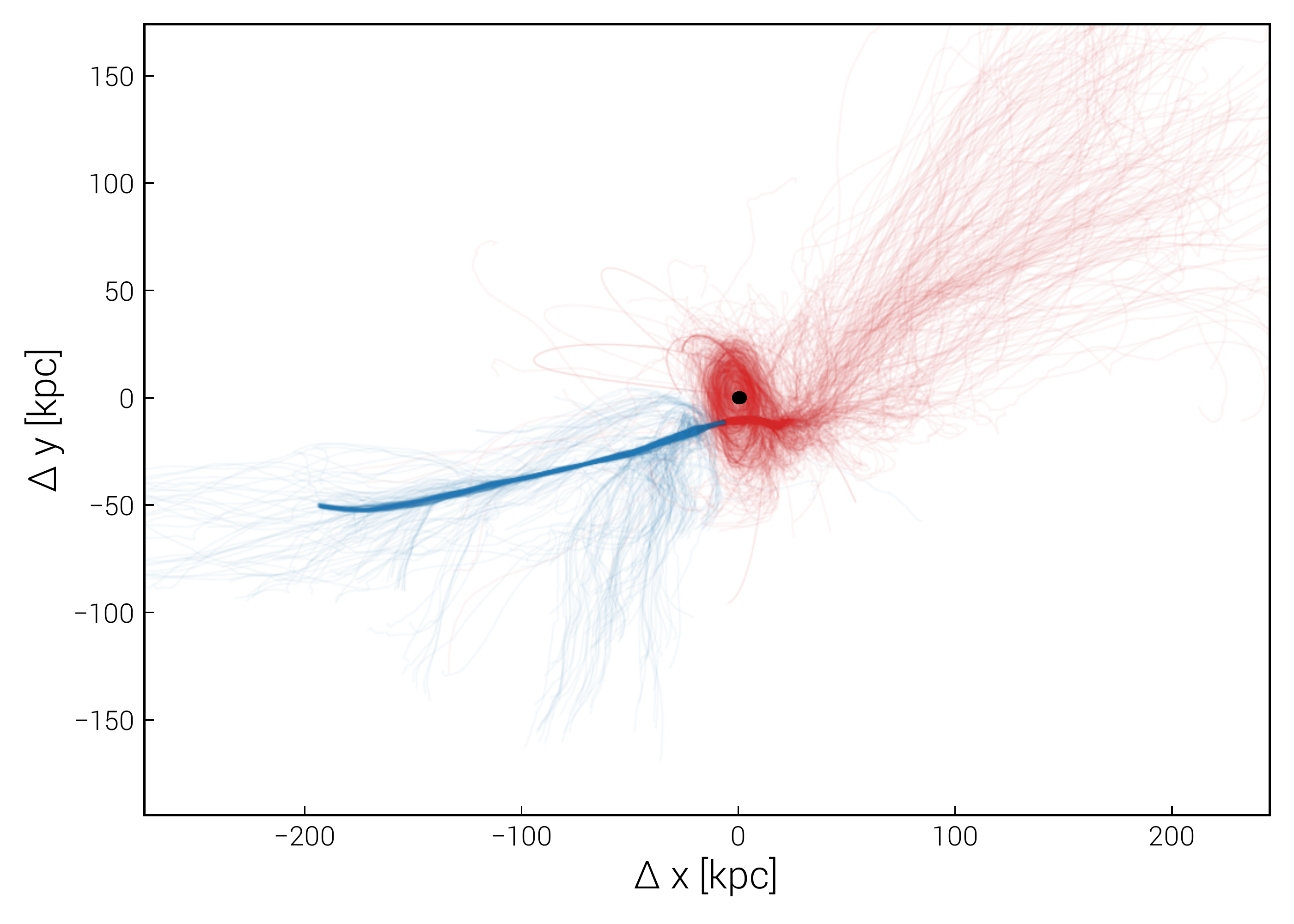}
\includegraphics[angle=0,width=3.45in]{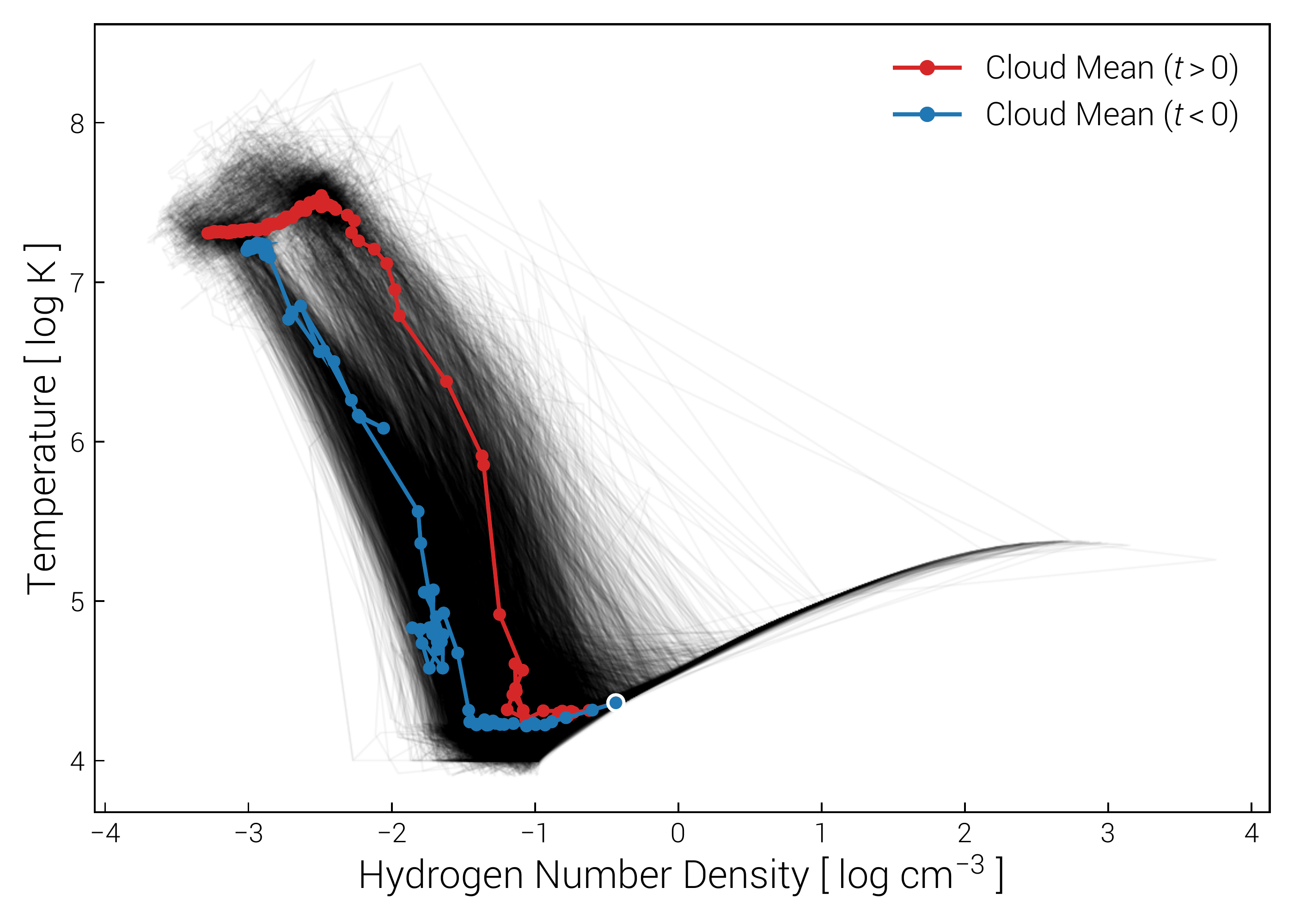}
\caption{ Time evolution of cloud properties. We follow all the member tracer particles in an identified structure, starting at $z=0.5$, going backwards in time to $z=0.6$ (`$t<0$', mean shown in blue; median dotted) as well as forwards in time to $z=0.4$ (`$t>0$', mean in red; median dotted). Circles mark the time intervals between subbox snapshots, which are approximately six million years. Individual tracers are shown in faint black lines. \textbf{Upper left:} distance from the halo center. The cloud is accelerating from the halo outskirts towards the center. The gas which accumulates to make this cloud is seeded by a core which is already coherent $500$ Myr ago. \textbf{Upper right:} physical gas density. The gas is densest around the time the cloud is identified, while tracers exist in both cold-dense and hot-diffuse phases at earlier and later times. \textbf{Center left:} temperature evolution, where the average drops rapidly over the past $\sim 200$ Myr, from $\sim 10^{6.5}$ K to $\sim 10^4$ K. \textbf{Center right:} size evolution, defined as the maximum separation between members. The dashed line shows the size evolution of gas which remains cold. Gas which makes up the cloud is spatially localized at $z=0.5$, but this mass disperses rapidly in time to occupy large fractions of the halo volume. \textbf{Lower left:} Spatial evolution, showing individual tracer tracks, relative to the evolving halo center (black dot). \textbf{Lower right:} evolution in the temperature-density phase diagram, with the $z=0.5$ mean position marked by the white circle. The narrow feature at high density is the star-forming EOS. 
 \label{fig_cloud_timeevo}}
\end{figure*}

\begin{figure*}
\centering
\includegraphics[angle=0,width=7.0in]{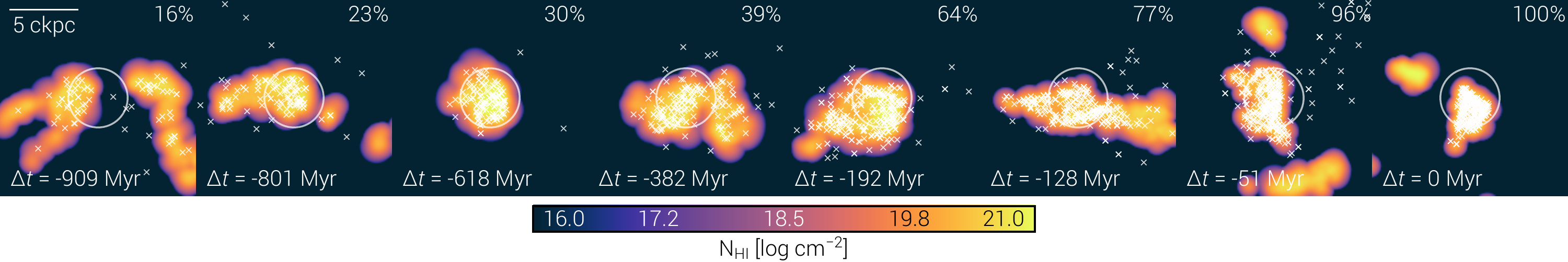}
\vspace{-0.3cm}
\caption{ Visualization of the time evolution of the gas structure of a cold cloud. Time progresses from left to right, spanning the 1 Gyr prior to $z=0.5$ (rightmost panel), and the color shows neutral HI column density. The shrinking center position of the cloud is marked with the circle, which has a constant size equal to the $z=0.5$ half mass radius. Individual white crosses mark the locations of all member tracers which remain in the field of view (fraction given in the upper right corner), which is 15 ckpc across and along the line of sight. This cloud undergoes numerous mergers and fragmentations, with a high cycling of gas between phases, into and out of the cloud. The binary merger at $\Delta t \sim -900$\,Myr combines two progenitor cloud seeds which exist at even earlier times.
 \label{fig_cloud_timevis}}
\end{figure*}

The right panel of Figure \ref{fig_radprof_tcool_tff} again shows the $t_{\rm cool} / t_{\rm ff}\,(r)$ plane, now colored according to the local radial velocity (negative denoting inflow). We emphasize that the distribution of mass, i.e. where most gas is, cannot be ascertained directly from this plot. The diffuse background medium typically has $t_{\rm cool} / t_{\rm ff} \gtrsim 100$ and cooling times $\sim t_{\rm H} - 10 t_{\rm H}$. Most gas is in this hot phase, and only denser gas inside, and in the overdense wakes behind, cold clouds reaches $t_{\rm cool} / t_{\rm ff} < 10$. What is clear is that there is a nonzero abundance of such gas at most radii, although this is very subdominant by mass. Furthermore, this gas has distinct kinematics: it is preferentially infalling (negative radial velocity). This is expected, as isobaric cold gas is denser than its surroundings and therefore falls inwards \citep{singh15}. An overdense cloud will naturally sink towards the center of the halo until the ratio of the background to cloud density approaches unity \citep{pizzolato05}, and will in general infall more slowly than the free-fall speed because of drag forces from the hot background \citep{afruni19}.

We note the important and obvious contrast: cold gas entrained by outflows, or produced within outflows, will instead have positive velocities, at least initially \citep{wang95,veilleux05}. We can see some cold gas with positive velocity in Figure \ref{fig_radprof_tcool_tff} (right panel), which may be pushed out by, or condensed in the periphery of \citep{russell19}, rising bubbles driven by AGN feedback \citep[seen in TNG50 outflows;][]{nelson19b}. In all cases, cold-phase gas is not generally settled within the gravitational potential.

\subsection{Time evolution of cold cloudlets}

To more explicitly study the time evolution of cold gas clouds, we turn to the Lagrangian tracer particles \citep{genel13,nelson13}. These allow us to follow the position and thermodynamical properties of all the gas mass which comprises a cloud at $z=0.5$, tracing it both forwards and backwards in time. We thus ask the question: what was its origin, and what will be its fate? 

Figure \ref{fig_cloud_timeevo} shows the time evolution tracks for a randomly selected cold cloud at $z=0.5$. The cloud has a radius of $\simeq 1$ kpc, containing 400 gas cells and 483 tracers. We contrast the median/mean behavior backwards ($t<0$; blue) and forwards ($t>0$; red) in time. Individual tracer trajectories are shown as thin black lines, and we track this cloud through a subbox volume, giving us a high time resolution of $\sim 6$\,Myr. Note that by identifying a set of tracers and tracking this fixed ensemble through time, we are following the Lagrangian history of gas which makes up the cloud at $z=0.5$, rather than re-identifying the cloud at different epochs to measure the evolution of its properties.

First, the distance from the center of the halo (upper left) shows that the cloud is infalling from the halo outskirts towards the center. The cloud traverses $\sim$ 100 kpc in $\sim$ 500 Myr prior to $z=0.5$, but the (radial) infall velocity of the cloud is not constant with time, as evidenced by the increasing slope of the halocentric distance versus time. At $t \sim -200$ Myr, the mean density of the cloud rises significantly (upper right), increasing the density contrast with respect to its local background. As a result, buoyant support is increasingly lost, and the cloud begins to sink rapidly towards the halo center. On similar timescales, the average temperature correspondingly plummets (center left) by a factor of one hundred, from $\sim 10^{6.5}$\,K to $\sim 10^{4.6}$\,K. The phase diagram (lower right) shows that tracers which make up the cloud at $z=0.5$, cold and dense at the time of identification, have cooled from the diffuse hot phase in the past, and will return to this hot phase in the future.\footnote{The narrow feature visible at $n_{\rm H} > 1$ cm$^{-3}$ is the equation of state for star-forming gas in the TNG model. Some tracers briefly occupy such high density gas during their evolution, however we have explicitly checked that clouds are dispersed by hydrodynamical processes and not by supernovae feedback, as no wind-phase outflows are generated.}

It is clear, however, that some of the tracers reside in cold/dense gas cells at earlier times, indicating that a progenitor cloud exists. Gas making up the cloud is only briefly spatially localized (center right); we show a size measure defined as the maximum pairwise separation between tracers. At $z=0.5$ this gas is confined to a size of $\lesssim 1$ kpc, but disperses both forwards and backwards in time, rapidly distributing among $\gtrsim 100$ kpc halo volumes.

Despite this apparent dispersal, for both hot and cold tracers, it is clear that a coherent `core' exists, particularly in the past. This progenitor cloud is visible in the thick bundle of trajectories in halocentric distance versus time (marked with the orange arrow), and in the spatial trajectories of the tracers themselves (lower left). Here, blue again denotes times prior to $z=0.5$, and red the behavior into the future, with the halo center marked by the black dot. This coherent core, which infalls from the left, is both cold and dense at much earlier times, and suggests that the cloud has formed through a stimulated cooling/accretion onto this initially overdense seed.

After $t=0$ the majority of the mass accretes towards the halo center and comes into small radii orbits, while some is ejected to larger distances, either dynamically or due to energy injection from the central supermassive black hole. The portion of this cold gas which reaches the center of the galaxy will fuel subsequent star-formation and supermassive black hole growth. 

Backwards in time the cloud has a complex assembly history, as evidenced in Figure \ref{fig_cloud_timevis}. Here we show the evolving gas structure of the cloud, from $z=0.5$ to 1 Gyr prior, through a series of time panels of the HI column density in a small $\sim$\,15 kpc field of view. The cloud undergoes both fragmentation and mergers with other clouds. It exists as an identifiable and discrete cloud for more than one billion years. Although almost all of the gas (tracers) which comprise this cloud at $z=0.5$ are completely in the hot phase 1 Gyr ago, the cloud has a clear cold progenitor at this time. The ultimate source of the initial perturbation is not entirely clear. By the time all the original tracers are widely distributed elsewhere in the hot gas, the progenitor cold cloud is associated with a large collection of other clouds, co-spatially related to a satellite galaxy which accreted at even earlier times. This implies that the seed cloud (i) is a fragment of a tidal or ram-pressure stripped tail, or (ii) cooled as a result of perturbations in, or near, satellite galaxy gas -- i.e. group environmental processes \citep{gauthier13,osullivan18}.

A detailed look at the motion of tracers into and out of the assembling cloud shows that there is a high cycling of material, whereby stripping and mixing of cold gas back into the hot phase competes with accretion and cooling into the cold phase. Cold gas clouds at any instant in time have been seeded by the existence of cold clouds in the past, but the gas which resides in the cold phase has become cold in the not so distant past. This strong cycling of mass between phases implies that tracers selected in the original cloud do not have much information about the original perturbation. Once a population of cold clouds exist, they also tend to self-perpetuate and form structures at a small scale, by continually breaking up and colliding with one another.

As the cloud falls inward, with appreciable velocity, it clearly accretes hot gas from its forward looking direction, i.e. the hot medium into which it moves. As its orbit becomes more stationary in space (e.g. at apocenter) and its motion with respect to the halo center/background gas is reduced, the cloud accretes hot gas much more isotropically. This cloud is a single case study, we have verified through inspection of numerous similar tracks that its evolution reflects the common behavior of cold clouds. Although clouds on less radially plunging orbits may experience less heating, the continual mixing in of new gas and the long-term survival and growth of initial cold cloud seeds is the norm.

To be quantitative we undertake a global tracer analysis of halo gas. For every tracer in the CGM we determine its accretion time, determined as the most recent $r_{\rm vir}$ crossing time, and the accretion `mode', segregating between (i) smooth from the IGM, (ii) stripped/ejected from a galaxy prior to its infall, or (iii) merger \citep[following][]{nelson15a}. The last is defined as all gas gravitationally bound to a satellite at its infall time. We contrast gas which comprises cold clouds versus the hot halo at $z=0.5$. First, cold cloud gas has accreted, on average, more recently, with a median accretion redshift of $z \sim 0.87$ versus $z \sim 0.96$, a difference of 400 Myr. This deviation increases for gas in the outer halo. Second, cold cloud gas has a somewhat smaller fractional contribution from smooth accretion, $\sim 5$\% versus $\sim 12$\%, and a correspondingly larger contribution from merging satellites, $\sim 78$\% versus $\sim 69$\%. Stripped origin gas makes up the remaining $\sim 20$\% in both cases. Together, these points support the strong role of gas stripped from infalling satellites as a dominant contributor to CGM cold clouds.

This seeded growth may be related to the mechanism for cool gas formation proposed by \cite{gronke18}, who investigated the growth of a pre-existing, overdense gas cloud within a constant velocity outflow. That work found that clouds can stimulate mixing with the surrounding hot phase and thereby increase the total cold gas mass \citep[see also][]{marinacci10}. However, unlike in a galactic outflow, velocity structure in the extended CGM is primarily driven by gravity, as is the relative velocity between phases. 

We therefore estimate if the cold cloud seeds seen in the TNG50 massive halos satisfy the relevant criterion, such that cold gas clouds/tails should grow by a similar mechanism. To do so we evaluate Eqn. 2 of \cite{gronke18}, which compares the cooling time of an intermediate (mixed) phase of gas at the cloud-background interface, $t_{\rm cool,mix}$, to the destruction timescale of such a cold cloud, $t_{\rm cc}$. These analytical arguments lead to a minimum cloud size $R_{\rm cloud}$, such that $t_{\rm cool,mix}/t_{\rm cc} < \alpha = 1$, given by

\begin{equation}
R_{\rm cloud} > \frac{v_{\rm wind} \,t_{\rm cool,mix}}{\chi^{1/2}}
  \simeq \rm{2 pc} \frac{T_{\rm cloud,4}^{5/2} \,\mathcal{M}_{\rm wind}}{P_{3} \,\Lambda_{\rm mix,-21.4}} \frac{\chi}{100}.
\end{equation}

\noindent where $T_{\rm cloud,4} = T / 10^4$\,K and $P_3 = nT / (\rm 10^3 \,cm^{-3}\, K)$. We plug in values for our regime, where the relative velocity between cold seeds and the CGM is a few 100 km/s, such that the mach number $\mathcal{M} \sim$ (100 km/s)/$c_{\rm s}$ $\sim 0.2$, where $c_{\rm s}$ is the sound speed of the background. The density contrast $\chi \sim T_{\rm hot} / T_{\rm cold} \sim 10^3$, and the pressure is $P = P_{\rm CGM,hot}$ such that $P_3 \sim 1$. Note that magnetic fields are neglected in this estimate. The result is $R_{\rm cloud} \sim 5 - 50$ pc, implying that a seed cold blob of size larger than this should entrain more cold gas as it falls in. Clearly, all our cold clouds resolved in the CGM of TNG50 halos are in this regime, and we would expect the mechanism of \cite{gronke18} to operate. However, whether or not cold clouds actually grow in this way remains unclear -- \cite{schneider20} show with high-resolution wind simulations that outflowing cold clouds satisfying this same criterion do not grow, and that the overall cold-phase gas mass decreases rather than increases. The origin of this discrepancy remains unresolved, but is likely due to different assumptions on turbulence and/or cooling in the background medium.

In our simulated halos there are additional cosmological perturbations which add to the realism, and the complexity, of the problem. Importantly, halos as massive as LRG hosts frequently host satellite galaxies, which act as strong perturbers to the background medium. These perturbations generate turbulence, facilitating cold cloud coalescence and fragmentation \citep{mohapatra19} and injecting gravitational energy \citep{subramanian06}. At the same time, ram pressure stripping removes cold interstellar gas, producing extended tidal tails and jellyfish galaxies \citep{yun19}, leading to observable cold gas in intragroup media \citep{bielby17,poggianti17,nielsen18,johnson18}. Stripped gas can also induce additional cooling of the hot halo onto a central galaxy \citep{zhu18}. It is not clear if the physical mechanisms of cold-phase growth, typically studied in the context of galactic outflows, also apply in the case of this gravitational problem.

Overall, condensation/precipitation appears to play an important role in producing the cold CGM gas in these halos. However, in contrast to more idealized investigations, we find that this is a local, rather than global, process, sourced ultimately by perturbations of cosmological origin.

We reiterate that the TNG model does not include thermal conduction, and in TNG50 we will not resolve the Field length of the cold phase, which in the absence of mitigating effects may be important for the long-term dynamics of a thermally unstable gas \citep{koyama04}. Future idealized studies of cold-phase evolution can take the cosmological outcomes we find here as the starting point for more detailed, high-resolution investigations of disruption, mixing, and survival.


\begin{figure*}
\centering
\includegraphics[angle=0,width=7.0in]{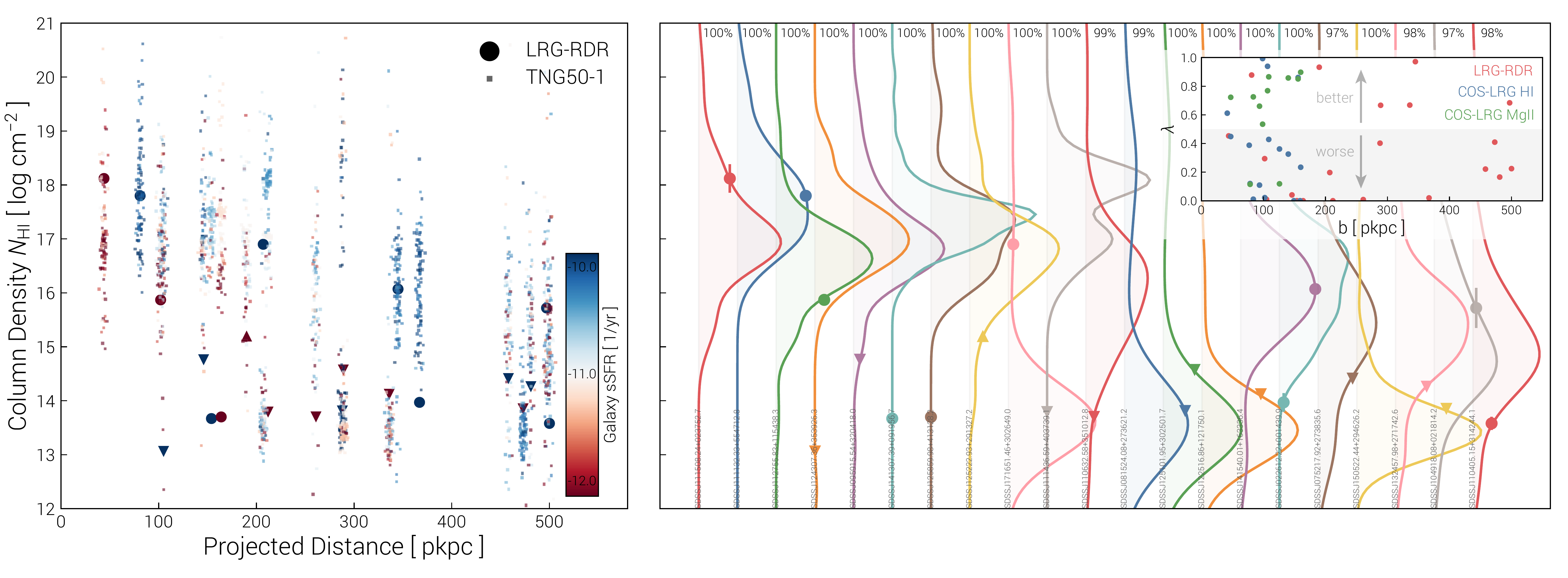}
\caption{ Comparison of neutral HI column densities around TNG50 LRG host halos and observational data of the LRG-RDR survey at $0.3 \lesssim z \lesssim 0.6$. The left panel shows $N_{\rm HI}$ as a function of impact parameter (b), where large circles are the 21 observed galaxies, and small squares are the 100 matched simulation analogs for each. Note that agreement along the x-axis is by construction, while agreement along the y-axis represents model validation. Observational limits are denoted by upward or downward triangles, and color corresponds to specific star formation rate (sSFR $> 10^{-11}$\,yr$^{-1}$ in blue, otherwise in red). The right panel shares the same y-axis, and repeats the same observations with errorbars, ordered in ascending impact parameter, horizontally spaced for visual clarity. Under each data point we show the PDF of simulated $N_{\rm HI}$ columns (colored lines) for the 100 matched TNG50 galaxies. The inset shows the $\lambda$ values (see text) versus impact parameter for each sightline of the three surveys, where higher values represent better agreement and lower values worse agreement.
 \label{fig_lrg_rdr_comparison}}
\end{figure*}

\section{Comparison with Observations} \label{sec_obs}

We have so far not addressed the question of whether or not the cold gas contents and structures observed in TNG50 halos are in any way consistent with available observational data. In this section we contrast against several datasets: the COS-LRG survey \citep{chen18,zahedy19}, the LRG-RDR survey \citep{berg19}, the COS-Halos survey \citep{tumlinson11,werk13}, and stacking results around the main SDSS LRG sample \citep{lan14,lan18}. We focus exclusively on absorption, in MgII as well as neutral HI, both of which trace cold $\sim 10^4$\,K gas.

We begin with Figure \ref{fig_lrg_rdr_comparison}, which contrasts against LRG-RDR with the observational comparison strategy of \cite{nelson18b}. In short, we construct a tailored `mock' survey consisting of a simulated galaxy sample which is statistically consistent with the observed galaxies. To do so, we create 100 survey realizations, matching the stellar mass (assumed $\sigma = 0.2$ dex) and specific star formation rate ($r < 30$ pkpc values) of each of the 21 observed galaxies, selecting from all central galaxies in the simulation snapshot consistent with the observed redshift (spanning $0.28 < z < 0.61$). For each of these 2100 simulated galaxies, we measure a $N_{\rm HI}$ column density at the \textit{same} impact parameter as the observed galaxy, taking a random position angle and random projection direction with respect to the galaxy through a depth of $\Delta v = \pm 1000$\,km\,$s^{-1}$.

The (dis)agreement varies between the observed LRGs. For instance, the first PDF (red) shows moderate agreement and the second (blue) demonstrates excellent agreement, since the observed column falls near the peak of the PDF of predicted columns. On the other hand, the fourth galaxy (orange) is entirely inconsistent, as the predicted PDF is compact near $N_{\rm HI} \sim 10^{17}$ cm$^{-2}$, while the observations suggest only an upper limit of $N_{\rm HI} < 10^{13}$ cm$^{-2}$. To be quantitative, we calculate a statistical measure $\lambda \in [0,1]$, similar to a p-value, defined either as (i) the integral of the normalized area of the PDF consistent with a limit, or (ii) twice the smaller of the two half-definite integrals bounded by the observed value, in the case of a detection. If an observation lies near the peak of the predicted PDF, $\lambda \rightarrow 1$ (indicating good agreement), whereas if it sits far into the tail, $\lambda \rightarrow 0$ (indicating poor agreement). We find, averaging across the observed sample, $\langle \lambda \rangle = 0.22^{+0.46}_{-0.21}$, indicating that the observed data and the simulations have on average tension at the $\sim 1\sigma$ level. This means the two are not inconsistent at the level sufficient to conclude that the observed points could not have been drawn from the simulated PDFs. The agreement is better for detections with $\langle \lambda \rangle = 0.26$, and worse for upper limits, $\langle \lambda \rangle = 0.19$. 

We note that this level of agreement is worse than that demonstrated in \cite{nelson18b} using the same procedure to compare OVI column densities versus COS-Halos data, where we found $\langle \lambda \rangle \simeq 0.52$, indicating rather good agreement. For the TNG model there is more tension in the cold-phase contents of the CGM than in the hotter $\sim 10^{5.5}$\,K phase traced by OVI, although these comparisons apply to different halo mass regimes. 

\begin{figure*}
\centering
\includegraphics[angle=0,width=3.45in]{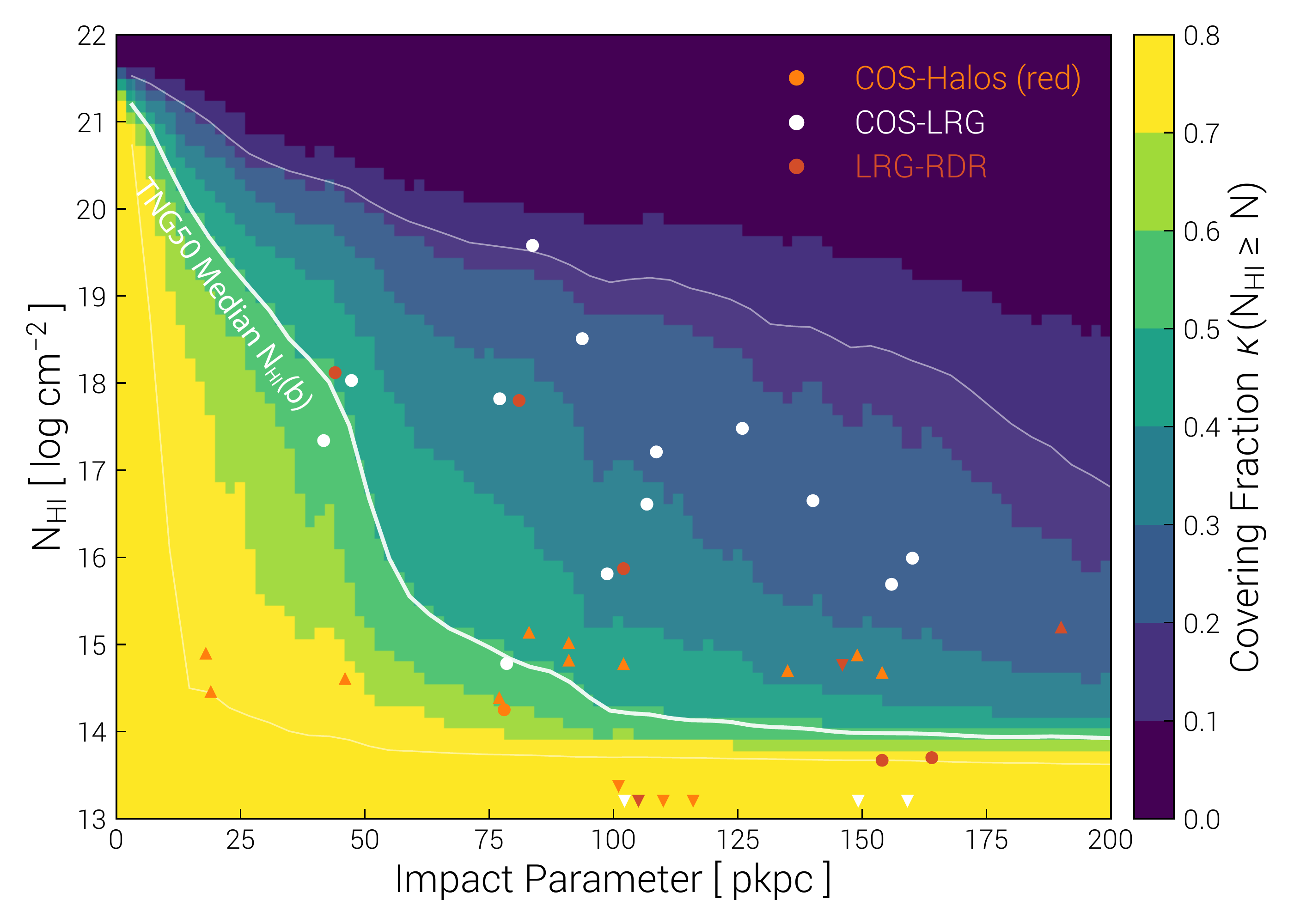}
\includegraphics[angle=0,width=3.45in]{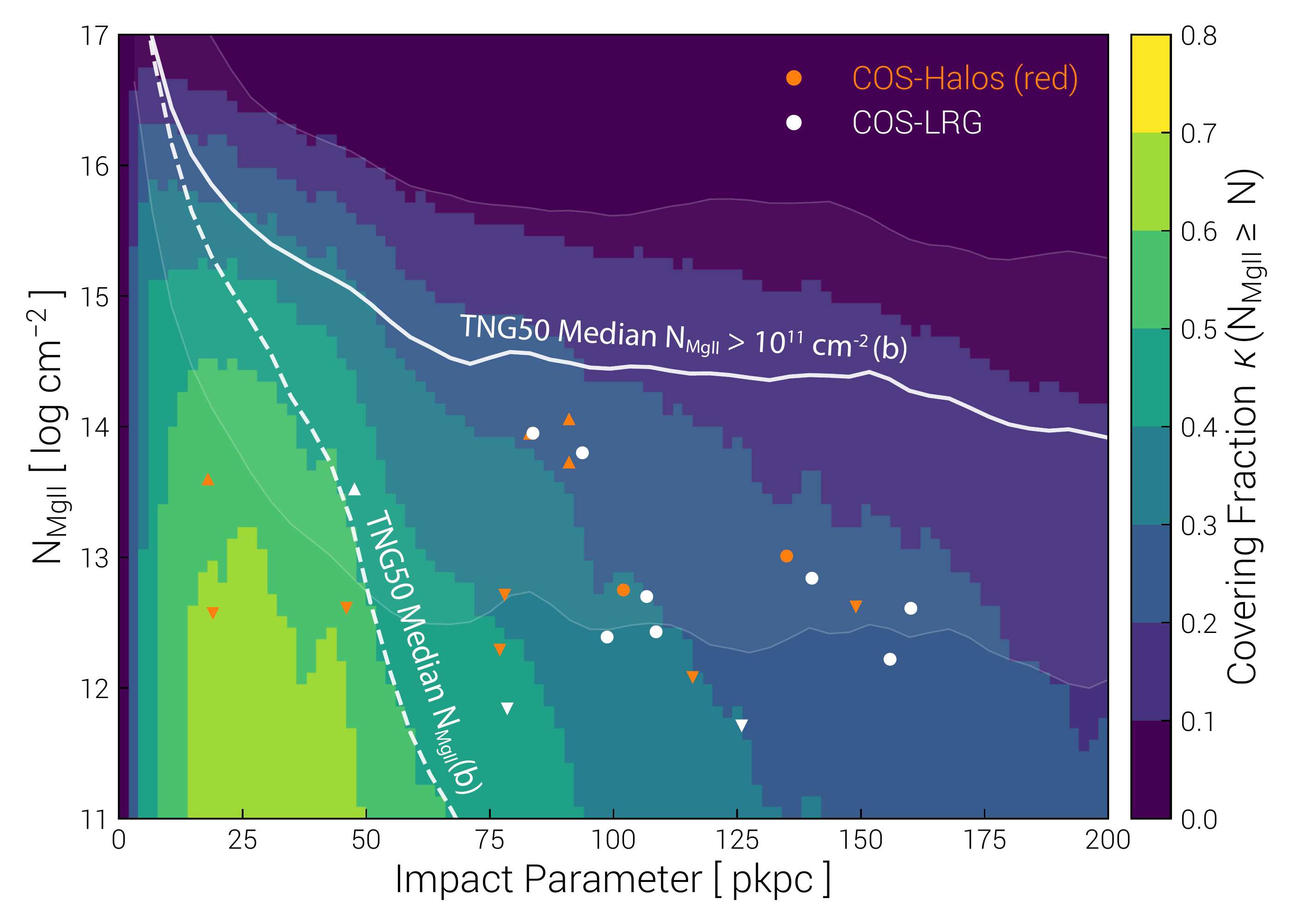}
\caption{ Cumulative covering fractions of HI (left) and MgII (right) for the halo mass bin $10^{13.2} - 10^{13.8}$\msun. In both cases, the background color shows the fraction of sightlines at that impact parameter which would intersect a column density $N_{\rm HI,MgII}$ or higher. We compare to COS-LRG \citep{zahedy19}, LRG-RDR \citep{berg19}, and COS-Halos \citep[red galaxies only;][]{werk13} observations. The solid white line shows the median $N_{\rm HI}(b)$ (left panel, with $1\sigma$ scatter band) from TNG50. The corresponding $N_{\rm MgII}(b)$ relation is shown in the right panel (dashed line), contrasted against the median MgII column density trend restricted to sightlines which intersect a nonzero abundance of MgII (solid line). We adopt a uniform projection depth of $\Delta v = \pm\,$500 km/s.
 \label{fig_coverfracs_hi_mg2}}
\end{figure*}

Repeating the same procedure for the COS-LRG survey of 16 systems (not shown for brevity), we find an average value of $\langle \lambda_{\rm HI} \rangle = 0.34^{+0.42}_{-0.32}$ in the case of the neutral HI column comparison. This implies similar if slightly better agreement. For 11 of the 16 LRGs we also have $N_{\rm MgII}$ measurements available and derive a corresponding $\langle \lambda_{\rm MgII} \rangle = 0.24^{+0.26}_{-0.02}$. Individual $\lambda$ values as a function of impact parameter are shown in the inset in Figure \ref{fig_lrg_rdr_comparison}. The distributions of MgII column densities also reveal an important point: a sightline either intersects a large column of cold gas, or no cold gas at all -- there is little room in-between. This is evident visually from the MgII and HI distributions around our prototypical halo, Figures \ref{fig_vis_mg2_single} and \ref{fig_vis_hi_single}, where a sightline either intersects (one or more) cloud(s), or it does not, passing only through diffuse hot halo gas \citep[which itself could have small though observable coronal Ly$\alpha$ absorption;][]{richter20}. This is consistent with \cite{berg19} who argue that the observed non-continuous $N_{\rm HI}$ distribution implies LRGs exhibit either high $N_{\rm HI}$ gas, or not much at all. We therefore expect that the covering fraction $\kappa_{\rm MgII}$ should be reasonably low, but that every detection should be strong \citep[see also][]{gauthier10}.

Overall, the HI and MgII column density measurements from the LRG-RDR and COS-Halos surveys are not in remarkable agreement with the simulation predictions, and mild to moderate tension between the two is clear. Unfortunately, the small sample sizes of $\sim$ tens of LRGs currently available in targeted surveys preclude a more concrete conclusion at present. Better statistics in future surveys will allow much more robust comparisons.

\subsection{Covering fractions of cold gas tracers}

Figure \ref{fig_coverfracs_hi_mg2} outlines the predicted covering fraction of cold gas around massive halos in TNG50. The left panel shows neutral HI, and the right panel MgII. In both cases we consider the plane of $(N,b)$, column density versus impact parameter, showing with the color the covering fraction $\kappa$ for the threshold $\geq N_{\rm HI,MgII}$ at that particular distance. We include the COS-LRG (white), LRG-RDR (red), and COS-Halos red/quiescent (orange) datasets, although the latter is only for reference, as these targets are much lower mass than LRGs ($M_\star \sim 10^{10.8}$\msun versus $\sim 10^{11.3}$\msun on average).

At a particular column, the covering fraction of neutral HI drops rapidly with distance, as we move horizontally across the left panel. For instance, we find that sightlines exceed the LLS boundary of $N_{\rm HI} = 10^{17}$ cm$^{-2}$ $\gtrsim 70$ percent of the time for $b \lesssim 20$ kpc, but this drops to $\kappa \simeq 50$ percent for $b \simeq 50$ kpc. A random sightline at an impact parameter of 100 kpc from the halo center will intersect a LLS with only $\sim 30$ percent probability. This value decreases to $\kappa \lesssim 10$ percent for a DLA $N_{\rm HI} > 10^{20.3}$ cm$^{-2}$ column density. Similarly, at a given impact parameter, the covering fraction decreases rapidly with increasing column density. The median curve of $N_{\rm HI}(b)$ for the simulated halos has three distinct regimes: neutral HI columns decrease (i) moderately fast with distance for $d < 50$ kpc, (ii) rapidly for 50 kpc $< d <$ 80 kpc, and (iii) plateau to a near constant $N_{\rm HI} \sim 10^{14}$ cm$^{-2}$ for $d > 100$ kpc. The variation, between different sightlines and different halos, is significant. The scatter indicated by the two thin white lines (the lower inside the yellow region) is large -- strong absorbers are frequently intersected even when the median column density at a given impact parameter may be orders of magnitude lower.

The behavior of $\kappa_{\rm MgII}$ is qualitatively similar, with the notable exception that there is no `minimum' column density encountered with high probability. Overall, the chance of intersecting an observable column (i.e. $N_{\rm MgII} \gtrsim 10^{12}$ cm$^{-2}$) is low, never exceeding $\kappa \sim 0.6$. Most of this $N_{\rm MgII}(b)$ plane has much smaller covering fractions, typically $20 - 50$ percent for the regimes where sightlines around LRGs exist. The actual median column density drops precipitously, reaching unobservably low values already by $b \sim 60$ kpc (dashed white line), and this is due to the large fraction of (effectively) zero column sightlines. As discussed above, this occurs because there are two distinct possibilities: a sightline either intersects one or more clouds and produces a strong absorber, or it does not, passing through the diffuse hot halo gas. If we consider only those sightlines which intersect cold gas, restricting to the column densities visible on this figure ($N_{\rm MgII} > 10^{11}$ cm$^{-2}$), the corresponding median profile differs substantially already for $d \gtrsim 20$ kpc (solid white line). In this case, $N_{\rm MgII}$ decreases relatively slowly out to $d \sim 80$ kpc and then flattens. If a sightline intersects a cold cloud, then the median $N_{\rm MgII} \gtrsim 10^{14}$ cm$^{-2}$ regardless of distance for $d < 200$ kpc. The scatter is again large -- roughly three orders of magnitude -- indicating the different absorption signatures when a sightline intersects the outskirts versus central core of a cold cloud, as well as a single, versus multiple, clouds. As a point of observational reference, \cite{zahedy19} make an estimate of $N_{\rm cl} = 3.7^{+0.6}_{-0.4}$ clouds intersected per sightline within $d < 100$ kpc, and $N_{\rm cl} = 2.2^{+0.9}_{-0.5}$ for larger impact parameters. 

\begin{figure*}
\centering
\includegraphics[angle=0,width=6.0in]{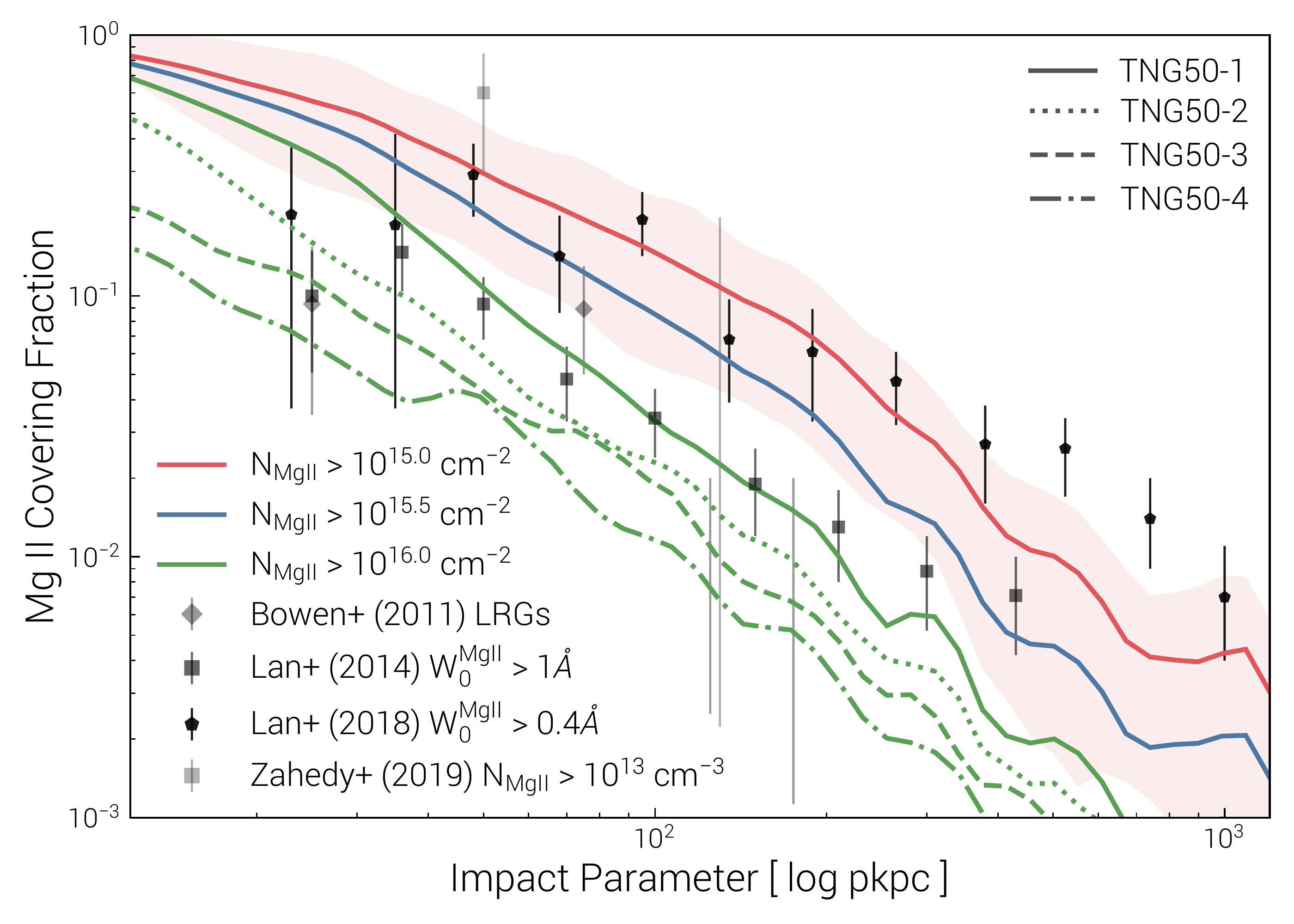}
\caption{ Covering fraction of `strong' MgII absorbers as a function of impact parameter, compared to various observational datasets. For simplicity we stack all halos in the mass range $10^{13.2} - 10^{13.8}$\msun (selecting only on halo mass and not on SFR, color, or other galaxy properties) at $z \in \{0.4, 0.5, 0.6\}$ (to boost statistics) and project MgII columns through a depth of $\Delta v = 1000$ km/s. We qualitatively compare against \protect\cite{lan18} with an equivalent width cut of $> 0.4\,\AA$ and \protect\cite{lan14} for EW $> 1.0\,\AA$. Additional points from \protect\cite{bowen11} and \protect\cite{zahedy19} are included for reference. At present, the level of (dis)agreement with data can only be assessed at face-value and is not intended to be quantitative, as we compare covering fractions based on column density versus equivalent width thresholds. The mapping between these two quantities depends on the MgII curve of growth, and only in the case of a single absorber per sightline could the $> 0.4\,\AA$ points be roughly compared against the green line, and the $> 1.0\,\AA$ points against the red line. Future work will improve this comparison by directly deriving equivalent widths from the simulations.
 \label{fig_coveringfrac_mgii}}
\end{figure*}

It is difficult to make a quantitative comparison to the data in this space. We note that the COS-LRG survey finds that strong MgII absorbers with $N_{\rm MgII} > 10^{13}$ cm$^{-3}$ are common at $d < 100$ kpc, while absorbers at $d \gtrsim 100$ kpc have significantly lower columns which are actually always \textit{lower} than this threshold. For $d < 100$ kpc COS-LRG estimates a covering fraction of $\kappa_{\rm MgII} = 0.60^{+0.25}_{-0.30}$, with $\kappa_{\rm MgII} \sim 0 - 0.2$ at $d = 100 - 160$ kpc, for the same column threshold, and both with large uncertainties \citep{zahedy19}. Considering the outer halo as a whole, the qualitative comparison with TNG50 implies that the simulated covering fraction may be lower -- there are eight detections, but only two upper limits -- whereas the actual $N_{\rm MgII}$ values are on average higher than COS-LRG detections. The poor number statistics are again limiting, and to be more definitive we consider the different regime of MgII detections in spectra from the Sloan Digital Sky Survey.

Figure \ref{fig_coveringfrac_mgii} makes a more quantitative comparison of the covering fraction of MgII around LRGs between TNG50 and the SDSS stacking results of \cite{lan14} and \cite{lan18}. For simplicity, and to boost the statistics of massive halos from the simulation side, we stack together all galaxies in the total halo mass range $10^{13.2} - 10^{13.8}$\msun at three redshifts partially covering the observed range, $z \in \{0.4, 0.5, 0.6\}$. We note that the statistics of the observations in this case far exceed those of the simulations, containing thousands to hundreds of thousands of LRG-absorber pairs. The observations show, surprisingly, that the passive LRG sample is surrounded by non-negligible covering fractions of cool MgII gas, even out to appreciable distances. In particular, for 100 $< d \lesssim$ 500 kpc (i.e. the inner $r_{\rm vir}/2$ CGM), the covering fraction remains at $\sim$ 1 percent for strong ($> 1\,\AA$) absorbers, and $2 - 5$ percent for weaker ($> 0.4\,\AA$) absorbers \citep{lan18}.

Unfortunately, we cannot exactly match the equivalent width thresholds of the data, because we directly compute mock column densities and do not produce synthetic absorption spectra \citep[e.g.][]{oppenheimer06}. The curve of growth of MgII 2796,2803 implies that the $1\,\AA$ threshold (dark squares) could be roughly compared to $N = 10^{16} - 10^{17}$ cm$^{-2}$ while the $0.4\,\AA$ cut (dark circles) could be roughly compared to a column density between $N = 10^{13} - 10^{16}$ cm$^{-2}$. However, these relationships depend on the assumption of a single absorber along each sightline, which is why relatively high columns are needed to reach the large equivalent width values. This may or may not be true in reality. For instance, lower redshift, lower mass halos with MgII absorption rarely show a single component, two or more being more common \citep[][although these systems do not reach the EWs probed by the SDSS data compared to here]{churchill07,werk13}. \cite{zahedy19} infer an average number of MgII absorption components of $\sim$ 3.7 at $d < 100$\,kpc, declining to $\sim$ 2.2 for $d > 100$\,kpc. 

\begin{figure*}
\centering
\includegraphics[angle=0,width=7.0in]{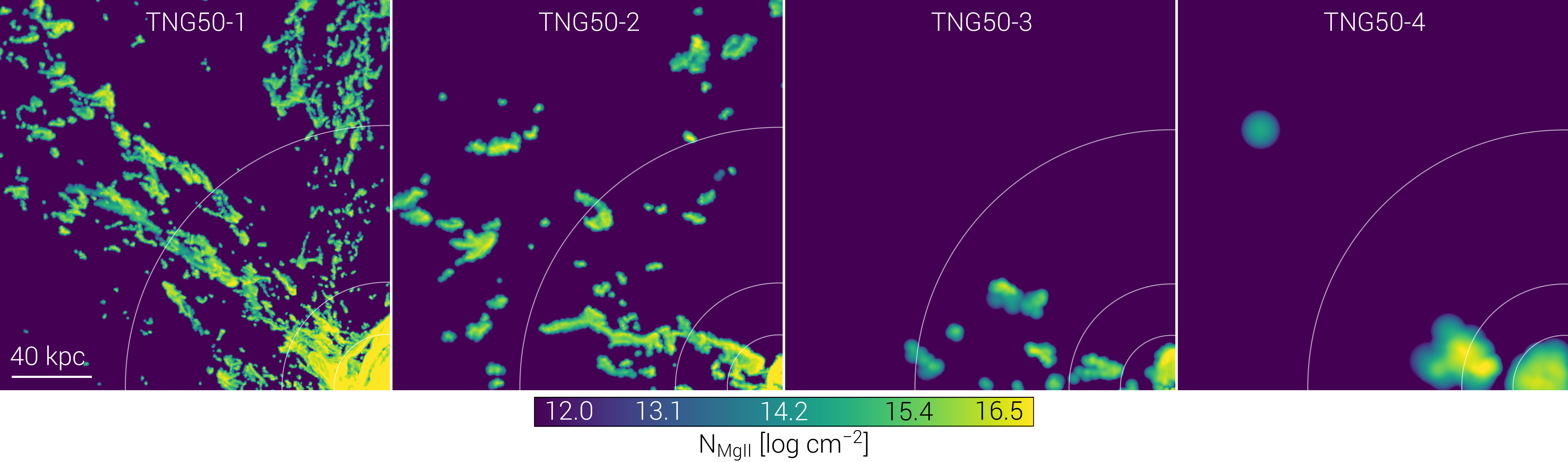}
\caption{ Resolution convergence: visual impression of the distribution and covering fraction of cold MgII-bearing gas in a $10^{13.5}$\msun halo at $z=0.5$, focused on the upper-left quadrant of this system. From left to right: TNG50-1 through TNG50-4, each simulation decreasing by a factor of eight (two) in mass (spatial) resolution. Note that this is exactly the same halo, matched between runs. The overall abundance of cold gas in the CGM of LRGs decreases rapidly with worse numerical resolution, while the individual clouds also grow larger and fewer in number.
 \label{fig_resolution_vis}}
\end{figure*}

By way of an explicit example, to reach an EW $\sim 1\,\AA$ with a single absorber a column of $N_{\rm MgII} \sim 10^{17}$\,cm$^{-2}$ is required. However, a much weaker absorber with $N_{\rm MgII} \sim 10^{15}$\,cm$^{-2}$ will produce an EW $\sim 0.5\,\AA$ ($b \sim 10$\,km/s). Two such absorbers sufficiently close in velocity space would then sum to a total EW of $\sim 1\,\AA$. These ranges cover reasonable Doppler $b$ parameters for the narrow linewidths of MgII absorbers around LRGs, $3 - 15$ km\,s$^{-1}$. It is therefore not clear which configuration may be giving rise to an observed EW. The best way to make this comparison is through synthetic absorption spectra and matching the equivalent width thresholds directly. However, as this is not yet possible in the current work, we present the above comparison in the qualitative, preliminary sense only. In the future, comparisons in spectral/wavelength space will allow us to draw more robust conclusions. They will also enable us to assess the thermal state of absorbing gas, probe line-of-sight kinematics, and enable a comparison of the number of discrete absorbers (i.e. clouds) found along sightlines. 

With these caveats in mind, we find overall that simulated covering fraction profiles are broadly consistent with the observations. In particular, they exhibit non-negligible covering fractions even out to $\sim$ 100s of kpc away from LRGs. There may be an excess of cold, MgII-traced gas in the very centers of halos, $d < 20$ kpc, but data is sparse. There may be a dearth of absorption at large distances, $d > 400$ kpc, but we reiterate that in this regime cosmological projection effects become relevant (i.e. a 2-halo term). Random background absorbers contribute a floor of $\kappa_{\rm MgII} \sim 0.01$ \citep{huang16}. The individual bumps and wiggles in the TNG50 curves reflect the limited statistics, where individual galaxies in projection modulate the results at large impact parameter.

These observational comparisons have an important consequence for the size-scale of cold gas. In particular,  theoretical arguments exist for very small characteristic cold gas cloud sizes, including the $c_{\rm s} \,t_{\rm cool} \sim 0.1$\,pc length-scale proposed by \citep{mccourt18}. As previously noted, any physical mechanism producing gas structure with very small physical scales ($\ll\,$100 pc) would not be resolved in TNG50. That is, our spatial resolution is roughly 1000 times worse than $0.1$\,pc. However, the broad agreement between TNG50 and observational constraints on the covering fractions of cold gas surrounding LRGs demonstrates that such sub-parsec gas structures are not required to explain absorption line data. That is, the `large' gas clouds investigated herein, with characteristic sizes $\sim 100\,\rm{pc} - 1\,\rm{kpc}$ and above, provide an explanation for observations. We conclude that either (i) the idea of this small characteristic length scale is not quantitatively correct, a `fog' of cloudlets remaining qualitatively supported by our resolution trends, or (ii) for observables such as column densities, covering fractions, and line-of-sight kinematics it is not necessary to resolve such a small characteristic scale, as they arise from or are dominated by larger gas structures.

We conclude by returning to the question of numerical resolution, the existence of small-scale cold gas clouds in TNG50, and the inferences available from observations. The three lower green lines (dotted, dashed, and dot-dashed) in Figure \ref{fig_coveringfrac_mgii} show the resolution progression of TNG50-2, TNG50-3, and TNG50-4. Each resolution step corresponds to a factor of eight in mass, or two in space, such that TNG50-1 has eight times higher spatial resolution than TNG50-4. We see that the covering fraction of MgII increases monotonically with increasingly better numerical resolution. This is similar to the finding that the covering fraction of cold (neutral) gas in the CGM of lower-mass galaxies also increases strongly with numerical resolution \citep{fg16}. Similar to the trend with the total number of cold CGM clouds seen in Figure \ref{fig_cloudprops_vs_halomass}, there is little indication that the result is converged even at the resolution of TNG50-1. The implication is that, while larger cold clouds may be realistically resolved at a given resolution, the size of the smallest structures will be set by numerical limits. 

Figure \ref{fig_resolution_vis} shows a visual representation of the resolution dependence of the cold CGM around LRGs. We emphasize the upper-left quadrant of a single $10^{13.5}$\msun halo, the three circles marking $\{0.05, 0.1, 0.25\} \times r_{\rm vir}$. The four panels, from left to right, show the same matched halo, at the same time, in the four runs of the TNG50 resolution series. The overall mass (abundance) of cold gas, as traced by MgII, drops rapidly with decreasing numerical resolution. The qualitative large-scale structure is similar between runs, implying a cosmological accretion origin. At the same time its small-scale morphology changes: individual cold clouds are larger, and fewer in number. Ultimately, even higher dynamic range cosmological simulations will be invaluable to unravel the cold-phase of the CGM within the context of the cosmic baryon cycle.


\section{Summary} \label{sec_conclusions}

In this paper we use the TNG50 cosmological magneto-hydrodynamical simulation \citep{nelson19b,pillepich19} to explore the abundance and origin of cold $\sim 10^4$\,K gas in the circumgalactic medium (CGM) of massive, $\sim 10^{13} - 10^{13.5}$\msun halos at $z=0.5$, the regime of observed luminous red galaxies (LRGs). The high resolution of TNG50 enables us to discover that these halos are filled with large amounts of cold gas, as traced by neutral HI as well as MgII. Our main findings are:

\begin{itemize}
    \item Although the $\sim 10^{13} - 10^{13.5}$\msun dark matter halos hosting massive galaxies are filled with a hot, virialized plasma at $\sim 10^7$\,K, we discover that they also contain a large abundance of cold gas with a characteristic temperature of $\sim 10^4$\,K. This cold-phase of the CGM takes the form of thousands of small, $\sim$\,kpc sized clouds, which fill the inner halo out to hundreds of kpc.
    \item The total amount of cold gas, both neutral HI and MgII, increases monotonically with increasing halo mass. The median column density $N_{\rm MgII}$ also rises, at all impact parameters, with increasing halo mass, particularly at large radii. As a result, the median radial profiles of $N_{\rm MgII}$ become more extended (flatter) moving up to the group-mass scale of $\sim 10^{13.5}$\msun halos.
    \item The hydrodynamical resolution of TNG50 within cold clouds, in contrast to the low-density background hot halo, is high: their cores are resolved by gas cells as small as 200 physical parsecs, increasing to $\sim 500$\,pc in the cloud-interface regions.
    \item Cold CGM clouds are not in (thermal) pressure equilibrium. They are severely thermally underpressurized, but have enormous magnetic pressure, with plasma $\beta = P_{\rm gas} / P_{\rm B} \ll 1$ on average, a natural consequence of flux freezing during collapse. The initial formation, non-linear evolution, and long-term morphology of cold gas may be strongly influenced by the presence of magnetic fields. 
    \item Cold CGM clouds in TNG50 have typical radii of $\sim 0.5 - 1$ kpc, and sub-solar metallicities of $\sim Z_\odot / 2$. They are predominantly infalling towards the halo center, not outflowing. Their internal structure reveals cold cores surrounded by a short $t_{\rm cool}$, inflowing, intermediate temperature gas phase. This pressure-driven, local cooling flow leads to cold-phase growth in the CGM.
    \item A Lagrangian tracer analysis shows that these cold clouds are seeded by strong local density perturbations with $\delta \rho / \bar{\rho} \gg 1$, triggering rapid cooling via thermal instability. These overdensities act as `seeds' and stimulate additional cooling as they accrete through the halo. These initial seeds commonly arise due to perturbations from infalling, tidally and ram-pressure stripped satellite galaxies.
    \item Although the classic $t_{\rm cool} / t_{\rm ff} < \{1, 10\}$ criteria are not globally satisfied in the hot halo because of weak entropy stratification/gradients, perturbations are \textit{locally} surrounded by an intermediate phase of mixed, rapidly cooling $t_{\rm cool} / t_{\rm ff} < 10$ gas.
    \item After formation, cold clouds undergo repeated collisions (mergers) with other cold clouds, as well as fragmentation. The gas which makes up the cold-phase at any moment of time has not been cold long: clouds continually mix with the background medium, leading to a rapid cycling between hot and cold phases. Chemical mixing is, however, incomplete -- cold clouds are typically metal enriched with respect to the local background of hot halo gas.
    \item Cold clouds are not efficiently disrupted or mixed away. In contrast, they are long-lived and typically survive over cosmological, $\gtrsim$\,Gyr timescales.
    \item The abundance of cold gas as well as its structure is a strong function of numerical resolution. The total number of cold clouds continues to increase for higher resolution simulations, resulting in higher covering fractions. It is unlikely that TNG50 has reached the convergent state of the cold-phase CGM.
    \item Contrasting against observational constraints we broadly find that TNG50 produces the sufficiently high covering fractions of extended, cold gas as seen in data. Preliminary comparisons of predicted HI column densities with the LRG-RDR survey, as well as HI and MgII columns from the COS-LRG survey, indicate moderate agreement, while more robust conclusions will require better statistics. The covering fractions of MgII around SDSS LRGs, from $\sim 20$ kpc to $\sim 500$ kpc, are also reasonably consistent, agreeing to 10\% or better even at large radii. In the future more robust comparisons will be enabled by realistic spectral mocks.
\end{itemize}

Modern high-resolution cosmological simulations like TNG50 appear to naturally reproduce the surprising observation that massive quenched galaxies are surrounded by large amounts of cold gas in their circumgalactic medium. At the same time, the issue of convergence remains difficult. TNG50 demonstrates that its $m_{\rm gas} \simeq 8 \times 10^4$\msun resolution is a minimum to resolve the observed cold-phase structure, but we are working at its limits. We must ultimately determine the convergent properties of the cold CGM, confirm the observed origin and survivability characteristics, and understand if small-scale structure is truly important, for the CGM and the galaxy itself.

Future simulations with even higher dynamic range will make it possible to access spatial scales previously reserved for idealized, non-cosmological cloud-scale experiments. They will also enable explorations of additional physical mechanisms neglected in the TNG model (including thermal conduction, cosmic rays, and radiative effects) to ultimately unravel the physics of the cold-phase of the circumgalactic medium.


\section*{Data Availability}

Data directly related to this publication and its figures is available on request from the corresponding author. The IllustrisTNG simulations themselves are publicly available and accessible at \url{www.tng-project.org/data} \citep{nelson19a}, where the TNG50 simulation will also be made public in the future.

\section*{Acknowledgements}

PS acknowledges a Swarnajayanti Fellowship from the Department of Science and Technology, India (DST/SJF/PSA- 03/2016-17), and a Humboldt fellowship for supporting his sabbatical stay at MPA Garching. FM acknowledges support through the Program “Rita Levi Montalcini” of the Italian MIUR. The primary TNG simulations including TNG50 were realized with compute time granted by the Gauss Centre for Supercomputing (GCS) under GCS Large-Scale Projects GCS-ILLU (2014; PI Springel) and GCS-DWAR (2016; PIs Nelson and Pillepich) on the GCS share of the supercomputer Hazel Hen at the High Performance Computing Center Stuttgart (HLRS). GCS is the alliance of the three national supercomputing centres HLRS (Universit{\"a}t Stuttgart), JSC (Forschungszentrum J{\"u}lich), and LRZ (Bayerische Akademie der Wissenschaften), funded by the German Federal Ministry of Education and Research (BMBF) and the German State Ministries for Research of Baden-W{\"u}rttemberg (MWK), Bayern (StMWFK) and Nordrhein-Westfalen (MIWF). Additional simulations and analysis were carried out on the supercomputers at the Max Planck Computing and Data Facility (MPCDF). 

\bibliographystyle{mnras}
\bibliography{refs}


\appendix
\section{Impact of Magnetic Fields} \label{sec_appendix}

\begin{figure*}
\centering
\includegraphics[angle=0,width=6.5in]{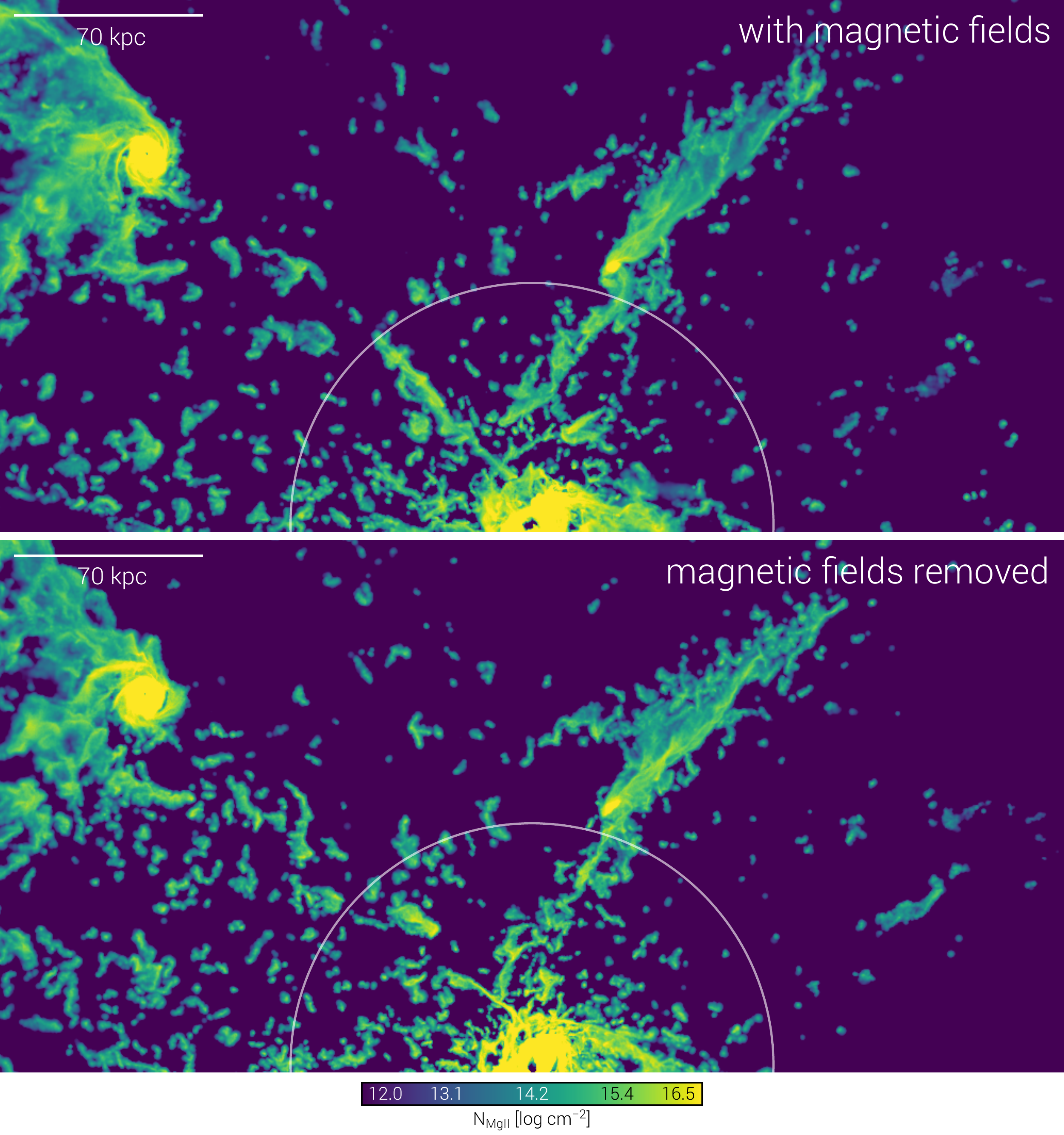}
\caption{ Two test re-simulations of a single halo, designed to assess the impact of magnetic fields on the cold cloud population and properties.
 \label{fig_nomhd}}
\end{figure*}

To assess the role of the MHD we run two additional test simulations of a single halo. Namely, one of the lower mass LRG hosts with $M_{\rm halo} = 10^{13}$\,\msun ($M_\star = 10^{11.3}$\msun) at $z=0.5$. We create multi-mass `zoom' initial conditions (ICs) for a halo from a TNG volume in the usual way, by progressively coarsening numerical resolution outside of the Lagrangian region occupied by this halo of interest. These new ICS are run twice: in the first case we allow the halo to evolve under the fiducial model to $z=0.5$, where it is indistinguishable from the corresponding halo in TNG50-1 from which the zoom ICs have been generated. In a second case, we instantaneously set the magnetic energy density of all gas cells to zero at $z=0.6$, and then evolve the system for the remaining $\sim$\,500\,Myr to $z=0.5$ in the absence of magnetic fields. This is roughly half the halo dynamical time, and sufficient for cold clouds to traverse large fractions of the halo volume in their orbits.

In practice, since total energy is conserved in the code, this `loss' of magnetic energy is compensated by (relabeled as) an increase of thermal energy, such that total pressure is largely invariant. This leads to an impulsive change in the state of the gas, the main impact being an increase in temperature in regions of the strongest magnetic fields. This additional energy should then be quickly lost via radiative cooling. We prefer this approach to any more contrived alternative, such as slowly decreasing the magnetic energy density to zero over some finite time scale.

Furthermore, we make this removal `on the fly' in the evolved halo, rather than in the initial conditions themselves. This could be done either by setting the primordial seed field to zero (amplification then being impossible for ideal MHD), or by switching off the MHD entirely (and thus changing the Riemann solver from the TNG default HLLD). However, magnetic fields have been shown to have non-trivial, indirect coupling to the growth of supermassive black holes and thus to their feedback energetics \citep{pillepich18a}, which we expect would propagate to significant differences in the final $z=0.5$ host halo.

Figure \ref{fig_nomhd} shows a comparison of the MgII column density at $z=0.5$ through the upper half of the halo, the central galaxy sitting at the center of the circle which marks $r_{\rm vir}/4$. The fiducial re-simulation with MHD is shown in the top panel, while the second re-simulation with the magnetic fields disabled at $z=0.6$ is shown in the bottom panel. The same large satellite galaxy is visible in both cases in the same position (upper left).

In the no MHD test case we observe no striking, qualitative differences in the properties or abundance of cold gas in the inner halo. Visual inspection shows that the total number of discrete cold clouds is similar, and that if anything, cold clouds are slightly smaller and denser, with less intermediate density gas, as would be expected if the clouds shrink slightly to reach higher thermal pressures and thus stay roughly in their previous equilibrium state. We conclude that, at TNG50-1 resolution, with the included physics, and over a halo dynamical timescale, magnetic fields do not appear to have a strong impact on the formation, evolution, or disruption of the cold gas cloud population. 

Recent and upcoming observations of the magnetic fields in the CGM of galaxies \citep{mao17,lan20b}, including characterization through fast radio bursts \citep{prochaska19}, will offer direct constraints on the magnetic properties of the circumgalactic medium.

\end{document}